\RequirePackage[l2tabu, orthodox]{nag}

\documentclass[a4paper,onecolumn,final,11pt]{book}

	\begingroup\expandafter\expandafter\expandafter\endgroup
	\expandafter\ifx\csname IncludeInRelease\endcsname\relax
	  \usepackage{fixltx2e}
	\fi

	\usepackage[a4paper]{geometry}
		
	\usepackage[stretch=10]{microtype}

	\usepackage[utf8]{inputenc}
	\usepackage[english	]{babel}
	\usepackage{ordinalpt}
	
	\usepackage{libertine}
	\usepackage{lmodern}
	\usepackage[T1]{fontenc}
	
	\usepackage{booktabs}
	\usepackage{array}
	\usepackage{tabu}

	\usepackage{graphicx}
	\usepackage[font={small}]{caption}
	\usepackage{subcaption}

	\usepackage{amsmath}
	\usepackage{amssymb}
	\allowdisplaybreaks[3]
	\usepackage{amsthm}
	\usepackage{mathtools}
	\usepackage{nicefrac}
	\usepackage{siunitx}
	
	\usepackage[svgnames]{xcolor}
	\definecolor{grayodd}{rgb}{0.92,0.92,0.92}
	\definecolor{grayeven}{rgb}{0.95,0.95,0.95}
				
	\usepackage{listings}
	
	\definecolor{commentgreen}{rgb}{.2,0.6,.2}
	\definecolor{stringmauve}{rgb}{0.58,0,0.82}
	
	\usepackage{inconsolata}

	\usepackage{epstopdf}  
	\usepackage[shortlabels,inline]{enumitem}  
	\usepackage{float}  
	\usepackage{multirow}
	\usepackage{flafter}  
	\usepackage{placeins}  
	\usepackage{physics}  

	\usepackage[makeroom]{cancel}  
	
	\usepackage[super]{nth}  
	\usepackage{amsfonts}

	\definecolor{gp1}{HTML}{e41a1c}
	\definecolor{gp2}{HTML}{377eb8}
	\definecolor{gp3}{HTML}{4daf4a}
	\definecolor{gp4}{HTML}{984ea3}
	\definecolor{gp5}{HTML}{ff7f00}
	\definecolor{gp6}{HTML}{ffff33}
	\definecolor{gp7}{HTML}{a65628}
	\definecolor{gp8}{HTML}{f781bf}
		
	\usepackage{tikz}

	\definecolor{linkblue}{HTML}{0066cc}
	\usepackage[
		unicode,
		colorlinks=true,
		linkcolor=MidnightBlue,
		citecolor=Green,
		urlcolor=linkblue,
	]{hyperref}

\usepackage{todonotes}
\usepackage{pdfpages}
\usepackage{dsfont}
\usepackage[toc,page]{appendix}
\usepackage[backend=biber,style=numeric-comp,sorting=none]{biblatex}
\bibliography{references}
\providecommand{\keywords}[1]{\textbf{Keywords: } #1}
\providecommand{\palavraschave}[1]{\textbf{Palavras-chave: } #1}
\usepackage{xr}
\graphicspath{
	{.} 
	{Z4_Average/}
	{analysis_results/Propagator/plots_extra/}
	{analysis_results/Propagator/plots/}
	{analysis_results/three_gluon/plots/}
	{analysis_results/four_gluon/plots/}
}

\usepackage{tikz-feynman}

\newcommand{\metrica}[2]{g_{\mu_#1\mu_#2}}
\newcommand{\mymom}[2]{{p_#1}_{\mu_#2}}
\newcommand{\mymomp}[1]{p_{\mu_#1}}
\newcommand{\invmom}[2]{{p_#1}^{[#2]}}
\newcommand{\invmomp}[1]{p^{[#1]}}
\newcommand{\gev}{~\si{GeV}}

\usepackage[titles]{tocloft}
\usepackage{cleveref}

\usepackage[acronym,nomain,nonumberlist]{glossaries}
\makenoidxglossaries
\newacronym{qcd}{QCD}{Quantum Chromodynamics}
\newacronym{lqcd}{LQCD}{Lattice Quantum Chromodynamics}
\newacronym{dse}{DSE}{Dyson-Schwinger equation(s)}
\newacronym{ir}{IR}{Infrared}
\newacronym{1pi}{1PI}{One particle irreducible}

\usepackage{etoolbox}
\makeatletter
\patchcmd{\titlepage}{\setcounter{page}\@ne}{}     {}{\PackageError{titlepage}{failed to patch \string\begin{titlepage} to not reset the page counter}{}}
	\patchcmd{\endtitlepage}{\setcounter{page}\@ne}{}  {}{\PackageError{titlepage}{failed to patch \string\end{titlepage} to not reset the page counter}{}}
\makeatother

\newcommand\abstractname{Abstract}  
\makeatletter
\if@titlepage
\newenvironment{abstract}{%
	\titlepage
	\null\vfil
	\@beginparpenalty\@lowpenalty
	\begin{center}%
		\bfseries \abstractname
		\@endparpenalty\@M
\end{center}}%
{\par\vfil\null\endtitlepage}
\else
\newenvironment{abstract}{%
	\if@twocolumn
	\section*{\abstractname}%
	\else
	\small
	\begin{center}%
		{\bfseries \abstractname\vspace{-.5em}\vspace{\z@}}%
	\end{center}%
	\quotation
	\fi}
{\if@twocolumn\else\endquotation\fi}
\fi
\makeatother

\makeatletter
\renewcommand{\maketitle}{
	\begin{titlepage}
		\begin{center}
			\bfseries \Huge
			\@title
		\end{center}
		\vspace{0.5cm}
		\begin{center}
			\@author
		\end{center}
		\vspace{0.5cm}
		\begin{center}
			\Large
			\textit{A thesis submitted for the degree of}\\
			\textit{Master in Physics} 
			\\[0.2cm]
			\includegraphics[scale=0.5]{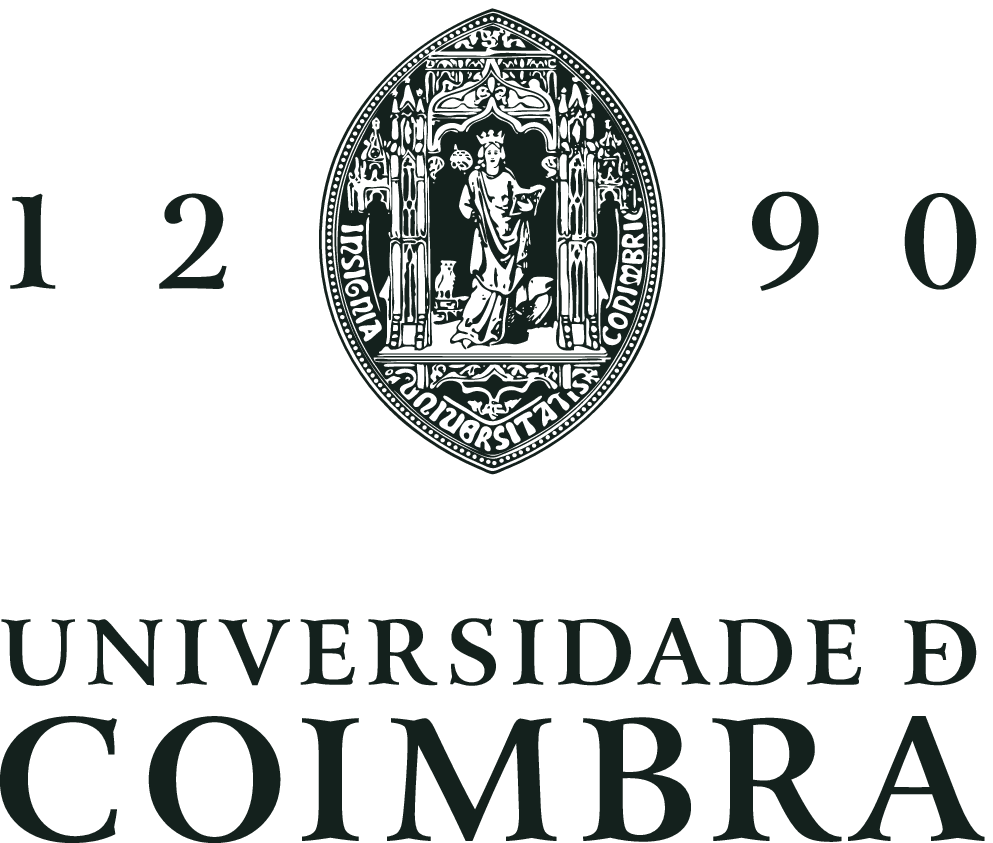}
			\\[0.2cm]
			\Large
			Departmento de Física \\
			Faculdade de Ciência e Tecnologia\\
			Universidade de Coimbra
		\end{center}
		\vspace{0.2cm}
		\begin{center}
			\Large
			\@date
		\end{center}
		\clearpage
	\end{titlepage}
}
\makeatother

\title{Gluon Correlation Functions from Lattice Quantum Chromodynamics}
\author{\LARGE Guilherme Telo Rodrigues Catumba\\[0.5cm]{\large Supervisors:}\\{\large Orlando Oliveira}\\{\large Paulo Silva}}
\date{October 2020}

\begin{document}


\frontmatter

\maketitle

\renewcommand{\abstractname}{Abstract}
\begin{abstract}
	
This dissertation reports on the work developed in the past year by the author and in collaboration with his supervisors, Prof. Dr. Orlando Oliveira and Dr. Paulo Silva.
The main topic of the thesis is the study of the gluon sector in pure Yang-Mills theories via the computation of two, three and four point Landau gauge gluon correlation functions evaluated using the lattice formalism of QCD.  
Monte-Carlo simulations reported herein use the Wilson gauge action for lattice QCD.


The first goal was to understand and quantify the deviations, relative to the usual continuum description of lattice correlation functions, introduced by using appropriate lattice tensors.	
To achieve this we rely on different lattice tensor representations for the gluon propagator in four dimensions to measure the deviations of the lattice propagator from its continuum form. 
We also identified classes of kinematic configurations where these deviations are minimal and the continuum description of lattice tensors is improved. 
Other than testing how faithful our description of the propagator is, these tensor structures also allow to study how the continuum Slavnov-Taylor identity for the propagator is verified on the lattice for the pure Yang-Mills theory. 
We found that the Slavnov-Taylor identity is fulfilled, with good accuracy, by the lattice data for the two point function.


A second goal was the lattice computation of the three gluon vertex using large ensembles of configurations. 
The so-called zero crossing, a property that is related with the ghost dominance at the infrared mass scales and puts restrictions on the behaviour of the three gluon vertex, was investigated.
In addition, we also explore the possible existence of a ghost mass preventing the infrared divergence of the vertex. 
In our study of the three gluon correlation function we used functional forms to model the lattice data and explore the two different possibilities for the behaviour of the function.
For the first case we provide an estimate of the mass scale associated with the zero-crossing and search for a possible sign of the divergence.
On the other hand, for the second case we study the possible occurrence of a sign change and the finite value of the three gluon vertex for vanishing momentum.


A last topic is the computation of the four gluon vertex. 
On the lattice this is a particularly difficult calculation that requires the subtraction of contributions from lower order correlation functions. 
A suitable choice of kinematics allows to eliminate such unwanted contributions.
Furthermore, large statistical fluctuations hinder the precise computation of this object.
Our investigation is a proof of concept, we show that the lattice computation of the four gluon correlation function seems to be feasible with reasonable computational resources.
Nonetheless, an increase in statistics is necessary to provide a clearer and precise signal on the complete correlation function and to compute the corresponding one particle irreducible function. 
		
%

\vspace{0.5cm}
\keywords{Lattice QCD, Gluon propagator, Gluon correlation functions, Lattice tensor representations, Three gluon vertex, Four gluon vertex}
\end{abstract}

\renewcommand{\abstractname}{Resumo}
\begin{abstract}

Esta dissertação é o resultado do trabalho desenvolvido ao longo do último ano pelo autor e juntamente com os seus orientadores, Prof. Dr. Orlando Oliveira e Dr. Paulo Silva. 
A dissertação consiste no estudo do sector gluónico em teorias de Yang-Mills através do cálculo de funções de correlação de dois, três e quatro gluões. Para isto utilizou-se o formalismo da QCD na rede usando simulações de Monte-Carlo com a ação de Wilson na \textit{gauge} de Landau.


O primeiro tópico de estudo passou por analisar os desvios, relativamente ao contínuo, introduzidos pela substituição do espaço-tempo por uma rede de quatro dimensões.
Para isso foram usadas representações tensoriais da rede para calcular o propagador de gluões e comparadas com a descrição tensorial do contínuo.
Com esta análise foram identificadas classes de configurações cinemáticas para as quais os desvios relativamente à descrição do contínuo são reduzidos.
Além de testar a integridade da descrição do propagador, é também possível investigar como a identidade de Slavnov-Taylor para o propagador é validada nas simulações de Monte-Carlo. 
Os resultados das diferentes representações tensoriais mostram que a identidade de Slavnov-Taylor é satisfeita na rede.


A função de correlação de três gluões também foi calculada usando dois conjuntos de configurações na rede.
O objetivo principal foi a análise do comportamento da função de correlação no infra-vermelho,
nomeadamente, a existência de uma possível troca de sinal da função para baixos momentos. 
Esta propriedade relaciona-se com o domínio dos campos \textit{ghost} para baixas escalas de momentos e que induz uma possível mudança de sinal assim como uma possível divergência.
Além desta hipótese, também a possibilidade da existência de uma massa para o campo \textit{ghost} que previne a divergência para baixos momentos foi estudada.
Com o objetivo de melhorar a análise, foram usadas formas funcionais para modelar o vértice de três gluões e estudar as duas possibilidades no infra-vermelho.
Em particular, através dos modelos, a escala para a mudança de sinal foi avaliada assim como o comportamento geral da função para baixos momentos.


O último objetivo foi o cálculo do vértice de quatro gluões, que representa uma dificuldade acrescentada, nunca tendo sido avaliado na rede.
A dificuldade deve-se à complexidade tensorial e às contribuições de vértices de ordem menor que surgem na computação da função de correlação completa de quatro gluões. Estas contribuições foram eliminadas através de uma escolha adequada da configuração cinemática.
Além disso, as flutuações estatísticas são grandes e dificultam a análise.
Os resultados demonstraram que o cálculo do vértice de quatro gluões é exequível com recursos computacionais acessíveis.
No entanto, é fundamental aumentar a precisão no cálculo para obter um sinal mais definido e calcular o vértice sem propagadores externos.
		
\vspace{0.5cm}
\palavraschave{QCD na rede, Propagador do gluão, Funções de correlação de gluões, Representações tensoriais na rede, Vértice de três gluões, Vértice de quatro gluões}
\end{abstract}

\renewcommand{\abstractname}{Acknowledgements}

\begin{abstract}

\textcolor{white}{`A  spectre  is  haunting  Europe...'}

\noindent	
I would like to begin by thanking my supervisors for their exceptional support over the past year. Both Prof. Dr. Orlando Oliveira and Dr. Paulo Silva were very patient and receptive towards my questions and their attentive guidance was certainly very important. I am grateful for their insight and improvements towards the construction of this dissertation.

Moreover, I would like to thank all my cherished friends whose company throughout the past years was fundamental to my growth and without whom this journey would have been much more tedious. 
A special thanks to all my friends in BiF for the company, affection and all the shared adventures. Likewise, to my childhood friends, thank you for being caring and for the company throughout this journey.

Finally, I wish to express my deepest gratitude to my mother for the strenuous care and dedication.
\\[0.1cm]

This work was granted access to the HPC resources
of the PDC Center for High Performance Computing at the KTH Royal
Institute of Technology,
Sweden, made available within the Distributed European Computing
Initiative by the PRACE-2IP,
receiving funding from the European Community’s Seventh Framework
Programme (FP7/2007–2013)
under grand agreement no. RI-283493. The use of Lindgren has been
provided under DECI-9 project
COIMBRALATT. The author acknowledges that the results of this research
have been achieved using
the PRACE-3IP project (FP7 RI312763) resource Sisu based in Finland at
CSC. The use of Sisu has
been provided under DECI-12 project COIMBRALATT2.

It is also important to acknowledge the Laboratory for Advanced Computing at
University of Coimbra for providing HPC resources that have contributed
to the research results reported within this thesis.

This work was supported with funds from Fundação para a Ciência e Tecnologia under the projects UID/FIS/04564/2019 and UIDB/04564/2020.

\end{abstract}


\tableofcontents

\addcontentsline{toc}{chapter}{\listfigurename}
\listoffigures

\addcontentsline{toc}{chapter}{\listtablename}
\listoftables

\addcontentsline{toc}{chapter}{Acronyms}
\glsaddall
\printnoidxglossary

\pagebreak

\mainmatter
\section*{Units and Conventions}
In this dissertation we use natural units
\begin{equation*}
	\hbar = c = 1
\end{equation*}
where $\hbar$ is the reduced Planck constant and $c$ the speed of light in the vacuum. In these units energy, momentum and mass have the same units -- expressed in $\si{MeV} ~(1.6022\times 10^{-13}~\si{\joule})$. Length and time also have common units, inverse of energy.
To re-establish units, the following conversion factor is considered
\begin{equation*}
	\hbar c = 197.326~\si{MeV}~\si{fm} = 1
\end{equation*}
and in SI units
\begin{align*}
	1~\si{MeV} &= 1.7827\times 10^{-30}~\si{\kilogram} \\
	1~\si{fm}  &= 3.3356\times 10^{-24}~\si{s}.
\end{align*}

Greek indices ($\mu,\nu,\rho,$ etc) are associated with space-time indices going through $(0,1,2,3)$ or $(1,2,3,4)$ for Minkowski and Euclidean space, respectively. 
The $g_{\mu\nu}$ symbol is reserved for the Minkowski metric tensor $g_{\mu\nu} = \text{diag}(1,-1,-1,-1)$ while the Kronecker symbol $\delta_{\mu\nu}$ is the Euclidean metric tensor.
Latin indices ($a,b,$ etc) are usually reserved for the colour degrees of freedom associated with the $SU(N)$ algebra.

The Einstein summation convention for repeated indices 
\begin{equation}
	a_\mu b^\mu \equiv \sum_{\mu} a_\mu b^\mu
\end{equation}
is used throughout the work, unless explicitly noted. This convention applies to both space-time and colour degrees of freedom. The position of the indices is irrelevant when considering colour, or Euclidean metric.

\chapter*{Introduction}\label{chap:intro}
\addcontentsline{toc}{chapter}{\nameref{chap:intro}}

The modern description of the fundamental interactions in nature considers four interactions: gravitational, electromagnetic, weak, and strong. Apart from the gravitational interaction which does not have a proper quantum formulation, the last three are described by quantum field theories. 
These three fundamental interactions define what is called the Standard Model, a gauge theory associated with the symmetry group $SU(3) \otimes SU(2) \otimes U(1)$ describing current particle physics. 

The $SU(2) \otimes U(1)$ sector of the Standard Model contemplates the electromagnetic and weak interactions (electroweak) \cite{halzen1984quarks}. Perturbation theory accounts for most of the phenomena occurring in this sector. 
When the physical processes involve hadrons through the strong force (e.g. protons, neutrons, pions) for low energy processes, perturbation theory fails. Hence, non-perturbative methods are necessary to study the $SU(3)$ sector which accounts for the dynamics of quarks and gluons. Quantum chromodynamics (QCD) is the current description of the strong interaction.
 
Lattice field theory is a possible non-perturbative approach to formulate QCD.
The formulation of the theory on a discretized lattice with finite spacing and volume provides a regularization, which renders the theory finite. 
When combined with the Euclidean space-time, lattice field theories become formally equivalent to classical statistical theories. Hence, other than serving as a regularized formulation of the theory it also serves as a computational tool. 
In lattice quantum chromodynamics (LQCD), physical quantities are computed using Monte-Carlo simulations that require large computational power. Current simulations can reach a satisfying level of precision in the computation of several quantities such as the strong coupling constant, hadron masses, and also the study of some properties such as confinement and chiral symmetry (see \cite{wuhan} for a summary of the current advances and investigations in the field). 
	
All of the work developed in this thesis uses the pure Yang-Mills theory, where the fermion dynamics is not taken into account -- \textit{quenched approximation}. This corresponds to disregarding quark loops in the diagrammatic expansion. 
Although this approximation seems too radical, the systematic errors involved are small \cite{Aoki_2003}.

A quantum field theory is defined by its  correlation functions \cite{bailin1993,Peskin:1995}, summarizing the dynamics and interactions among fields. 
Despite not being physical observables and not experimentally detectable, due to its gauge dependency, correlation functions are important for they can be related to various phenomena of the theory. 
Indeed, in supposedly confining theories such as QCD whose quanta (quarks, gluons, and the unphysical ghosts) do not represent physically observable states, correlation functions should encode information on this phenomenon \cite{kugo_confinement,Zwanziger_2004}. Vertices can also serve to compute the coupling constant and define a static potential between colour charges \cite{Bali_coulingconstant,parrinelo_3gluon}, and also explore properties of bound states \cite{Eichmann_2016}. Correlation functions are also the building blocks of other non-perturbative continuum approaches such as the Dyson-Schwinger equations (DSE) \cite{Fischer_2006}. These frameworks usually partially rely on lattice data, and thus a good comprehension of these objects is important.

This thesis addresses three different topics. Firstly, we investigate the lattice gluon propagator relying on lattice tensor representations with the aim to understand the deviations of correlation functions relative to the continuum theory \cite{vujinovic2019tensor, Vujinovi_2019}. 
This has become a relevant topic as modern computations of the gluon propagator use large statistical ensembles of configurations.

The second objective is to compute the three gluon vertex and study its infrared (IR) behaviour. 
The purpose of this analysis is to search for evidences and shorten the estimated interval of the \textit{zero-crossing}, corresponding to a possible sign change of the three gluon one particle irreducible (1PI) function for low momentum. This property can be traced back to the fundamental dynamics of the pure Yang-Mills theory, namely the ghost dynamics as predicted by the DSEs \cite{Aguilar_2014,DSE_3gluon_binosi}. 
In this framework, the sign change is necessary for the finiteness of the equations assuming a tree level form of the ghost-gluon, and four gluon vertex \cite{Eichmann_3gluon}.
Various DSE investigations \cite{Blum_3gluonDSE,Eichmann_3gluon} as well as other methods \cite{pelaez_3gluon,Campagnari_3gluon_coulombgauge} found the zero-crossing for the deep IR.
Recent lattice $SU(3)$ studies \cite{anthony_2016,ATHENODOROU2016444,bocaud_refining_zerocrossing} as well as $SU(2)$ \cite{Cucchieri_2008,maas2020threegluon} predict the zero crossing for the deep infrared region, around $150-250~\si{MeV}$.
Moreover, the exact momentum of the crossing seems to be dependent on the group symmetry and dimensionality, being generally lower for the four-dimensional case \cite{Aguilar_2014}. 
Additionally, general predictions come from pure Yang-Mills theories and thus unquenching the theory could spoil this behaviour. 
However, several DSE based references \cite{pelaez_3gluon,Cyrol_threegluon,Eichmann_3gluon} argue this is a pure gluon phenomenon, and that the presence of light mesons \cite{Unquenching_zerocrossing,Aguilar_2020} only shifts the zero-crossing momentum to a lower IR region.

From the point of view of continuum frameworks, this property is highly dependent on the approximations employed and thus should always be validated by lattice simulations. The latter usually suffer from large fluctuations, or from difficult access to IR momenta.
Furthermore, a recent analytical investigation on both the gluon and ghost propagators found evidence of the existence of a non-vanishing ghost mass which could regularize the three gluon vertex, thus removing the divergence \cite{alex2020analytic}.
While the existence of a dynamical gluon mass is properly established in previous investigations \cite{papavassiliou2013effective}, the case of the ghost field is undetermined. The existence of a finite dynamical ghost mass would in principle remove the logarithmic divergence and thus we also explore this possibility.

The last objective of this work is to perform a first lattice computation of the four gluon correlation function. General predictions for the IR structure of this vertex exist only from continuum formulations \cite{Binosi_2014,Huber_2015}.
These are dependent on truncation schemes and other approximations and again lattice results are needed to validate the predictions. 
The four gluon vertex has four Lorentz indices and four colour indices, therefore its tensor structure is rather complex, allowing for a large number of possible tensors. The increased statistical fluctuations are related to it being a higher order correlation function, involving fields at four distinct lattice sites.	
Besides, as a higher order function, its computation requires the removal of unwanted contributions from lower order correlation functions. These can be eliminated by a suitable choice of kinematics.	
	
The outline of this dissertation begins with a general introduction to the necessary tools and theoretical basis to understand the lattice formulation and results. Chapter \ref{chap:qft} begins with a brief description of the formalism for a general quantum field theory with the QCD theory being introduced and its properties briefly reviewed. Correlation functions and other objects of the theory are introduced.

The lattice formulation of QCD is presented in chapter \ref{chap:lqcd}. We motivate and construct the discretization procedure and present the lattice version of various fundamental objects. This chapter also includes some computational aspects needed to  perform lattice simulations. 

In \cref{chap:tensorbases} the main work of this dissertation begins with an analysis of the correct lattice symmetries and the construction of lattice adequate tensor bases. 
Additionally, details about discretization effects, possible correction methods and tensor bases for the three and four gluon correlation functions are introduced.

Results are shown in chapter \ref{chap:results} which is divided in three main sections, dedicated to each of the three main objectives of this work.  
This is followed by final conclusions and possible extensions for this work.

Finally, the results obtained in this thesis regarding the tensor structure of the propagator were summarized in \cite{Catumba2021hcx}.

\chapter{Quantum Field Theory}\label{chap:qft}
Quantum Chromodynamics is a $SU(3)$ gauge theory. Historically, the colour quantum number was introduced in order to reconcile Fermi statistics with the observed ground state of strongly interacting particles. A new quantum number was needed to guarantee the anti-symmetry of the wave-function \cite{halzen1984quarks}. Later, these new degrees of freedom were found to be associated with a gauge theory.

In this chapter we give a brief overview of QCD and how the theory arises from the principle of gauge invariance. Some important concepts in a quantum field theory are also presented. Quantum field theories are well described in \cite{Peskin:1995,ramond1997field,Srednicki:2007}, and QCD is thoroughly exposed in \cite{muta2010foundations}.

\section{QCD Lagrangian -- Gauge invariance}

The Lagrangian of QCD involves the matter, quark fields $\psi$ and the gluon fields $A_\mu$. The first form a representation of the group symmetry, namely the fundamental representation of $SU(3)$, while the latter are in the adjoint representation of the group (see \cref{apend:liegroups}).

The classical QCD Lagrangian arises when we impose gauge invariance to the Dirac Lagrangian 
\begin{equation}
	\mathcal{L}_\text{Dirac} = \bar{\psi}\left(i\gamma^{\mu}\partial_\mu  - m\right)\psi.
\end{equation}
where $\bar\psi = \psi^\dagger\gamma^0$ with $\gamma^0$ being the zeroth Dirac matrix, $\gamma^\mu$.
For a general $SU(N)$ theory, the gauge principle requires the invariance of the Lagrangian under a local group transformation 
\begin{equation}
\psi(x) \rightarrow \psi'(x) = V(x)\psi(x)
\end{equation}
with $V(x)$ an element of the fundamental representation of the group. When performing a local transformation, the kinetic term of the Lagrangian  breaks the invariance since it compares fields at different points with distinct transformation laws
\begin{equation}
\psi(y) - \psi(x) \rightarrow V(y)\psi(y) - V(x)\psi(x).
\end{equation}
In order to make comparisons at different points we introduce the group valued comparator $U(x,y)$ satisfying $U(x,x) = \mathds{1}$ and the gauge transformation
\begin{equation}
U(x,y) \rightarrow V(x)U(x,y)V^\dagger(y).
\label{eq:U_transformation}
\end{equation}
With this object we may define the \textit{covariant derivative}, using the following difference,
\begin{equation}
D_\mu\psi(x) \equiv \lim\limits_{\varepsilon_\mu \rightarrow 0} \frac{1}{\varepsilon}\left[U(x,x+\varepsilon)\psi(x + \varepsilon) - \psi(x)\right].
\end{equation}
with $y=x+\varepsilon$, and $\varepsilon$ an infinitesimal.
With this definition, the new derivative transforms similarly to the fields,
\begin{equation}
D_\mu\psi(x) \rightarrow V(x)D_\mu\psi(x).
\end{equation}
Introducing a new field, the connection $A_\mu(x)$, by
\begin{equation}
U(x,x+\varepsilon) = \mathds{1} - ig\varepsilon^\mu A_\mu(x) + \order{\varepsilon^2}.
\label{eq:comparator_expansion}
\end{equation}
where $g$ is the bare strong coupling constant, we write the covariant derivative as
\begin{equation}
D_\mu\psi(x) = (\partial_\mu - igA_\mu(x))\psi(x).
\end{equation}
The transformation law for the newly introduced field $A_\mu(x)$ is
\begin{equation}
A_\mu(x) \rightarrow V(x)A_\mu(x) V^{-1}(x) - \frac{i}{g}(\partial_\mu V(x))V^{-1}(x).
\label{eq:gaugetransform_finite}
\end{equation}

An arbitrary group element $V(x)$ can be expressed by the Lie algebra elements through the exponentiation mapping
\begin{equation}
V(x) = \exp(i\alpha^a(x)t^a)
\end{equation}
with the algebra generators $t^a$ defined in appendix \ref{apend:liegroups} and $\alpha^a(x)$ a set of functions parametrizing the transformation. The connection $A_\mu(x)$ is thus an element of the algebra which can be written in terms of the fields $A_\mu^a(x)$ 
\begin{equation}
A_\mu(x)=A_\mu^a(x)t^a.
\end{equation}

Hence, to guarantee gauge invariance of the Dirac Lagrangian we replace normal derivatives by the covariant.
Furthermore, we need to introduce a kinetic term for the new field that must depend only on the gauge fields $A_\mu$ and its derivatives. The usual construction is the field-strength tensor
\begin{equation}
F_{\mu\nu} = \frac{i}{g}\left[D_\mu,D_\nu\right] = (\partial_\mu A_\nu - \partial_\nu A_\mu) - ig\left[A_\mu,A_\nu\right]
\label{eq:fieldstrengthtensor}
\end{equation}
which can be written in terms of its components $F_{\mu\nu} = F_{\mu\nu}^at^a$ using the structure constants of the group $f^{abc}$,
\begin{equation}
F_{\mu\nu}^a = (\partial_\mu A_\nu^a - \partial_\nu A_\mu^a) + gf^{abc}A_\mu^bA_\nu^c.
\label{eq:fieldstregthtensor_explicit}
\end{equation}
The first equality in \ref{eq:fieldstrengthtensor} gives a geometrical interpretation of the tensor, as it can be seen as the comparison of the field around an infinitesimal square loop in the $\mu - \nu$ plane, indicating how much it rotates in the internal space when translated along this path \cite{Peskin:1995}. To obtain a gauge invariant scalar object from this tensor, we consider the trace operation over the algebra elements and the following contraction
\begin{equation}
	\Tr\left[(F_{\mu\nu}^at^a)^2\right] = (F_{\mu\nu}^a)^2/2.
\end{equation} 

With these elements we write the classical QCD Lagrangian
\begin{equation}
\mathcal{L}_{\text{QCD}} = -\frac{1}{4}F_{\mu\nu}^{a}F^{a\mu\nu} + \bar{\psi}\left(i\gamma^{\mu}(\partial_\mu - igA_\mu^at^a) - m\right)\psi
\label{eq:qcd-lagrangian}
\end{equation}
whose form, namely the gluon-quark interaction is restricted by gauge invariance\footnote{Gauge invariance also restricts the gauge fields to be massless since the term $A_\mu^aA_\mu^a$ is not gauge invariant.}.
The matter field $\psi(x)$ is a vector of spinors for each flavour of quark ($f = u,d,s,c,t,b$). Each quark flavour has an additional colour index $a = 1,2,3$ in a three dimensional representation of the $SU(3)$ group. $m$ is a diagonal matrix in flavour space containing the bare quark masses for each flavour. 
The eight independent gluon fields associated with the group generators are the gauge fields $A_\mu^a(x)$ which also carry a Lorentz index, labelling the corresponding directions in space-time, $\mu = 0,1,2,3$.

For the present work, we are interested in the pure Yang-Mills Lagrangian involving the gluon dynamics only
\begin{equation}
	\mathcal{L}_{\text{YM}} = -\frac{1}{4}F_{\mu\nu}^{a}F^{a\mu\nu}.
\end{equation}

\section{Quantization of the theory}\label{sec:quantization of the theory}

In the path integral quantization  for a general quantum field theory \cite{Peskin:1995,Schwartz2014,bailin1993}, described by a set of fields $\phi_a$\footnote{The index $a$ may represent independent fields, different members of a set of fields related by some internal symmetry, or the components of a field transforming non-trivially under Lorentz transformation, e.g., a vector.}, the theory is defined by the generating functional
\begin{equation}
\mathcal{Z}[J]	= \int\mathcal{D}\phi e^{i\int d^4x\left(\mathcal{L} + J_a(x)\phi_a(x)\right)}
\end{equation}
where $J_a(x)$ is an external source, and the condensed notation was employed
\begin{equation}
\mathcal{D}\phi \equiv \prod_{x,a}d\phi_a(x).
\end{equation}
A quantum field theory is completely determined by its Green's functions \cite{bailin1993,Peskin:1995} defined as
\begin{equation}
G_{i_1,...,i_n}^{(n)}(x_1,...,x_n) = \bra{0} T\left[\hat\phi_{i_1}(x_1)...\hat\phi_{i_n}(x_n)\right] \ket{0}
\end{equation}
i.e. by a time ordered vacuum expectation value of the product of $n$ field operators at distinct points. In this quantization procedure, Green's functions are computed from the generating functional by functional differentiation with respect to the sources
\begin{equation}
\bra{0} T\left[\hat\phi_{i_1}(x_1)...\hat\phi_{i_n}(x_n)\right] \ket{0} = \eval{\frac{1}{i^n\mathcal{Z}[J]}\frac{\delta^n\mathcal{Z}[J]}{\delta J_{i_1}(x_1)...\delta J_{i_n}(x_n)}}_{J = 0}.
\label{eq:correlation_function}
\end{equation}
This vacuum expectation value can thus be written as
\begin{equation}
\expval{\hat\phi_{i_1}(x_1)...\hat\phi_{i_n}(x_n)} = \frac{1}{\mathcal{Z}[0]}\int\mathcal{D}\phi \left(\phi_{i_1}(x_1)...\hat\phi_{i_n}(x_n)\right)e^{iS}
\label{eq:correlation_function_integral}
\end{equation}
with the notation $\expval{\hat\phi_{i_n}(x_n)...\hat\phi_{i_n}(x_n)} \equiv \bra{0} T\left[\hat\phi_{i_n}(x_n)...\hat\phi_{i_n}(x_n)\right] \ket{0}$. \Cref{eq:correlation_function_integral} shows that Green's functions are accessed by performing a weighted average over all possible configurations of the system.

The path integral quantization carries some problems when applied to gauge theories. The generating functional
\begin{equation}
\mathcal{Z} = \int\mathcal{D}A e^{iS[A]}.
\end{equation}
involves the integral over the gauge fields $A_\mu^a(x)$. For any field configuration $A_\mu$ we may define a gauge orbit to be the set of all fields related to the first by a gauge transformation $\alpha$.
All these configurations have the same contribution to the functional integral, and so constitute an infinite contribution.

The over counting of these degrees of freedom need to be eliminated in order to have a well defined theory. Faddeev and Popov \cite{Faddeev1967} suggested the use of a hypersurface to restrict the integration in configuration space. This is achieved by a gauge fixing condition of the form $F^a[A] - C^a(x) = 0$\footnote{$F[A]$ is a field dependent term. $C^a(x)$ is a set of functions also determining the gauge fixing condition. $F[A] = \partial_\mu A^\mu(x)$ and $C^a(x)=0$ in the Landau gauge.}. This way we isolate the contribution over repeated configurations by factorizing it as $\int \mathcal{D}\alpha \int \mathcal{D}A_\mu \exp^{iS[A]}$, being eliminated by the normalization.

To impose this integration restriction we insert the following expression in the generating functional,
\begin{equation}
1 = \int\mathcal{D}\alpha\delta(F^a[A^\alpha]- C^a(x))\det\left(\frac{\delta F^a[A^\alpha]}{\delta\alpha}\right)
\label{eq:identity faddeev}
\end{equation}
where $A^\alpha$ represents the gauge transformed field $A$, $\delta(F[A^\alpha])$ is a Dirac $\delta$ over each space-time point, and the determinant is due to the change of variables. The generating functional reads
\begin{equation}
\mathcal{Z} = \int \mathcal{D}A\int\mathcal{D}\alpha\delta(F^a[A^\alpha]- C^a(x))\det\left(\frac{\delta F[A^\alpha]}{\delta\alpha}\right)e^{iS[A]}.
\end{equation}
Performing a gauge transformation from $A_\mu^\alpha$ to $A_\mu$ we can eliminate the dependence on the gauge transformation from the integrand. For this we use the gauge invariance of the action and of the volume element in group space $\mathcal{D}\alpha$ \cite{Gattringer2010}. Also, an unitary transformation leaves the measure $\mathcal{D}A$ and the determinant unchanged
\begin{equation}
\mathcal{Z} = \int\mathcal{D}\alpha \int\mathcal{D}A \delta(F^a[A]- C^a(x))\det\left(\frac{\delta F[A]}{\delta\alpha}\right)e^{iS[A]}.
\end{equation}
This way we factorized the infinite factor, which is eliminated by normalization. In addition, we may multiply $\mathcal{Z}$ by a constant factor
\begin{equation}
\int\mathcal{D}C\exp\left[-\frac{i}{2\xi}\int d^4x {C^a}^2\right]
\end{equation}
corresponding to a linear combination of different Gaussian weighted functions $C^a$. The generating functional now reads
\begin{equation}
\mathcal{Z} =\int\mathcal{D}A \det\left(\frac{\delta F[A]}{\delta\alpha}\right)\exp{iS[A]- \frac{i}{2\xi}\int d^4x F[A]^2}.
\end{equation}
The Faddeev-Popov determinant is defined as
\begin{align}
\det M = \det \left( \frac{\delta F([A],x)}{\delta\alpha(y)} \right), &&
M_{ab}([A], x, y) = \frac{\delta F^a([A],x)}{\delta\alpha^b(y)} \label{eq:ghostmatrix}
\end{align}
Using Grassmann, anti-commuting variables it is possible to define the Faddeev-Popov determinant as a functional integral over a set of anti-commuting fields -- ghost fields $\bar\eta,\eta$
\begin{equation}
\det M = \int\mathcal{D}\bar\eta\mathcal{D}\eta \exp\left(-i\int d^4x \bar\eta^aM_{ab}\eta^b\right).
\end{equation}
With this, we have a final form for the generating functional,
\begin{equation}
\mathcal{Z} = \int\mathcal{D}A_\mu\mathcal{D}\bar\eta\mathcal{D}\eta e^{i\int d^4x\mathcal{L}_\text{eff}},
\end{equation}
expressed with an effective Lagrangian
\begin{equation}
\mathcal{L}_\text{eff} = \mathcal{L} - \frac{F^2}{2\xi} - \bar\eta M\eta.
\end{equation}
These new anti-commuting fields can be interpreted as new particles contributing to the dynamics of the system.
However, being scalars under Lorentz transformations while anti-commuting fields, ghosts do not respect the spin-statistics theorem \cite{weinberg1995quantum} and cannot be interpreted as physical particles -- only contributing to closed loops in Feynman diagrams and never as external fields. They are a mathematical artifact resulting from the gauge fixing procedure.

\section{Propagator and vertices}

The effective Yang-Mills Lagrangian is
\begin{align}
\mathcal{L} &= \frac{1}{2}( \partial^\mu A^{a\nu}\partial_\nu A_\mu^a - \partial^\mu A^{a\nu}\partial_\mu A_\nu^a ) - \frac{1}{2\xi}(\partial^\mu A_\mu)^2 \nonumber \\
&-\frac{1}{2}gf^{abc}A^{b\mu}A^{c\nu}(\partial_\mu A_\nu^a - \partial_\nu A_\mu^a) \nonumber \\
&- \frac{1}{4}g^2f^{abc}f^{ade} A^{b\mu}A^{c\nu}A_\mu^dA_\nu^e \nonumber \\
&- \bar\eta^a\partial^\mu(\partial_\mu - gf^{abc}A_\mu^a)\eta^b.
\end{align}
Analytically, the computation of the complete correlation functions (Green's functions) is not possible. However, perturbation theory can provide some information on the form of these functions. For this we need to know the Feynman rules for the theory, which can be read off from the Lagrangian at tree level and are summarized in this section. Its derivation can be consulted in \cite{Peskin:1995, ryderquantum}.

The gluon propagator is read off from the quadratic terms in the gluon fields in the Lagrangian. In momentum space, the propagator reads
\begin{equation}
D_{\mu\nu}^{ab}(p^2) = \frac{\delta^{ab}}{p^2}\left[g_{\mu\nu} + (\xi - 1)\frac{p_\mu p_\nu}{p^2}\right].
\end{equation}
Note that $\xi = 0$ in the Landau gauge. 

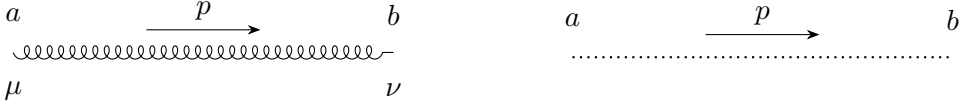
\begin{figure}[!htb]
	\begin{minipage}{.5\columnwidth}
		\begin{center}
			\begin{tikzpicture}
			everyblob={draw=green!40!black, patterncolor=green!40!black}
			\begin{feynman}
			\vertex (a1);
			\vertex[right=5cm of a1] (a2);
			
			\diagram* {
				(a1) -- [gluon, momentum ={[arrow shorten = 0.35]\(p\)}] (a2)
			};
			\end{feynman}
			\node at (0,0.5) {$a$};
			\node at (5,0.5) {$b$};
			\node at (0,-0.5) {$\mu$};
			\node at (5,-0.5) {$\nu$};
			\end{tikzpicture}
		\end{center}
	\end{minipage}
	\begin{minipage}{.5\columnwidth}
		\begin{center}
			\begin{tikzpicture}
			\begin{feynman}
			\vertex (a1);
			\vertex[right=5cm of a1] (a2);
			
			\diagram* {
				(a1) -- [ghost, momentum ={[arrow shorten = 0.35]\(p\)}] (a2)
			};
			\end{feynman}
			\node at (0,0.5) {$a$};
			\node at (5,0.5) {$b$};
			\node at (0,-0.5) {$~$};
			\node at (5,-0.5) {$~$};
			\end{tikzpicture}
		\end{center}
	\end{minipage}
	\caption{Gluon and ghost propagators.}
	\label{fig:Propagators}
\end{figure}

\begin{figure}[!htb]
	\begin{center}
		\begin{tikzpicture}
		\begin{feynman}[small]
		\vertex (a1);
		\vertex[right=4cm of a1] (a2);
		\vertex[below right = 2cm and 2cm of a1] (a3);
		\vertex[below=2 cm of a3] (a4);
		
		\diagram* {
			(a1) -- [ghost, momentum = {[arrow shorten = 0.35]\(p\)}, edge label=\(a\), near start] (a3),
			(a2) -- [ghost, momentum = {[arrow shorten = 0.35]\(q\)}, edge label=\(b\), near start] (a3),
			(a4) -- [gluon, edge label=\(c~\mu\), near start] (a3)
		};
		\end{feynman}
		
		\end{tikzpicture}
	\end{center}
	\noindent
	\begin{minipage}{.5\columnwidth}
		\begin{center}
			\begin{tikzpicture}
			\begin{feynman}[small]
			\vertex (a1);
			\vertex[right=4cm of a1] (a2);
			\vertex[below right = 2cm and 2cm of a1] (a3);
			\vertex[below=2 cm of a3] (a4);
			
			\diagram* {
				(a1) -- [gluon, edge label=\((p_1~a_1~\mu_1)\), near start] (a3),
				(a2) -- [gluon, edge label=\((p_2~a_2~\mu_2)\), near start] (a3),
				(a4) -- [gluon, edge label=\((p_3~a_3~\mu_3)\), near start] (a3)
			};
			\end{feynman}
			\end{tikzpicture}
		\end{center}
	\end{minipage}
	\begin{minipage}{.5\columnwidth}
		\begin{center}
			\begin{tikzpicture}
			\begin{feynman}[small]
			\vertex (a1);
			\vertex[right= 4cm of a1] (a2);
			\vertex[below= 4cm of a1] (a3);
			\vertex[right= 4cm of a3] (a4);
			\vertex[below right= 2cm and 2cm of a1] (a5);
			
			\diagram* {
				(a1) -- [gluon, edge label=\((a_1~\mu_1)\), near start] (a5),
				(a2) -- [gluon, edge label=\((a_2~\mu_2)\), near start] (a5),
				(a3) -- [gluon, edge label=\((a_3~\mu_3)\), near start] (a5),
				(a4) -- [gluon, edge label=\((a_4~\mu_4)\), near start] (a5)
			};
			\end{feynman}
			\end{tikzpicture}
		\end{center}
	\end{minipage}
	\caption{Ghost-gluon coupling vertex (top) and three and four gluon vertices with all momenta defined inwards.}
	\label{fig:vertices}
\end{figure}
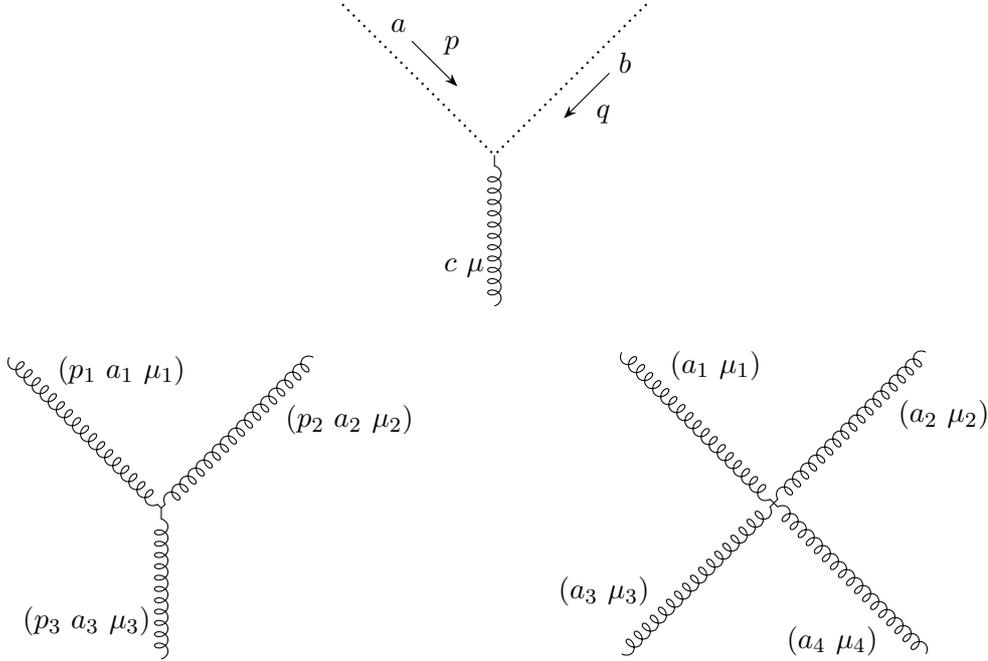

The ghost fields also have associated Feynman rules. In the chosen gauge the functional derivative \eqref{eq:ghostmatrix}, obtained with the infinitesimal version of \eqref{eq:gaugetransform_finite},
\begin{equation}
A'^a_\mu = A_\mu^a + f^{abc} A^b_\mu\alpha^c + \partial_\mu\alpha^a,
\label{eq:gaugetransform_infinitesimal}
\end{equation}
is of the form $M_{ab} = \partial^\mu D_\mu$\footnote{Note that $D_\mu$ here is written in the adjoint representation with the generators $(t^a)_{bc} = -if^{abc}$.}, resulting in a lagrangian contribution
\begin{equation}
\mathcal{L}_\text{ghost} = -\bar \eta^a\partial_\mu\partial^\mu\eta^a + gf^{abc}\bar\eta^a\partial^\mu(A_\mu^b\eta^c).
\end{equation}
The ghost will have an associated tree-level propagator, fig. \ref{fig:Propagators},
\begin{equation}
\Delta^{ab}(p^2) = \frac{\delta^{ab}}{p^2}
\end{equation}
and a ghost-gauge field coupling vertex $-gf^{abc}p_\mu$ represented in figure \ref{fig:vertices}.

The gluon self `interaction' vertices result from the second and third line of the Lagrangian. Their form, however, is written considering the Bose symmetry of the objects, which allow us to interchange each particle $(p_i,a_i,\mu_i)$ without affecting its form. The Feynman rule for the three gluon vertex in momentum space, shown schematically in fig. \ref{fig:vertices}, reads
\begin{equation}
{\Gamma^{(0)}}_{\mu_1\mu_2\mu_3}^{a_1a_2a_3}(p_1,p_2,p_3) = gf^{a_1a_2a_3}[g_{\mu_1\mu_2}(p_1-p_2)_{\mu_3} + g_{\mu_2\mu_3}(p_2-p_3)_{\mu_1} + g_{\mu_3\mu_1}(p_3-p_1)_{\mu_2}]
\end{equation}
whereas for the four gluon vertex the corresponding tree level expression is given by
\begin{align}
{\Gamma^{(0)}}_{\mu_1\mu_2\mu_3\mu4}^{a_1a_2a_3a_4}(p_1,p_2,p_3,p_4) = -g^2\big[&f^{a_1a_2m}f^{a_3a_4m}(g_{\mu_1\mu_3}g_{\mu_2\mu_4} - g_{\mu_1\mu_4}g_{\mu_2\mu_3}) \nonumber \\
&f^{a_1a_3m}f^{a_2a_4m}(g_{\mu_1\mu_2}g_{\mu_3\mu_4} - g_{\mu_1\mu_4}g_{\mu_2\mu_3}) \nonumber \\
&f^{a_1a_4m}f^{a_2a_3m}(g_{\mu_1\mu_2}g_{\mu_3\mu_4} - g_{\mu_1\mu_3}g_{\mu_2\mu_4})
\big].
\end{align}

\section{Complete vertices}

In a non-perturbative framework, we aim to have access to the complete correlation functions whose tensor structure ought to be different from the simple bare vertices obtained at zero order in perturbation theory. 
Hence, we must build the most general structure for each correlation function under the symmetries of the theory.

The tensor structure for the gluon propagator is completely defined by the Slavnov-Taylor identity\footnote{These are relations between the correlation functions which come from the gauge invariance of the theory. They express the symmetries of the classical theory through the quantum expectation values. Also called generalized Ward identities.} and the gauge condition -- see \cite{Peskin:1995,ryderquantum}.
The Landau gauge Slavnov-Taylor identity for the gluon propagator reads \cite{Slavnov1972fg}
\begin{equation}
	\partial^\mu_x\partial^\nu_y \expval{T\{ A_\mu^a(x) A_\nu^b(y) \}} = 0
	\label{eq:slavnov-taylor}
\end{equation}
which fixes the orthogonal form of the propagator. 
Therefore, in the Landau gauge, this results in
\begin{equation}
D_{\mu\nu}^{ab}(p) = \delta^{ab}D(p^2)\left[g_{\mu\nu} - \frac{p_\mu p_\nu}{p^2}\right]
\label{eq:full_continuum_propagator}
\end{equation}
with its coefficient differing from the tree-level form by a form factor $D(p^2)$.

For higher order correlation functions we distinguish the gluon correlation functions $G_{\mu_1...\mu_n}^{a_1...a_n}$ obtained with \eqref{eq:correlation_function} from the pure gluon vertex $\Gamma_{\mu_1...\mu_n}^{a_1...a_n}$ obtained with the removal of the external propagators. For the three gluon vertex we thus define
\begin{align}
&\expval{A_{\mu_1}^{a_1}(p_1)A_{\mu_2}^{a_2}(p_2)A_{\mu_3}^{a_3}(p_3)} = (2\pi)^4\delta(p_1 + p_2 + p_3)G_{\mu_1\mu_2\mu_3}^{a_1a_2a_3}(p_1,p_2,p_3) \\
& G_{\mu_1\mu_2\mu_3}^{a_1a_2a_3}(p_1,p_2,p_3) = D_{\mu_1\nu_1}^{a_1b_1}(p_1)D_{\mu_2\nu_2}^{a_2b_2}(p_2)D_{\mu_3\nu_3}^{a_3b_3}(p_3)\Gamma_{\nu_1\nu_2\nu_3}^{a_1a_2a_3}(p_1,p_2,p_3).
\label{eq:three_gluon_greens_definition}
\end{align}
Analogous expressions can be considered for the four gluon vertex.
\begin{figure}[htb!]
	\begin{minipage}{.5\columnwidth}
		\begin{center}
			\begin{tikzpicture}
			\begin{feynman}[small]
			\vertex (a1);
			\vertex[right=4cm of a1] (a2);
			\vertex[below right = 2cm and 2cm of a1] (a3);
			\vertex[below=2 cm of a3] (a4);
			
			\diagram* {
				(a1) -- [scalar] (a3),
				(a2) -- [scalar] (a3),
				(a4) -- [scalar] (a3)
			};
			\end{feynman}
			\draw[fill=white,thick] (2,-2) circle (0.4cm);
			\draw[pattern=dots,pattern color=black!30] (2,-2) circle (0.4cm);
			\node at (4,-2) {$\Gamma_{\nu_1\nu_2\nu_3}^{a_1a_2a_3}(p_1,p_2,p_3)$};
			\end{tikzpicture}
		\end{center}
	\end{minipage}
	\begin{minipage}{.5\columnwidth}
		\begin{center}
			\begin{tikzpicture}
			\begin{feynman}[small]
			\vertex (a1);
			\vertex[right= 4cm of a1] (a2);
			\vertex[below= 4cm of a1] (a3);
			\vertex[right= 4cm of a3] (a4);
			\vertex[below right= 2cm and 2cm of a1] (a5);
			
			\diagram* {
				(a1) -- [scalar] (a5),
				(a2) -- [scalar] (a5),
				(a3) -- [scalar] (a5),
				(a4) -- [scalar] (a5)
			};
			\end{feynman}
			\draw[fill=white,thick] (2,-2) circle (0.4cm);
			\draw[pattern=dots,pattern color=black!30] (2,-2) circle (0.4cm);
			\node at (4.5,-2) {$\Gamma_{\nu_1\nu_2\nu_3\nu_4}^{a_1a_2a_3a_4}(p_1,p_2,p_3,p_4)$};
			\end{tikzpicture}
		\end{center}
	\end{minipage}
	\caption{Three and four gluon vertices with external propagators removed.}
	\label{fig:vertices_1PI}
\end{figure}
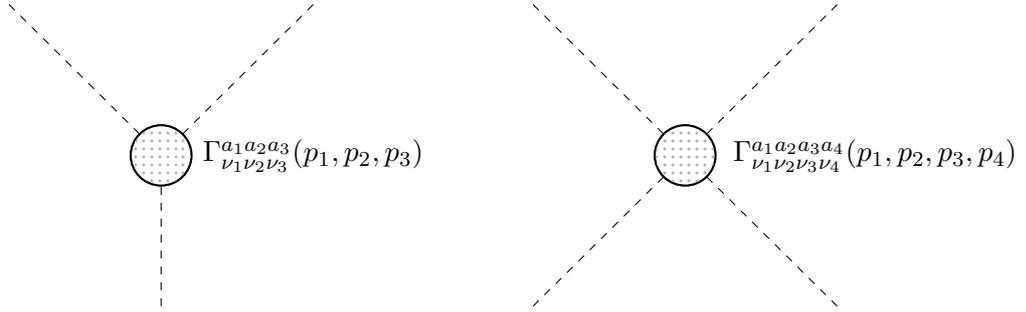
Notice that the average for the three gluon correlation function is computed as 
\begin{equation}
\expval{A_{\mu_1}^{a_1}(x_1)A_{\mu_2}^{a_2}(x_2)A_{\mu_3}^{a_3}(x_3)} = \frac{\int\mathcal{D}AA_{\mu_1}^{a_1}(x_1)A_{\mu_2}^{a_2}(x_2)A_{\mu_3}^{a_3}(x_3)e^{i\int d^4x\mathcal{L}}}{\int\mathcal{D}Ae^{i\int d^4x\mathcal{L}}}.
\end{equation}
To compute these higher order correlation functions we construct their tensor structures by taking into account the symmetries of the system, namely Bose symmetry allowing to freely exchange each pair of indistinguishable particles and their associated quantum numbers. Proceeding this way we construct the most general form for these objects. This construction will be presented in \cref{chap:tensorbases}.

It is also important to make a further distinction between the pure (gluon) vertices $G$ and the one particle irreducible (1PI) functions, $\Gamma$ which do not have the contribution from disconnected diagrams and cannot be reduced to other diagrams by removing a propagator -- see \cite{Peskin:1995,ryderquantum}. 
These are the objects we are interested in obtaining from the lattice -- further details will be given when considering the four gluon vertex in \cref{sec:intro_fourgluon_bases}.

\section{Regularization and Renormalization}

In general, quantum field theories involve divergences other than the ones solved by the Faddeev-Popov method.
These divergences need to be taken care of.

The theory is first regularized, making it finite. 
This is done, in general, by introducing parameters in the theory which absorb the divergences.
In a perturbative approach, this could be done by an ultraviolet momentum cut off or dimensional regularization for example. 
The introduction of a finite space-time lattice with spacing $a$ is a common regularization procedure with the advantage of allowing to perform numerical simulations.

The theory is then renormalized by rescaling the parameters and fields of the theory in a way that the removal of the divergences is not spoiled when the regularization parameter is eliminated. 
 
The rescaling is performed on a finite number of parameters such as the fields, and the fundamental constants of the theory. Following \cite{bailin1993} a possible rescaling procedure for QCD would be
\begin{align}
&A_\mu^a \rightarrow Z_A^{1/2}A_\mu^a,    &   &m\rightarrow Z_mZ_\psi^{-1}m, \\
&\psi \rightarrow Z_\psi^{1/2}\psi,       &   &g\rightarrow Z_g g, \\
&\eta^a \rightarrow Z_\eta^{1/2}\eta^a,   &   &\xi^{-1}\rightarrow Z_\xi Z_A^{-1}\xi^{-1}
\end{align}
where the various $Z_i$ are the necessary renormalization constants to render the theory finite. 

Green's functions have associated rescaling rules constructed from the ones above. 
Considering gauge fields only, the Green's functions renormalization involve $Z_A$. For instance, the renormalized gluon propagator $G^{(2)}_r$ relates to the bare object as $G^{(2)}_r = Z_A G^{(2)}$.

Performing a renormalization procedure involves choosing a point where the quantities are fixed by some given, standard values. The momentum subtraction MOM scheme is a usual choice, it fixes the renormalized Green's function to match the tree level value for a given momentum scale $\mu$. Again, using the gluon propagator, the constant $Z_A$ is found from
\begin{equation}
D(p^2=\mu^2) = Z_A D_L(\mu^2) = \frac{1}{\mu^2}
\end{equation}
where $D(p^2)$ is the renormalized form factor and $D_L(p^2)$ the non-renormalized form factor.
See \cite{zinn2002quantum} for more details, and \cite{Rothe} for a lattice dedicated description.

\chapter{Lattice quantum chromodynamics}\label{chap:lqcd}
In this chapter the formulation of quantum chromodynamics on a finite discretized lattice will be presented.
Lattice QCD provides a formulation which allows to study the non-perturbative regime of QCD and a regularization of the theory. This framework preserves gauge invariance and serves as an explicit computational tool.

This chapter begins with the introduction of the lattice formalism, constructing all objects in the discretized framework. After this, attention will be given to some computational aspects of this work which are necessary to compute lattice quantities. Lattice theories, with emphasis on LQCD are presented in \cite{Wilson1974,Rothe,Gattringer2010}.

\section{Euclidean formulation}\label{sec:euclidean formulation}

The Minkowski space-time is not convenient to study functional path integrals due to the oscillatory behaviour of the exponential in the action. We use imaginary time thus becoming an Euclidean space. This is accomplished by a Wick rotation, where the real time $t$ is rotated by $\pi/2$ into the complex plane, $\tau=it$.
The exponential becomes similar to the Boltzmann factor on the partition function of statistical mechanics,
\begin{equation*}
\int\mathcal{D}\phi e^{iS[\phi]} \rightarrow \int\mathcal{D}e^{-S_E[\phi]}.
\end{equation*}
The object $S_E$ is the Euclidean version of the action, obtained by performing the change of variables above.
This transformation establishes the formal connection with statistical mechanics, allowing its methods to be applied on lattice field theories, notably Monte-Carlo methods to obtain correlation functions. 
In the forthcoming analysis we consider the Euclidean formulation of QCD and the metric is thus equivalent to $\delta_{\mu\nu}$.

%

\section{Discretization}

In the lattice formulation the continuous space-time is replaced by a 4-dimensional Euclidean lattice $\Lambda$ with spacing $a$ whereby each point is labelled by four integers, $n = (n_1,n_2,n_3,n_4)$. We consider $n_4$ to be the imaginary time direction. 
In this work we consider hypercubic lattices, each side having the same number of points, $n_i\in[0,N-1]$.

All objects appearing in the continuum theory must be rewritten on the lattice formulation. For a general quantum field theory with fields $\phi$, the degrees of freedom are the classical fields $\phi(an)$ in the discrete lattice sites. The lattice action must be built in a way that preserves all possible properties of the continuum theory. 
However, the discretization procedure is not unique which can be seen by the structure of the discrete derivative, taking various possible forms,
\begin{align}
&\partial_\mu\phi(x) = \frac{1}{a}\left(\phi(x+\hat\mu a) - \phi(x)\right) + \order{a}\\
&\partial_\mu\phi(x) = \frac{1}{2a}\left(\phi(x+\hat\mu a) - \phi(x-\hat\mu a)\right) + \order{a^2}.
\end{align}
This freedom in obtaining the lattice form can be used to minimize the appearance of lattice artifacts\footnote{This freedom opens the possibility for \textit{improvement schemes} which modify the action in a way to reduce lattice artifacts \cite{HASENFRATZ199853} -- these are not considered in this work.}.  

On the lattice, all possible space translations are restricted to be at least one lattice unit in size. This results in the discretization of the allowed momenta. To see this, consider the usual continuum Fourier transform,
\begin{equation*}
\phi(x) = \int\frac{d^4p}{(2\pi)^4}\tilde\phi(p)e^{ipx}.
\end{equation*}
Since $x = an$ is an integer multiple of the spacing $a$ we get
\begin{equation*}
e^{ip_\mu x_\mu} = e^{i(p_\mu x_\mu + 2\pi n_\mu)} = e^{i(p_\mu + 2\pi/a)x_\mu},
\end{equation*}
hence the momentum $p_\mu$ is equivalent to $p_\mu + 2\pi/a$, allowing us to restrict the momentum integration to the Brillouin zone, $-\pi/a < p_\mu \leq \pi/a$. This removes high frequency modes and regularizes the theory.
Thus, in infinite volume we would write
\begin{equation*}
\phi(x) = \int_{-\pi/a}^{\pi/a}\frac{d^4p}{(2\pi)^4}\tilde\phi(p)e^{ipx}.
\end{equation*}
To perform numerical simulations, however, the volume of the lattice is finite, where we impose boundary conditions, $\phi(x + \hat\mu N_\mu a) = e^{i\theta_\mu}\phi(x)$. The finite volume imposes the additional discretization of momentum. Applying the Fourier transform to this condition
\begin{align*}
\int_{-\pi/a}^{\pi/a}\frac{d^4p}{(2\pi)^4}\tilde\phi(p)e^{ip_\mu(x_\mu + \hat\mu N_\mu a)} &= \int_{-\pi/a}^{\pi/a}\frac{d^4p}{(2\pi)^4}\tilde\phi(p)e^{ip_\mu x_\mu+i\theta_\mu} \\
\Leftrightarrow e^{ip_\mu N_\mu} &= e^{i\theta_\mu}~ (\text{no sum})
\end{align*}
where $\hat\mu$ is an unitary lattice vector in the direction $\mu$.
We work with periodic boundary conditions, thus $\theta_\mu = 0$ and we get the discrete momentum values,
\begin{equation}
p_\mu = \frac{2\pi n_\mu}{aN_\mu}, ~n_\mu\in\{-N_\mu/2+1, ..., N_\mu/2\}.
\label{eq:lattice_momenta}
\end{equation}
Notice how the use of a finite volume relates to the lowest non-zero momentum accessible on a given lattice and also to its resolution.
Having a finite number of available momenta, the discrete Fourier transform becomes the sum,
\begin{equation*}
\phi(x) = \frac{1}{V}\sum_{n\in\Lambda}\tilde\phi(p_n)e^{ip_n\cdot x}
\end{equation*}
where $V = N^4$ is the volume of the space-time grid for the hypercubic lattice.

Other than the discretized momentum \eqref{eq:lattice_momenta}, in this work we will also consider the lattice perturbation theory \cite{CAPITANI_perturb} improved momentum defined by
\begin{equation}
\hat p_\mu = \frac{2}{a}\sin\left(\frac{ap_\mu}{2}\right) = \frac{2}{a}\sin\left(\frac{\pi n_\mu}{N}\right).
\label{eq:improved_momenta}
\end{equation}
This form comes from the tree-level propagator of a massless scalar field on the lattice. 

The general path integral quantization scheme is built analogously to the continuum formulation. The partition function is constructed 
\begin{equation}
\mathcal{Z} = \int \mathcal{D}\phi e^{-S_E(\psi)}
\end{equation}   
with the field measure replaced by a finite product
\begin{equation}
\mathcal{D}\phi = \prod_{n\in\Lambda} d\phi(n)
\end{equation}
and the expectation value of an observable is computed as
\begin{equation}
\expval{\mathcal{O}} = \frac{1}{\mathcal{Z}}\int \mathcal{D}\phi e^{-S_E(\phi)}\mathcal{O}(\phi).
\end{equation}

\section{Lattice Quantum Chromodynamics}\label{sec:Lattice Chromodynamics}

We consider the discretization of the pure Yang-Mills sector of the QCD Lagrangian.
On the lattice the gluon fields appear in order to preserve gauge invariance in local gauge transformations, $\psi(n) \rightarrow V(n)\psi(n)$, where $V(n)$ are $SU(3)$ group elements on the lattice sites.
In the continuum, we considered the covariant derivative to ensure the gauge invariance of the action, and this was implemented such that the comparison of fields at different points was properly defined. 
To this end, we used the concept of a comparator.

On the lattice, two fields in neighbouring points have corresponding transformations $V(n)$ and $V(n+a\hat\mu)$.
We define the \textit{link variables} as a comparator $U_\mu(n)$, connecting both points.
These oriented group elements live in the links between sites and are the fundamental fields in this framework.
These satisfy an analogous gauge transformation as the continuum counterpart
\begin{equation}
U_\mu(n) \rightarrow V(n)U_\mu(n)V^\dagger(n+a\hat\mu).
\end{equation}
The inverse link from the same lattice point is given by the adjoint operator $U_\mu^\dagger(n-a\hat\mu)$ -- see figure \ref{fig:comparators}.
\tikzset{->-/.style={decoration={
			markings,
			mark=at position #1 with {\arrow{>}}},postaction={decorate}}}
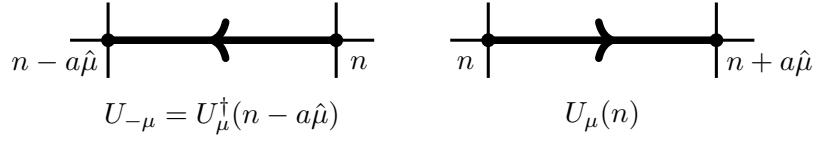
\begin{figure}[H]
	\begin{center}
		\begin{tikzpicture}
		\draw[->-=.56,line width=1mm] (-1,0) to (-4,0);
		\draw[->-=.56,line width=1mm] (1,0) to (4,0);
		\filldraw [black,very thick] (-1,0) circle (2pt);
		\filldraw [black,very thick] (-4,0) circle (2pt);
		\filldraw [black,very thick] (4,0) circle (2pt);
		\filldraw [black,very thick] (1,0) circle (2pt);
		\draw[-,line width=.4mm] (-1,-0.5) to (-1,0.5);
		\draw[-,line width=.4mm] (-4,-0.5) to (-4,0.5);
		\draw[-,line width=.4mm] (1,-0.5) to (1,0.5);
		\draw[-,line width=.4mm] (4,-0.5) to (4,0.5);
		\draw[-,line width=.4mm] (-4.5,0) to (-4,0);
		\draw[-,line width=.4mm] (4,0) to (4.5,0);
		\draw[-,line width=.4mm] (1,0) to (.5,0);
		\draw[-,line width=.4mm] (-1,0) to (-0.5,0);
		\node[align=center] at (2.5,-1) {$U_\mu(n)$};
		\node[align=center] at (-2.5,-1) {$U_{-\mu} = U_\mu^\dagger(n-a\hat\mu)$};
		\node[align=center] at (-4.7,-0.3) {$n-a\hat\mu$};
		\node[align=center] at (4.7,-0.3) {$n+a\hat\mu$};
		\node[align=center] at (0.7,-0.3) {$n$};
		\node[align=center] at (-0.7,-0.3) {$n$};
		\end{tikzpicture}
	\end{center}
	\caption{Link variables between $n$, $n+a\hat\mu$ and $n-a\hat\mu$. }
	\label{fig:comparators}
\end{figure}

The simplest lattice action, such that the Yang-Mills form is restored when the limit $a\rightarrow0$ is taken, can be built from the product of comparators in a closed loop.
Namely, we consider the \textit{plaquette}, \cref{fig:plaquette}, which is the simplest loop on the lattice
\begin{equation}
U_{\mu\nu}(n) = U_\mu(n)U_\nu(n+a\hat\mu)U_\mu^\dagger(n+a\hat\nu)U_\nu^\dagger(n).
\label{eq:plaquette}
\end{equation}
The gauge transformation of this product depends on a single lattice point,
\begin{equation}
U_{\mu\nu}(n) \rightarrow V(n)U_{\mu\nu}(n)V^\dagger(n).
\end{equation}
Hence, applying the trace we obtain a gauge invariant term
\begin{equation}
\Tr U'_{\mu\nu}(n) = \Tr\left(V(n)U_{\mu\nu}(n)V^\dagger(n)\right) = \Tr U_{\mu\nu}(n),
\label{eq:gauge invariant trace loop}
\end{equation}

{
	\vspace{1em}
	\begin{minipage}{.45\textwidth}
			\begin{tikzpicture}[scale=0.9, every node/.style={scale=0.9}]
			\draw[->-=.56,line width=.5mm] (-1,1) to (-1,-1);
			\draw[->-=.56,line width=.5mm] (1,1) to (-1,1);
			\draw[->-=.56,line width=.5mm] (1,-1) to (1,1);
			\draw[->-=.56,line width=.5mm] (-1,-1) to (1,-1);
			
			\draw[-=.56,line width=.1mm] (-1,-1.5) to (-1,1.5);
			\draw[-=.56,line width=.1mm] (1,-1.5) to (1,1.5);
			\draw[-=.56,line width=.1mm] (-1.5,-1) to (1.5,-1);
			\draw[-=.56,line width=.1mm] (-1.5,1) to (1.5,1);
			
			\filldraw [black,very thick] (-1,1) circle (2pt);
			\filldraw [black,very thick] (-1,-1) circle (2pt);
			\filldraw [black,very thick] (1,1) circle (2pt);
			\filldraw [black,very thick] (1,-1) circle (2pt);
			
			\node[align=center] at (-1.5,-1.5) {$n$};
			\node[align=center] at (-1.67,1.5) {$n+a\hat\nu$};
			\node[align=center] at (2.1,1.5) {$n+a\hat\mu+a\hat\nu$};
			\node[align=center] at (2,-1.5) {$n+a\hat\mu$};
			
			\node[align=left] at (-2,0) {$U_\nu^\dagger(n)$};
			\node[align=left] at (0,-1.5) {$U_\mu(n)$};
			\node[align=left] at (2.2,0) {$U_\nu(n+a\hat\mu)$};
			\node[align=left] at (0,1.5) {$U_\mu^\dagger(n+a\hat\nu)$};
			\end{tikzpicture}
	\end{minipage}~
	\begin{minipage}{.45\textwidth}
		
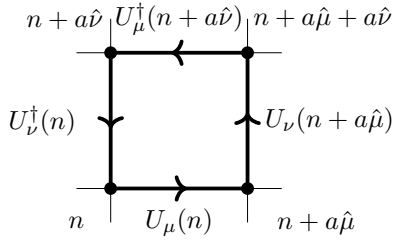
\captionof{figure}{Schematic representation of the minimal planar lattice loop, \textit{plaquette} in the plane $\mu - \nu$.}
		\label{fig:plaquette}
	\end{minipage}
	\vspace{1em}
}

Due to the form of the continuum action we need a relation between the link variables and the continuum gauge fields $A_\mu(x)$. Hence we establish a relation between lattice and continuum comparators $U_\mu(n) = U(n,n+\hat\mu) + \order{a}$. For this purpose, we introduce algebra valued lattice gauge $A_\mu$ fields by
\begin{equation}
U_\mu(n) = e^{iagA_\mu(n+a\hat\mu/2)} + \order{a}.
\label{eq:gaugelink_gaugefield}
\end{equation}
We rewrite\footnote{Using the Baker-Campbell-Hausdorff formula for the product of exponentials of matrices
	\begin{equation*}
	e^Ae^B = e^{A+B+\frac{1}{2}[A,B]+ ...}.
	\end{equation*}}
 \cref{eq:plaquette} using \eqref{eq:gaugelink_gaugefield} to relate the \textit{plaquette} with $F_{\mu\nu}(n)$
\begin{align}
U_{\mu\nu} &= e^{ia^2(\partial_\mu A_\nu(n) + \partial_\nu A_\mu(n) + i[A_\mu(n),A_\nu(n)]) + \order{a^3}}  \nonumber \\
&= e^{iga^2F_{\mu\nu}(n) + \order{a^3}}.
\label{eq:lattice_comparator_fieldtensor}
\end{align}
Hence, the Wilson Landau gauge action is obtained by
\begin{align}
S_\text{G}[U] =& \frac{\beta}{2N_c}\sum_{n}\sum_{\mu,\nu}\Re\Tr(\mathds{1} - U_{\mu\nu}(n))\\
=&\frac{a^4}{2g^2}\sum_{n}\sum_{\mu,\nu}\Tr(F_{\mu\nu}^2(n))+\order{a^2}
\label{eq:Wilson action}
\end{align}
where we defined the inverse bare lattice coupling $\beta = 2N_c/g^2$.
This action was formulated by Wilson in 1974 -- see \cite{Wilson1974}.

In this work we consider only the gauge part of the QCD action.
This approximation, disregarding the quarks dynamics is called \textit{quenched approximation}.  
Fermions are represented by Grassmann variables and its contribution to the generating functional can be written as a fermion determinant. 
The quenched approximation consists in replacing the determinant by a constant which diagrammatically consists in neglecting fermion loops contributions.
Typically, quenched lattice calculations of the hadronic spectra shows differences around $10$ to $20\%$ relative to experimental data \cite{Aoki_2003}. 

\section{Gauge fixing}\label{sec:gauge fixing}

While physical observables are gauge independent, the computation of correlation functions requires to choose a gauge. In fact, they can be shown to vanish if no gauge is fixed -- Elitzur's theorem \cite{Elitzur}.

In this work we consider the Landau gauge which in the continuum reads $\partial_\mu A^\mu(x) = 0$, or equivalently $p_\mu A^\mu(p) = 0$ in momentum space. 
On the lattice, it can be shown \cite{Gattringer2010} that this is equivalent to finding a stationary point of the following functional
\begin{equation}
F_U[V] = \frac{1}{VN_dN_c}\sum_{n,\nu}\Tr\left[V(n)U_\mu(n)V^\dagger(n+\hat\mu)\right],
\label{eq:gauge functional lattice}
\end{equation}
where $N_d$ and $N_c$ the dimensions and colour number, respectively, and $V$ is the volume of the lattice -- not to be confused with the gauge transformation $V(n)$.

However, in general the functional \cref{eq:gauge functional lattice} has many extrema -- this problem arises already in the continuum formulation. Ideally, we want the gauge condition (hypersurface defined in \cref{sec:quantization of the theory}) to intersect each gauge orbit uniquely, and thus a single representative is chosen from each gauge orbit.

However, Gribov \cite{Gribov_1977} found\footnote{Gribov considered non-abelian gauge theories in the Coulomb gauge $\partial_i A_i = 0$. This was later generalized for a 4-dimensional hypercubic and periodic lattice for any $SU(N_c)$ gauge theory \cite{Killingback_gauge}.} that the Faddeev-Popov procedure alone is not sufficient,
and that there are multiple solutions for the gauge condition still related by a gauge transformation. 
These multiple solutions due to the multiple intersections of the hypersurface within each orbit are the so called \textit{Gribov copies}.


The presence of the copies implies the existence of various stationary points of the functional. Gribov suggested additional constraints to the gauge field configuration space, restricting the region to the maxima of \eqref{eq:gauge functional lattice}. 
However, this \textit{Gribov region}\footnote{This subspace contains all local maxima of the functional. $$\Omega = \{A: \partial_\mu A_\mu = 0, M[A] \geq 0\}$$
where $M$ is the Faddeev-Popov matrix \cref{eq:ghostmatrix}.} 
is still not free of Gribov copies. Further restrictions define a subspace containing only the global maxima of $F_U$ -- called \textit{fundamental modular region}.
It can be shown that on the lattice this restriction guarantees the absence of Gribov copies in this region \cite{zwanziger_gribov}. 
Numerically, the search is limited to a local maximum -- in this work we used the steepest descent method, described in \cite{Fourier_acceleration_method}. The computer code uses both the Chroma \cite{EDWARDS2005_chroma} and PFFT \cite{PFFT_libraries} libraries.

A review of the gauge fixing on the lattice can be found in \cite{GIUSTI_2001}. It is worth referring that the effect of the Gribov copies was studied for the gluon propagator on the lattice \cite{Silva_2004,CUCCHIERI1998841} concluding that its effect are small -- less than $10\%$. In this work we do not consider the effect of the Gribov copies.

\section{Correlation functions from the lattice}

We are interested in computing correlation functions involving gauge fields $A_\mu$. On the lattice, the gluon field can be computed from the links \cref{eq:gaugelink_gaugefield}
\begin{equation}
agA_\mu(x+\hat\mu/2) = \frac{1}{2i}\left[U_\mu(n) - U_\mu^\dagger(n)\right] - \frac{1}{6i}\Tr\left[U_\mu(n) - U_\mu^\dagger(n)\right]
\label{eq:gaugefield_lattice}
\end{equation}
up to $\order{a^2}$ corrections. The second term ensures that the field is traceless, $\Tr A_\mu = 0$.
The momentum space lattice gauge field is obtained with the discrete Fourier transform defined before,
\begin{equation}
A_\mu(p) = \sum_{x}e^{-ip\cdot(x+\hat\mu/2)}A_\mu(x+\hat\mu/2)
\label{eq:gauge transform lattice}
\end{equation}
with $p=2\pi n/aN$ and $x=an$ where $n_\mu\in[-N/2+1,N/2]$. 

The gluon two point function is extracted from the average over gauge field configurations by
\begin{equation}
\expval{A_{\mu_1}^{a_1}(p_1)A_{\mu_2}^{a_2}(p_2)} = D_{\mu_1\mu_2}^{a_1a_2}(p_1)V\delta(p_1+p_2).
\end{equation}
In our numerical framework, we have access to algebra valued gauge fields $A_\mu(p)$ from \cref{eq:gaugefield_lattice,eq:gauge transform lattice}. 
To form a scalar in the colour sector we consider a trace and a suitable Lorentz contraction for the space-time indices. 
Considering the usual continuum tensor description for the gluon propagator \cref{eq:full_continuum_propagator}, the form factor $D(p^2)$ is obtained by
\begin{equation}
D(p^2) = \frac{2}{(N_c^2-1)(N_d-n)}\sum_{\mu}\expval{\Tr\left[A_{\mu}(p)A_{\mu}(-p) \right]}
\label{eq:continuumtensor_extraction}
\end{equation}
where $n=0$ if $p=0$, or $1$ otherwise.

For the gluon propagator, the analysis of the colour indices is simple, since only $\delta^{ab}$ can be used.
For the three and four gluon vertices we again access the product of gauge fields to which we apply the trace to obtain a scalar in colour space,
\begin{align}
&\expval{\Tr\left[A_{\mu_1}(p_1)A_{\mu_2}(p_2)A_{\mu_3}(p_3)\right]} = V\delta(\sum_{i}p_i)G_{\mu_1\mu_2\mu_3}(p_1,p_2,p_3) 
\label{eq:3gluon_correlation_trace}\\
&\expval{\Tr\left[A_{\mu_1}(p_1)A_{\mu_2}(p_2)A_{\mu_3}(p_4)A_{\mu_4}(p_4)\right]} = V\delta(\sum_{i}p_i)G_{\mu_1\mu_2\mu_3\mu_4}(p_1,p_2,p_3,p_4).
\label{eq:4gluon_correlation_trace}
\end{align}
The $G$'s represent the Green's functions with colour indices absorbed by the trace operation and whose form depends on the Lorentz tensor basis considered -- these will be properly defined in \cref{chap:tensorbases}. 

\section{Computational aspects}

\subsection{Expectation values on the lattice}\label{subsec:expectation values in the lattice}

In the Euclidean formulation of the theory, the expectation value of some field dependent operator is given by
\begin{equation}
\expval{\mathcal{O}} = \frac{1}{\mathcal{Z}}\int\mathcal{D}U \mathcal{O}(U)e^{-S_E[U]}.
\label{eq:euclidean_expect_value}
\end{equation}
To obtain numerical results we  consider only a finite number of field configurations. 
This is done by importance sampling considering the weight of the Boltzmann factor in the Euclidean action, and the integrals estimated by Monte-Carlo methods, \cite{morningstar2007monte}.

A set of gauge field configurations\footnote{ By a gauge field configuration we mean that each site of the lattice is attributed a value of the field $U$, i.e. a Lorentz vector of $SU(3)$ matrices.} $\{U_i\},~i=1,...,n$ is generated according to the probability distribution 
\begin{equation}
	 P(U) = e^{-S_E(U)}/\mathcal{Z}.
\end{equation}
The sequence is obtained by a Markov chain which generates the configurations, one after another according to a transition amplitude $P(U_i\rightarrow U_j)$\footnote{The precise form of the amplitudes depends on the chosen method \cite{Rothe}.} depending solely on the predecessor configuration. This transition amplitude should create a sequence distributed according to $P(U)$ in the large $n$ limit. 

When the set $\{U_i\},~i=m,...,n$ is distributed according to $P(U)$, it is said to be \textit{thermalized}. 
From the thermalized set we chose $N$ configurations, each separated from the former by $k$ Markov steps in order to reduce correlations among them. 
The set $\{U_i\},~i=1,...,N$ is the one used for the computation.
The configurations considered in this thesis \cite{anthony_2016} were obtained using a combination of the over-relaxation and the heat bath methods according to \cite{Gattringer2010}.

Having a finite number of configurations following the $\exp(-S_E(U))/\mathcal{Z}$ probability distribution, the expectation value \eqref{eq:euclidean_expect_value} is estimated by the sample mean
\begin{equation}
\bar{\mathcal{O}} = \frac{1}{N}\sum_{i=1}^{N}\mathcal{O}(U_i),
\end{equation} 
which corresponds to the correct average $\expval{\mathcal{O}}$ in the large $N$ limit.

If all configurations in the sample are statistically independent, having no correlations, then the sample average is normally distributed around the true expectation value, and the error estimate would be $\expval{\mathcal{O}} = \bar{\mathcal{O}}+1/\sqrt{N}$. 
To estimate the uncertainty of an average over the configurations without assuming a statistical distribution inherent to the variables, we use the \textit{Bootstrap method} defined below.

\subsection*{Setting the scale}

Lattice quantities are, in general, dimensionless with the values given in terms of the lattice spacing $a$. To obtain physical values we need to set this scale by choosing a suitable value for $a$ which is not an input parameter of the formulation. 

To do this we match a given dimensionless lattice object, $am_g$, with an experimental value ($m_{g,\text{phys}}$). The lattice spacing is then obtained by
\begin{equation}
a = \frac{am_g}{m_{g,\text{phys}}}.
\end{equation} 

The lattice spacing of the configuration ensembles used in this work were computed from the string tension data in  \cite{Bali_coulingconstant}. 
The string tension is defined from the quark-antiquark potential which is related to the large $n_4$ behaviour of the lattice expectation value of a planar rectangular loop (analogous to the square loop, \cref{eq:plaquette}), see \cite{Gattringer2010}.

\subsection{Bootstrap method}\label{sec:bootstrap}

In this thesis, all statistical errors from the simulations are estimated using the bootstrap method. The bootstrap is a distribution independent method that can be used to estimate the statistical error of any quantity $\mathcal{S}$. A review of the method can be found in \cite{bootstrap_Efron}. 

Considering a given initial sample of $N$ elements $\{U_i\},~i=1,...,N$ obtained from an unknown distribution (in our case the sample is the set of gauge field configurations). We are interested in obtaining the statistical error associated to a quantity $\mathcal{S}(U)$ which in this work corresponds to a mean value of some quantity over the configurations.

The method considers the \textit{empirical distribution} for the original sample, assigning the probability $1/N$ to each of the observed elements. A bootstrap sample is constructed by random sampling with replacement from this probability distribution. We obtain $N_b$ random samples $U_b^j=(U^j_1,...,U^j_N)$ from the original, of the same size $N$. For each sample $j$, the quantity is computed to be $\mathcal{S}^j \equiv \mathcal{S}(U^j)$. The idea of the method is that now, we have a proper random variable $\mathcal{S}^j$ with a known distribution -- the empirical. 

To obtain confidence intervals without assuming the underlying distribution, the bootstrap method provides asymmetric boundaries around the expectation value.
Having $N_b$ values $\mathcal{S}^j$, from which we obtain $\bar{\mathcal{S}}$, the upper and lower errors are estimated using confidence intervals,
\begin{align}
\sigma_\text{up} = \mathcal{S}_\text{up}-\bar{\mathcal{S}}, && \sigma_\text{down} = \bar{\mathcal{S}} - \mathcal{S}_\text{down}
\end{align}
where $\mathcal{S}_\text{up}$ and $\mathcal{S}_\text{down}$ are found in a way that they satisfy
\begin{align}
\frac{\#\{\mathcal{S}^j < \mathcal{S}_\text{up} \}}{N_b} = \frac{1+C}{2}, &&
\frac{\#\{\mathcal{S}^ j < \mathcal{S}_\text{down} \}}{N_b} = \frac{1-C}{2}
\end{align}
where $C$ is the coefficient chosen for the confidence interval, $C\in[0,1]$ and $\#\{\}$ represents the cardinality of a given set. 


In this work, $C$ was chosen to be $C = 0.675$ representing a $67.5\%$ probability of the true estimator falling in the interval. The uncertainty was taken to be the largest of the two errors. 


\chapter{Gluon tensor bases}\label{chap:tensorbases}

In this chapter we describe how the discretization of space-time affects the tensor representations of the gluon propagator.
Although we consider these structures for the gluon propagator, we will find that there are special kinematic configurations for which the lattice structures provide similar results as those obtained using the continuum tensor basis. 

Some general aspects of discretization effects and possible corrections methods will be also introduced.
Finally, the three and four gluon vertices will be discussed, and corresponding tensor bases will be shown.



\section{Tensor representations on the lattice}\label{sec:tensor representations of the lattice}

The $O(4)$ symmetry of the Euclidean continuum theory is replaced by the $H(4)$ group when space-time is discretized using an hypercubic group.
This group consists of powers of $\pi/2$ rotations around the coordinate axes and parity transformations of the whole lattice, i.e. inversions of the axes (corresponding operators are shown in \cref{apend_chap:lattice_tensors}). 

The definition of a tensor has an underlying group of transformations that for the lattice is the $H(4)$. 
Gluon correlation functions are tensors with respect to the $H(4)$ group and, therefore, identifying the tensor bases for this group is crucial to achieve a proper description for the gluon Green's functions.
These tensor structures differ from the continuum tensors due to lessened symmetry restrictions.

To see how this affects the construction of tensors we consider an $N_d$-dimensional vector space with a given transformation having matrix representation $M$.
A given vector $p$ in this space transforms as\footnote{The summation convention over repeated indices is used throughout this chapter.}
\begin{align}
	p' = Mp, && p'_\mu = M_{\mu\nu}p_\nu.
\end{align}
with components $p_\mu$ defined with respect to a given coordinate basis.
The generalization to higher order vector spaces is given by the definition of tensors with respect to the given transformation. A $k$-rank tensor is a quantity described in general by $N_d^k$ components $T_{\mu_1...\mu_k}$ in a given coordinate basis with the following transformation law
\begin{equation}
	T'_{\mu_1...\mu_k} = M_{\mu_1\nu_1}...M_{\mu_k\nu_k}T_{\nu_1...\nu_k}.
	\label{eq:tensor transform general}
\end{equation}
This definition includes vectors ($k=1$), as well as scalars ($k=0$) which are unchanged by the group transformations.

In an $O(N_d)$ symmetric space, scalar products of vectors are unchanged under the group transformations, employed by orthogonal $N_d\times N_d$ matrices, $M_{\mu\nu}=M_{\nu\mu}^{-1}$. 
To see how the definition \eqref{eq:tensor transform general} restricts the form of tensors, we consider the case of a scalar quantity $S$ depending on a vector $p$.
As a  scalar, it remains unchanged by the transformation, $S(p') = S(p)$. These two transformations restrict the dependence of $S$ on $p$ through the scalar product, $S(p^2)$,  since $p^2$ is an $O(N_d)$ group invariant.

If instead of a scalar we consider a vector valued function $\vec{V}(p)$ also depending on the vector $p$. By using its transformation law $V'_\mu(p') =  M_{\mu\nu}V_\nu(p)$ we conclude that the most general form for its components is
\begin{equation}
V_\mu(p) = V(p^2)p_\mu
\end{equation}
where $V(p^2)$ is a scalar of the vector $p$, \cite{hamermesh2012group}.

An important case for this work are second rank tensors $D_{\mu\nu}(p)$ depending on a single vector $p$. From \eqref{eq:tensor transform general} its transformation law is $D'_{\mu\nu}(p') = M_{\mu\rho}M_{\nu\sigma}D_{\rho\sigma}(p)$.
Hence, the most general form for this quantity is of the form
\begin{equation}
	D_{\mu\nu}(p) = A(p^2)\delta_{\mu\nu} + B(p^2)p_\mu p_\nu.
	\label{eq:general basis continuum}
\end{equation}
This tensor will be considered for the description of the gluon propagator to evaluate how the Landau gauge Slavnov-Taylor identity, \cref{eq:slavnov-taylor}, acts on the lattice.
With these three examples we see that continuum vectors have a simple, linear structure imposed by the continuum symmetry. 
We are interested in performing a similar construction considering the lattice symmetry.

The $H(N_d)$ group is a discrete subgroup of $O(N_d)$ in an $N_d$-dimensional space. It consists of $\pi/2$ rotations as well as parity inversions for each of the axes. 
However, it can be shown \cite{vujinovic2019tensor} that each group transformation can be written as a composition of permutations and inversions of the components -- signed permutations\footnote{This is seen by considering a 2-dimensional example: performing a clockwise $\pi/2$ rotation of a vector $c=(c_1,c_2)$ to $c'=(c_2,-c_1)$ can be achieved by the composition of the inversion of the first component followed by a permutation of both components. Generalizations for higher dimensional spaces are straightforward since these transformations may be independently applied to each hyperplane.}. 
The reason why it is worth to decompose the $H(N_d)$ group into these two smaller subgroups is that they are disjoint\footnote{In fact, permutations correspond to transformations with determinant $+1$ while inversions to transformations with determinant $-1$.}, and thus can be analysed independently. Hence, to find objects transforming properly under the $H(N_d)$ group it is sufficient to find those which transform properly according to both permutations and inversions.

\subsection{Scalars under the hypercubic group}

Proceeding as for the continuum case, we start with the scalar functions on the lattice depending on a single momentum vector $p$. We inspect the vector dependence of these objects which must be invariant under
permutations and inversions of components. It can be easily seen that the class of objects
\begin{equation}
p^{[2n]} \equiv  \sum_{\mu}p_\mu^{2n},~n\in\mathbb{N}
\end{equation}
satisfies this property, and each of them is an hypercubic invariant\footnote{The case $\invmomp{2}=p^2$ is the only invariant in the continuum, i.e. for $O(N_d)$.}. 
Hence, we would think that in general a momentum dependent scalar function would depend on all of these objects.  It was shown in \cite{weylgroups}, however, that only $N_d$ invariants are linearly independent, thus creating a minimal set of invariants. 

The interesting cases for this work are the scalar functions depending on a 4-dimensional vector $p$ which will generally change to
\begin{equation}
	S(p^2) \rightarrow S_L(p^2, p^{[4]}, p^{[6]}, p^{[8]})
\end{equation}
when passing to the lattice. The choice of the four lowest mass dimension independent invariants is done for practical reasons, but is nonetheless arbitrary. 

\subsection{Hypercubic vectors}\label{sec:hypercubic vectors}

We now generalize the vector notion for the hypercubic symmetric space.
As referred, we find its properties by analysing the permutations and inversions independently.

Starting with the permutations, and given that any general transformation of this kind can be written as a product of exchanges of only two components -- transpositions \cite{hamermesh2012group} -- we focus on those. Hence, an object transforming as a vector under arbitrary transpositions will also transform as a vector under a general permutation. 
Performing a transposition of components $\sigma \leftrightarrow \rho$, the transformation for the vector components $p_\mu$ in an $N_d$-dimensional space is 
\begin{align}
&p'_\nu = p_\nu,~\nu \neq \sigma,\rho \nonumber \\
&p'_\sigma = p_\rho, \nonumber \\
&p'_\rho = p_\sigma.
\end{align}
This is the fundamental transformation rule for a vector, however we are interested in finding the most general structure satisfying this rule. Indeed, any polynomial of the vector, $(p_\mu)^n$ also transforms as a vector under transpositions (a brief proof is shown in \cref{append_sec:polynomial transform})

However, to be a proper vector under $H(N_d)$ it also needs to satisfy the transformation under inversions. Taking the same $N_d$-dimensional vector $p$ and applying an inversion on its $\sigma$-th component, the transformed components are 
\begin{align}
	&p'_\mu = p_\mu, ~\mu\neq\sigma, \nonumber \\
	&p'_\sigma = -p_\sigma.
	\label{appen_eq:inversions}
\end{align}
To be a vector, the polynomial should transform exactly as \eqref{appen_eq:inversions}
\begin{align}
&(p'_\mu)^n = (p_\mu)^n, ~\mu\neq\sigma, \nonumber \\
&(p'_\sigma)^n = -(p_\sigma)^n,
\end{align}
and for this to be true, $n$ is necessarily an odd integer, otherwise an even integer would spoil the transformation by eliminating the minus sign of the inversion.
Therefore the most general structure satisfying the vector transformation is
\begin{equation}
v_\nu^n = p_\nu^{2n+1},~n\in\mathbb{N}.
\label{appen_eq:general_vector}
\end{equation}
Moreover, we also note that any linear combination of these vectors is also a vector (by linearity) and thus any function whose Taylor expansion includes only odd powers of a vector also constitutes a lattice vector. We now see that the sinusoidal, improved momentum 
\begin{equation}
	\hat p_\mu = 2\sin\left(\frac{ap_\mu}{2}\right)
\end{equation}
arising from lattice perturbation theory is a proper lattice vector, since it transforms correctly under the $H(4)$ group. 

A general lattice vector is then composed of a linear combination of $N_d$ vectors from the infinite possible vectors of the form \eqref{appen_eq:general_vector}
\begin{equation}
	V_\mu(p) = \sum_{n = 1}^{N_d}V_n v_\nu^{2n+1}
\end{equation}
where $V_n(p^2)$ are lattice scalar functions. The sum is limited by the dimension of space since in a $N_d$-dimensional space only $N_d$ linearly independent basis vectors can be constructed. 

\section{Lattice basis -- Gluon propagator}

We now consider the gluon propagator -- a second order tensor depending on a single vector, the momentum $p$. 
In colour space the lattice gluon propagator is a two dimensional tensor having the same form as in the continuum formulation. 
Indeed, $\delta^{ab}$ is the only second order $SU(3)$ tensor available. 
Thus we focus on the space-time structure of the propagator. 
Being a second order tensor depending on a single momentum $D_{\mu\nu}(p)$, the gluon propagator transforms as
\begin{equation}
	D'_{\mu\nu}(p) = M_{\mu\sigma}M_{\nu\rho}D_{\sigma\rho}(p).
\end{equation}
where $M\in H(4)$ is a matrix representation of an arbitrary group element.

Following \cite{vujinovic2019tensor} we consider the splitting of the tensor basis in the diagonal and off-diagonal terms. This is related with the way the hypercubic transformations act on the lattice tensors, not mixing the aforementioned groups of elements $D_{\mu\mu}$ and $D_{\mu\nu},~\mu\neq\nu$ (see \cref{appen_sec:Second order tensors under $H(4)$ symmetry} for a proof of this property). Accordingly, the diagonal and off-diagonal tensor elements will be parametrized differently, i.e. by different form factors.

The most general objects to construct the tensor basis are $\{\delta_{\mu\nu}, p_\mu^m p_\nu^n\}$. However, for the second element, since the transformation rule for the tensor applies independently for each momentum, a similar argument as the one used for the vectors in \cref{sec:hypercubic vectors} restricts $m$ and $n$ to be odd integers. Thus, we obtain a set of the most general possible tensor basis elements
\begin{equation}
	\{\delta_{\mu\nu}, p_\mu^{2k+1}p_\nu^{2s+1}\}, ~k,s\in\mathbb{N}.
\end{equation}

For the propagator itself, notice that a symmetric second order tensor has only $N_d(N_d+1)/2$ free parameters, i.e. for 4-dimensional space it is fully described by 10 form factors\footnote{In principle, however, further conditions implied by the Slavnov-Taylor identity and gauge fixing further reduce the number of independent parameters.}. However, for reasons that will be evident when analysing the results, we consider only two reduced bases for the propagator with three and five form factors. 

Consider the case of approximating the tensor by three form factors. The possible choices for diagonal and off-diagonal terms are $\{\delta_{\mu\mu}, p_\mu^2, p_\mu^4, ...\}$, and $\{p_\mu p_\nu, p_\mu^3p_\nu, ...\}$, respectively. Choosing the parametrization with the lowest mass dimension terms we obtain the form
\begin{align}
&D_{\mu\mu}(p) = J(p^2)\delta_{\mu\mu} + K(p^2)p_\mu^2, ~(\text{no sum}) \nonumber \\
&D_{\mu\nu}(p) = L(p^2)p_\mu p_\nu,~\mu\neq\nu.
\label{eq:partial_lattice_basis}
\end{align}

We also consider an extended tensor basis using five form factors. Performing the same construction as before and considering an explicit symmetrization on the space indices for the higher order non-diagonal terms, we obtain
\begin{align}
&D_{\mu\mu}(p) = E(p^2)\delta_{\mu\mu} + F(p^2)p_\mu^2 + G(p^2)p_\mu^4, ~(\text{no sum}) \nonumber \\
&D_{\mu\nu}(p) = H(p^2)p_\mu p_\nu + I(p^2)p_\mu p_\nu(p_\mu^2 + p_\nu^2),~\mu\neq\nu ~(\text{no sum}).
\label{eq:full_lattice_basis}
\end{align}

The extraction of the form factors involves the computation of its projectors, these are built in \cref{append_sec:projectors_latticebasis}. 
In \cref{chap:results} these form factors will be obtained from the lattice and there we will introduce continuum relations among them that follow from both the Slavnov-Taylor identity and gauge condition on the lattice.

Notice that the tensor basis can be built with normal momentum $p_\mu$ or the lattice perturbation theory improved momentum $\hat p_\mu$ which may serve as a further improvement. However, structures mixing both types of momenta are not considered.

Notice that the tensor parametrization by the bases is independent of the chosen gauge, however this choice will entail different relations among the form factors. We work with the Landau gauge, implying orthogonality of the gauge fields in the continuum, $p_\mu A_\mu(p)= 0$.

\subsubsection*{Generalized diagonal kinematics}

Having the general form of the lattice basis, it is important to consider configurations for which the basis is reduced to a simpler form, closer to the continuum tensor basis. 
To those we call generalized diagonal kinematics and its form is specified by a single scale or vanishing components. Of this group belong the full diagonal, $(n,n,n,n)$, the mixed configurations $(n,n,n,0)$ and $(n,n,0,0)$, and on-axis momenta $(n,0,0,0)$. 

For these configurations, the inclusion of certain tensor elements is redundant for they become linearly dependent, thus reducing the possible independent terms. Namely, for diagonal momenta $(n,n,n,n)$ we get $p_\mu^2 = n^2\delta_{\mu\mu}$. 
Therefore only a reduced number of form factors is extracted.
Details on the changes of the lattice basis for these kinematics and how the form factors are extracted are shown in appendix \ref{apend_chap:lattice_tensors}. 


\section{Reconstruction of tensors}\label{sec:reconstruction of tensors}

To analyse how accurately a tensor basis describes the correlators from the lattice, we perform a reconstruction procedure \cite{vujinovic2019tensor, Vujinovi_2019}. 
This consists in extracting a given set of form factors, associated to the corresponding basis element, from the lattice correlation function and with these functions rebuild the original tensor. 
If the rebuilt function is different from the original we can infer that the basis is not complete and information was lost during the projection process.  
To do this we consider the following quotient
\begin{equation}
\mathcal{R} = \frac{\sum_{\mu\nu}|\Gamma^\text{\tiny orig}_{\mu\nu}|}{\sum_{\mu\nu}|\Gamma^\text{\tiny rec}_{\mu\nu}|}
\label{eq:ratio}
\end{equation}
given by the sum of absolute values\footnote{The absolute value was considered in order to prevent possible unintentional cancellations among the tensor components.} of the original tensor and the reconstructed one.
A value of $\mathcal{R}=1$ indicates that the basis is complete. 

The procedure follows by assuming that the correlator is described by its basis elements $\tau^j$ with corresponding form factor $\gamma^j$
\begin{equation}
	\Gamma = \sum_{j = 1}^N \gamma^j\tau^j.
	\label{eq:general_vertex}
\end{equation}
One starts by computing each form factor $\gamma^j$ using the respective projector -- this step is the one where information may be lost if the basis is not complete, since in this case there are not enough form factors to fully represent the object. This extraction is performed on the original vertex $\Gamma^\text{orig}$, which in the case of this work comes from the lattice simulation. Using \cref{eq:general_vertex} we reconstruct the vertex and obtain $\Gamma^\text{rec}$.


\section{Z4 averaging}\label{sec:z4 averaging}

In the continuum formulation, having rotational invariance means that the form factors depend only on the magnitude of the momenta, i.e., that exists some sort of rotational `degeneracy' on the contribution from those points of the momentum space. 
On the lattice, the continuum symmetry is broken into a discrete subgroup, more generally, the Poincaré invariance is reduced to $\pi/2$ rotations, inversions and also fixed length translations (considering periodic boundary conditions) \cite{gupta1998introduction}. 

All points connected by these symmetry transformations have the same $H(4)$ invariants which label the orbits of the group, and are invariant under the transformations.
Therefore, these points should have the same contribution when computing lattice correlation functions\footnote{The contribution of these points may not be exactly the same due to statistical fluctuations.}. 

Hence, to help suppressing statistical fluctuations we consider equally the contribution from all points in the subspace defined from all possible group transformations on a given lattice point.
This is accomplished by averaging all computed quantities over all points in the same orbit which amounts to $4!\times 4^2 = 384$ points for each momentum configuration in four dimensions.


\section{Lattice artifacts and Correction methods}\label{sec:latticeartifacts}

In order to properly evaluate the form factors that characterize the correlation functions it is necessary to account for the artifacts arising from the discretization of space. 
These systematic errors become noticeable when the precision associated with a computation becomes high enough such that the statistical errors are small compared with these `defects'. 
Since the gluon propagator is computed with a good degree of precision, the removal of these artifacts becomes relevant. 

We distinguish two types of artifacts related to the introduction of the lattice. Firstly, finite size effects due to the use of a finite spacing $a$ as well as volume $V$. These were studied in \cite{Oliveira_2012} where it was found for the gluon propagator that the interplay between these two effects were far from trivial. 
Secondly, what we call hypercubic artifacts arise from the breaking of $O(4)$ symmetry, and the appearance of multiple $H(4)$ orbits from each $O(4)$ orbit. 
We consider the latter in this section.

Since we are interested in extracting scalar form factors, we consider the behaviour of lattice scalar functions and how they relate to the corresponding continuum objects.
Any scalar function with respect to a given symmetry group is invariant along the orbit generated by the corresponding group symmetry applied to a given point. 
For the $H(4)$ group each orbit is specified by the four group invariants
\begin{equation*}
	\{\invmomp{2},~ \invmomp{4},~  \invmomp{6},~  \invmomp{8}\}.
\end{equation*}
The simplest example of this is given by comparing with the continuum symmetry. In this case, an orbit is simply labelled by the invariant $p^2$. For instance, both momenta $p_1=(2,0,0,0)$ and $p_2=(1,1,1,1)$ have $p_1^2 = p_2^2 = 4$ in the same $O(4)$ orbit. However, these two points have different $H(4)$ invariants, $\invmom{1}{4} = 16$ and $\invmom{2}{4}=4$ belonging to distinct $H(4)$ orbits, thus should not be averaged equivalently.
We see that the dependence of the scalars on the $\invmomp{4}$ invariant spoils the continuum symmetry.

Clearly, hypercubic artifacts would be eliminated if all higher order invariants $n>2$ vanished since we would only have a $p^2$ dependence as in the continuum\footnote{Note that finite size effects still affect the result after this correction.}.
Another way to understand why the finiteness of the higher order invariants relates to hypercubic artifacts is seen by considering the improved momentum arising from lattice perturbation theory. 
By looking at the improved invariant $\hat p^2$ expanded in orders of $a$
\begin{equation}
	\hat p^2 = \left(2\sin(ap/2)\right)^2 =  p^2 - \frac{a^2}{12}\invmomp{4} + \frac{a^4}{360}\invmomp{6} + ...
\end{equation}
we see that it differs from the naively discretized continuum momentum by terms which are proportional to the invariants.
Therefore, we can minimize the lattice invariants in order to suppress hypercubic artifacts depending on non $O(4)$ group invariants, i.e. by reducing the first higher order invariant $\invmomp{4}$ we are effectively reducing the artifacts. To perform this correction two distinct methods are considered.

\subsection{Momentum cuts}

The simplest method consists in applying cuts to the momenta. This arises by noticing that the further a momentum is from the diagonal, the higher are its non $O(4)$ invariants for a fixed $O(4) $ invariant $p^2$. This was seen for the example considered before with $(2,0,0,0)$ being on-axis momentum with higher $\invmomp{4}$.

An empirical way to deal with  higher invariants coming from these kinematics is to directly discard these momenta from the data. 
The usual choice is to consider only momenta inside a cylinder directed along the diagonal of the lattice as defined in \cite{skullerud_momcuts}.
This selects the largest momenta with the smallest components, i.e. with the lowest $H(4)$ invariants.
The radius of the cylinder is chosen as to maintain a good amount of data while reducing the artifacts, and in general a radius of one momentum unit ($ap = 2\pi/N$) is considered.

This cut, however, does not remove low momentum on-axis points.
To improve the method we consider further conical cuts, i.e. we consider only momenta falling inside a conical region around the diagonal of the lattice $(1,1,1,1)$. 
Throughout the work we consider an angle of $20\si{\degree}$.

In addition, the cuts may be applied only to momentum above a given threshold since for the IR region most of the data falls far from the diagonal and some information should be kept.
The main problem with this method is that it only keeps a small fraction of the original data.


\subsection{H4 method}\label{sec:H4 method}

The H4 method \cite{Soto_2007,Beirevi1999} is more involved as it attempts to entirely eliminate the contribution of the invariants $\invmomp{n}$ with $n>2$ by performing an extrapolation. 
In this work we consider only the extrapolation for the first invariant $\invmomp{4}$, however, this method can be improved with higher order corrections (given that enough data is available).
Examples of the applications, improvements and general considerations on the method can be found in \cite{Soto_2007,Oliveira_2019,Boucaud_2003}. 

We consider a given scalar function under the lattice symmetry $\Gamma_L(\invmomp{n}),~n=2,4,6,8$ obtained by a proper averaging over the whole group orbit $O(\invmomp{n})$,
\begin{equation}
\Gamma_L\left(p^2,p^{[4]},p^{[6]},p^{[8]}\right) = \frac{1}{N_O}\sum_{p\in O(\invmomp{n})}\Gamma(p)
\end{equation}
where $N_O$ corresponds to the cardinality of the orbit.
We want to study how it relates to the continuum counterpart $\Gamma(p^2)$. 

Assuming that the scalar is a smooth function of the invariants, we may extrapolate to the continuum by
\begin{equation}
	\Gamma(p^2) \equiv \lim\limits_{\invmomp{4}\rightarrow 0} \Gamma_L(p^2,\invmomp{4})
\end{equation}
neglecting higher order invariants which vanish as $\order{a^4}$. In fact, to $\order{a^4}$ the same extrapolation is possible for the improved momentum
\begin{equation}
	\lim\limits_{\invmomp{4}\rightarrow 0} \Gamma_L(p^2,\invmomp{4}) = \lim\limits_{\hat p^{[4]} \rightarrow 0} \Gamma_L(\hat p^2,\hat p^{[4]})
\end{equation}
although in practice this extrapolation is not easily feasible.

To implement the extrapolation in practice, we assume that the dependence on the invariants is smooth, and also that the lattice is close to the continuum limit (small $a$) to use the expansion
\begin{equation}
\Gamma_L\left(p^2,p^{[4]},p^{[6]},p^{[8]}\right) = \Gamma_L(p^2,0,0,0) + \frac{\partial\Gamma_L}{\partial p^{[4]}}(p^2,0,0,0)p^{[4]} + \order{a^4}.
\end{equation}
Thus we may identify $\Gamma_L(p^2,0,0,0)$ as the continuum function $\Gamma(p^2)$ in finite volume and up to higher order lattice artifacts.
In practice this is applied only when several $H(4)$ orbits exist with the same $O(4)$ invariant $p^2$. The extrapolation is done by a linear regression in $\invmomp{4}$ at fixed $p^2$, taking the results as $\invmomp{4}\rightarrow 0$.

Since several $H(4)$ orbits should exist, this restricts the range of momentum to which the method is applicable. Normally, only the mid range of momentum contains enough data to perform the extrapolation, thus the deep infrared and high ultraviolet are not considered in this correction. 
The H4 method can be generalized for cases with more than a single independent momentum. In this work, both for the propagator and three gluon vertex, the simplest case of a single scale momentum is considered.

\section{Three gluon vertex}\label{subsec:three gluon continuum basis}

While the gluon propagator in the continuum is described by a single scalar function, $D(p^2)$, under the symmetries of the theory, higher order correlation functions admit an increased number of form factors for a general kinematic configuration.
Thus we must consider the most general form under the required symmetries.

For the three gluon vertex the colour structure is restricted to be antisymmetric
\begin{equation}
\Gamma_{\mu_1\mu_2\mu_3}^{abc}(p_1,p_2,p_3) = f^{abc}\Gamma_{\mu_1\mu_2\mu_3}(p_1,p_2,p_3)
\end{equation}
due to the charge invariance of the QCD Lagrangian \cite{Blum_2015,Smolyakov1982}. This guarantees the vanishing contribution from the symmetric term $d^{abc}$. 
We then require that the complete object obeys Bose symmetry, and since the colour structure is established by the anti-symmetric structure constants, this requires $\Gamma_{\mu_1\mu_2\mu_3}(p_1,p_2,p_3)$ to be anti-symmetric to the interchange of any pair $(p_i,\mu_i)$.

For the space-time part of the tensor representing the three gluon vertex we consider a continuum basis which consists of 14 independent tensors.
Throughout the work we use the basis constructed in \cite{Ball_chiu1980} which considers a separation between terms orthogonal to all momenta, and longitudinal terms.  
The general tensor is given by the transverse and longitudinal terms
\begin{equation}
	\Gamma_{\mu_1\mu_2\mu_3}(p_1,p_2,p_3) = \Gamma_{\mu_1\mu_2\mu_3}^{(T)}(p_1,p_2,p_3) + \Gamma_{\mu_1\mu_2\mu_3}^{(L)}(p_1,p_2,p_3).
\end{equation}
The first consists of four tensors
\begin{align}
&\Gamma_{\mu_1\mu_2\mu_3}^{(T)}(p_1,p_2,p_3) = F(p_1^2,p_2^2;p_3^2)\big[g_{\mu_1\mu_2}(p_1\cdot p_2) - {p_1}_{\mu_2}{p_2}_{\mu_1}\big]B_{\mu_3}^3 \nonumber \\
&+ H(p_1^2,p_2^2,p_3^2)\big[-g_{\mu_1\mu_2}B_{\mu_3}^3 + \frac{1}{3}({p_1}_{\mu_3}{p_2}_{\mu_1}{p_3}_{\mu_2} - {p_1}_{\mu_2}{p_2}_{\mu_3}{p_3}_{\mu_1})\big] \nonumber \\
& + \text{cyclic permutations,}
\label{eq:ballchiu_transverse}
\end{align}
with the definition,
\begin{equation}
B_{\mu_3}^3 = {p_1}_{\mu_3}(p_2\cdot p_3) - {p_2}_{\mu_3}(p_1\cdot p_3).
\end{equation}
The scalar form factors $F(p_1^2,p_2^2;p_3^2)$ are symmetric under interchange of the first two arguments, evidenced by the used of the semi-colon, while $H(p_1^2,p_2^2,p_3^2)$ is symmetric under the interchange of any of its arguments.
The remaining 10 longitudinal elements are of the form
\begin{align}
\Gamma_{\mu_1\mu_2\mu_3}^{(L)}(p_1,p_2,p_3) = &A(p_1^2,p_2^2;p_3^2)g_{\mu_1\mu_2}(p_1-p_2)_{\mu_3} \nonumber \\
&+ B(p_1^2,p_2^2;p_3^2)g_{\mu_1\mu_2}(p_1 + p_2)_{\mu_3} \nonumber \\
&+ C(p_1^2,p_2^2;p_3^2)({p_1}_{\mu_2}{p_2}_{\mu_1} - g_{\mu_1\mu_2}p_1\cdot p_2)(p_1-p_2)_{\mu_3} \nonumber \\
&+ \frac{1}{3}S(p_1^2,p_2^2,p_3^2)({p_1}_{\mu_3}{p_2}_{\mu_1}{p_3}_{\mu_2} + {p_1}_{\mu_2}{p_2}_{\mu_3}{p_3}_{\mu_1}) \nonumber \\
&+ \text{cyclic permutations}
\label{eq:ballchiu_longitudinal}
\end{align}
where both $A(p_1^2,p_2^2;p_3^2)$ and $C(p_1^2,p_2^2;p_3^2)$ are symmetric in their first two arguments while $B(p_1^2,p_2^2;p_3^2)$ is anti-symmetric. $S(p_1^2,p_2^2,p_3^2)$ is completely anti-symmetric.

With this form we have a proper description of the correlation function extracted from the lattice, with the right hand side of \eqref{eq:3gluon_correlation_trace} being replaced by 
\begin{align}
	G_{\mu_1\mu_2\mu_3}(p_1,p_2,p_3) = \frac{N_c(N_c^2-1)}{4}& D_{\mu_1\nu_1}(p_1)D_{\mu_2\nu_2}(p_2)D_{\mu_3\nu_3}(p_3)\times \nonumber \\
	 &\times(\Gamma_{\nu_1\nu_2\nu_3}^{(L)}(p_1,p_2,p_3) + \Gamma_{\nu_1\nu_2\nu_3}^{(T)}(p_1,p_2,p_3))
	 \label{eq:correlation_function 3gluon}
\end{align}
where the colour factor comes from the trace operation and $N_c=3$. The extraction of a general form factor is done by suitable projectors built analogously to those considered for the propagator. 

\subsection*{Kinematical configuration $(p,0,-p)$}

The kinematics used in this work is defined by $(p_1,p_2,p_3)=(p,0,-p)$ which due to having a single scale $p$ allows only the longitudinal terms. This is because contractions with external propagators eliminate the transverse terms with
\begin{equation}
	\mymomp{i}\Gamma_{\mu_1\mu_2\mu_3}^{(T)}(p_1,p_2,p_3)=0
\end{equation}
for any $i=1,2,3$.
The explicit expression for \cref{eq:correlation_function 3gluon} becomes
\begin{equation}
G_{\mu_1\mu_2\mu_3}(p,0,-p) = V\frac{N_c(N_c^2-1)}{4}D(p^2)^2D(0)\Gamma(p^2)\mymomp{2}\left(\delta_{\mu_1\mu_3}-\frac{\mymomp{1}\mymomp{3}}{p^2}\right)
\end{equation}
with
\begin{equation}
	\Gamma(p^2) = 2\left(p^2C(p^2,p^2;0) - A(p^2,p^2;0) \right)
\end{equation}
a dimensionless form factor. We see that for this specific configuration, only a combination of form factors can be extracted.
Finally, the 1PI form factor $\Gamma(p^2)$ can be projected by the following contraction 
\begin{equation}
	\Gamma(p^2)p^2 = \frac{4 \mymomp{2}\delta_{\mu_1\mu_3}G_{\mu_1\mu_2\mu_3}(p,0,-p)}{V N_c(N_c^2-1)D(p^2)^2D(0)(N_d-1)}
	\label{eq:3gluon_gamma}
\end{equation}
for non-vanishing momentum.

\section{Four gluon vertex}\label{sec:intro_fourgluon_bases}

The four point correlation function in QCD is the most complex elementary correlation function arising in the Yang-Mills theory. Having three independent momenta, four Lorentz and colour indices, it generates a large amount of possible structures \cite{Huber_2020}. 
On the other hand, being a higher order correlation function, its signal from the Monte-Carlo simulations is strongly affected by noise. 
This last problem justifies the absence of previous four gluon lattice studies. 

\begin{figure}[htb!]
	\centering
		\begin{tikzpicture}[baseline={([yshift=-.5ex]current bounding box.center)}]
	\begin{feynman}[small]
	\vertex (a1);
	\vertex[right= 2cm of a1] (a2);
	\vertex[below= 2cm of a1] (a3);
	\vertex[right= 2cm of a3] (a4);
	\vertex[below right= 1cm and 1cm of a1] (a5);
	
	\diagram* {
		(a1) -- [gluon] (a5),
		(a2) -- [gluon] (a5),
		(a3) -- [gluon] (a5),
		(a4) -- [gluon] (a5)
	};
	
	\end{feynman}
	\draw[fill=black!70,thick] (1,-1) circle (0.3cm);
	\end{tikzpicture}
	=
	\begin{tikzpicture}[baseline={([yshift=-.5ex]current bounding box.center)}]
	\begin{feynman}[small]
	\vertex (a1);
	\vertex[right= 2cm of a1] (a2);
	\vertex[below= 2cm of a1] (a3);
	\vertex[right= 2cm of a3] (a4);
	\vertex[below right= 1cm and 1cm of a1] (a5);
	
	\diagram* {
		(a1) -- [gluon] (a5),
		(a2) -- [gluon] (a5),
		(a3) -- [gluon] (a5),
		(a4) -- [gluon] (a5)
	};
	\end{feynman}
	\draw[fill=black!40,thick] (1,-1) circle (0.23cm);
	\end{tikzpicture}
	+3
	\begin{tikzpicture}[baseline={([yshift=-.5ex]current bounding box.center)}]
	\begin{feynman}[small]
	\vertex (a1);
	\vertex[right= 4cm of a1] (a2);
	\vertex[below= 2cm of a1] (a3);
	\vertex[right= 4cm of a3] (a4);
	\vertex[below right= 1cm and 1cm of a1] (a5);
	\vertex[right= 2cm of a5] (a6);
	
	\diagram* {
		(a1) -- [gluon] (a5),
		(a2) -- [gluon] (a6),
		(a3) -- [gluon] (a5),
		(a4) -- [gluon] (a6),
		(a5) -- [gluon] (a6)
	};
	
	\end{feynman}
	\draw[fill=black!40,thick] (1,-1) circle (0.23cm);
	\draw[fill=black!40,thick] (3,-1) circle (0.23cm);
	\end{tikzpicture}
	+3
	\begin{tikzpicture}[baseline={([yshift=-.5ex]current bounding box.center)}]
	\begin{feynman}[small]
	\vertex (a1);
	\vertex[right= 2cm of a1] (a2);
	\vertex[below= 2cm of a1] (a3);
	\vertex[right= 2cm of a3] (a4);
	
	\diagram* {
		(a1) -- [gluon] (a2),
		(a3) -- [gluon] (a4)
	};
	
	\end{feynman}
	\end{tikzpicture}
	\caption{Diagrammatic representation of the connected and disconnected terms contributing for the full, four-gluon correlation function.}
	\label{fig:scheme 4 gluon}
\end{figure}
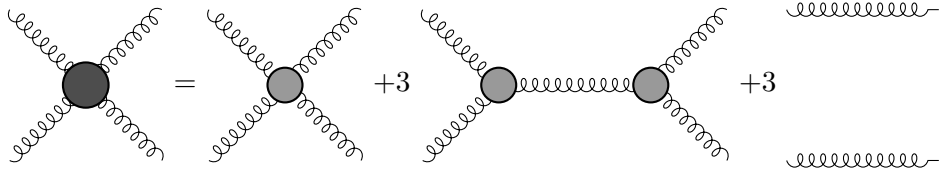

A further complication arises for this higher order correlation function. 
We are interested in computing the four gluon 1PI function, i.e. the pure four gluon vertex. While for the three gluon vertex this is simply obtained by the removal of external propagators from the complete correlation function, the four gluon correlation function carries additional contributions from lower order Green's functions.
Namely, disconnected terms and the three gluon vertex enter in the computation of the complete correlation function -- see \cref{fig:scheme 4 gluon}.
Thus the object we have access in the lattice for a general momentum configuration reads
\begin{align}
G^{(4)a_1a_2a_3a_4}_{\mu_1\mu_2\mu_3\mu_4}&(p_1,p_2,p_3,p_4) =  \nonumber \\
&D_{\mu_1\nu_1}(p_1)D_{\mu_2\nu_2}(p_2)D_{\mu_3\nu_3}(p_3)D_{\mu_4\nu_4}(p_4)\bar \Gamma^{(4)a_1a_2a_3a_4}_{\nu_1\nu_2\nu_3\nu_4}(p_1,p_2,p_3,p_4) \nonumber \\
& -iD_{\mu_1\nu_1}(p_1)D_{\mu_4\nu_4}(p_4)\Gamma^{(3)ma_1a_4}_{\sigma\nu_1\nu4}(p_1+p_4,p_1,p_4)\times \nonumber \\ 
&~~~\times D_{\sigma\rho}(p_1+p_4)\Gamma^{(3)ma_2a_3}_{\rho\nu_2\nu3}(p_2+p_3,p_2,p_3)D_{\mu_2\nu_2}(p_2)D_{\mu_3\nu_3}(p_3) \nonumber \\
& +D^{a_1a_3}_{\mu_1\mu_3}(p_1)D^{a_2a_4}_{\mu_2\mu_4}(p_2)\delta(p_1 + p_3)\delta(p_2 + p_4) \nonumber \\
& + \text{cyclic permutations.}
\label{eq:4gluon disconnected}
\end{align}
Only the first term, that includes the four gluon 1PI function is of interest to us and the remaining ought to be removed. 

We wish to remove lower order contributions without affecting the quality of the signal. 
Hence, we do not directly subtract the unwanted contributions in the simulations since other than requiring a heavier computation, the statistical fluctuations would be increased.
To carry out this extraction we consider a suitable choice of kinematics.
 
To see how this removes the unwanted contributions we notice that momentum conservation constrains the possible kinematic configuration for each vertex. Moreover, the orthogonality of external gluon propagators eliminates terms when contracted with the corresponding momentum
\begin{equation}
	p_\mu D_{\mu\nu}(p) = 0.
\end{equation}

The disconnected terms without interaction (last line in \cref{eq:4gluon disconnected}) are eliminated by a suitable kinematic configuration, that while allowed by momentum conservation for the four gluon vertex, it is not permitted for the two propagators.
Whereas the cancellation of disconnected terms is straightforward, the three gluon contributions requires to notice that the most general rank-3 continuum tensor necessarily involves a momentum factor. They are either linear, $\metrica{1}{2}\mymom{1}{3}$ or cubic in the momenta $\mymom{1}{2}\mymom{2}{3}\mymom{3}{1}$ -- see \cref{subsec:three gluon continuum basis}. Therefore we can eliminate the three gluon contribution by eliminating each of these terms appearing in $\Gamma^{(3)}$ above. If we choose a single scale momentum configuration $(p_1,p_2,p_3,p_4)=(ap,bp,cp,dp)$\footnote{Of the coefficients $a,b,c,d$ only three are independent, by momentum conservation.}, each external propagator will be of the form  $D_{\mu\nu}(p)$ thus eliminating each of the three gluon tensor structures by orthogonality. 

We see that a proper choice of kinematic configuration provides access to the pure four gluon vertex in the lattice 
\begin{align}
	G^{(4)a_1a_2a_3a_4}_{\mu_1\mu_2\mu_3\mu_4}&(ap,bp,cp,dp) =  \nonumber \\
	&D_{\mu_1\nu_1}(ap)D_{\mu_2\nu_2}(bp)D_{\mu_3\nu_3}(cp)D_{\mu_4\nu_4}(dp)\bar \Gamma^{(4)a_1a_2a_3a_4}_{\nu_1\nu_2\nu_3\nu_4}(ap,bp,cp,dp))
	\label{eq:pure four gluon vertex}
\end{align}
using the complete correlation function only, i.e. without additional operations involving lower order functions.

\subsection{Tensor bases}

Having access to the four gluon 1PI function we need to construct a tensor basis in which this function will be projected.
This basis involves a large number of possible structures. At the level of Lorentz tensors, there are three types of structures allowed that are built with the metric tensor and momenta. These are linear, quadratic or quartic in momenta,
\begin{equation}
	\{g_{\mu_1\mu_2}g_{\mu_3\mu_4},~g_{\mu_1\mu_2}p_{\mu_3}q_{\mu_4}, ~p_{\mu_1}q_{\mu_2}r_{\mu_3}k_{\mu_4}\}.
	\label{eq:4gluon lorentz possible tensors}
\end{equation}
which for a general momentum configuration make up 138 possible structures \cite{Gracey_2014}.
However, due to practical reasons, in the present work we consider a reduced basis limited to the first elements using the metric tensor only\footnote{Although this approximation cuts a large number of possible tensor structures, previous investigations found that the tree-level tensor seems to  provides the leading contribution in comparison with the rest of tensor structures \cite{Huber_2015}. This behaviour is also found in the three gluon correlation function \cite{Eichmann_3gluon}.}.  With this choice, only a smaller number of independent tensors will contribute to the vertex. 

For the colour sector we can use the $SU(3)$ antisymmetric structure constants $f^{abc}$, the symmetric terms $d^{abc}$ as well as $\delta^{ab}$ to construct all possible structures
\begin{equation}
	\{f^{ma_1a_2}f^{ma_3a_4},~d^{ma_1a_2}d^{ma_3a_4},~d^{ma_1a_2}f^{ma_3a_4},~\delta^{a_1a_2}\delta^{a_3a_4}\}.
	\label{eq:4gluon colour possible tensors}
\end{equation}
However, various group identities reduce the number of possible terms, see \cref{apend:liegroups}.

Due to the complexity associated with the tensor basis for a general kinematic configuration, in the following we restrict the construction to a specific, single scale configuration.

\subsection*{Kinematical configuration $(p,p,p,-3p)$}

We work with the configuration $(p,p,p,-3p)$ which was considered in the continuum investigations \cite{Binosi_2014,Huber_2015}. The most complete basis within our approximation to metric structures consists of three possible Bose symmetric tensors. These are the tree-level tensor, written again for convenience
\begin{align}
		{\Gamma^{(0)}}_{\mu_1\mu_2\mu_3\mu4}^{a_1a_2a_3a_4} = -g^2\big[&f^{a_1a_2m}f^{a_3a_4m}(g_{\mu_1\mu_3}g_{\mu_2\mu_4} - g_{\mu_1\mu_4}g_{\mu_2\mu_3}) \nonumber \\
		&f^{a_1a_3m}f^{a_2a_4m}(g_{\mu_1\mu_2}g_{\mu_3\mu_4} - g_{\mu_1\mu_4}g_{\mu_2\mu_3}) \nonumber \\
		&f^{a_1a_4m}f^{a_2a_3m}(g_{\mu_1\mu_2}g_{\mu_3\mu_4} - g_{\mu_1\mu_3}g_{\mu_2\mu_4})
		\big],
		\label{eq:4gluon tensor tree level}
\end{align}
a fully symmetric tensor (in both colour and Lorentz sectors)
\begin{equation}
	G^{a_1a_2a_3a_4}_{\mu_1\mu_2\mu_3\mu_4} = (\delta^{a_1a_2}\delta^{a_2a_3} + \delta^{a_1a_3}\delta^{a_2a_4} + \delta^{a_1a_4}\delta^{a_2a_3})(\metrica{1}{2}\metrica{3}{4} + \metrica{1}{3}\metrica{2}{4} + \metrica{1}{4}\metrica{2}{3})
	\label{eq:4gluon tensor G}
\end{equation}
which is orthogonal to $\Gamma^{(0)}$ in both spaces
\begin{align}
	{\Gamma^{(0)}}_{\mu_1\mu_2\mu_3\mu4}^{b_1b_2b_3b_4}	 G^{a_1a_2a_3a_4}_{\mu_1\mu_2\mu_3\mu_4} = 0, &&
	{\Gamma^{(0)}}_{\nu_1\nu_2\nu_3\nu4}^{a_1a_2a_3a_4}	 G^{a_1a_2a_3a_4}_{\mu_1\mu_2\mu_3\mu_4} = 0.
	\label{eq:orthogonality 4gluon tensors}
\end{align}
And finally, the third independent tensor is
\begin{align}
	X^{a_1a_2a_3a_4}_{\mu_1\mu_2\mu_3\mu_4} = &\metrica{1}{2}\metrica{3}{4}\left(\frac{1}{3}\delta^{a_1a_2}\delta^{a_3a_4} - d^{ma_1a_2}d^{ma_3a_4}\right) \nonumber \\
	+&\metrica{1}{3}\metrica{2}{4}\left(\frac{1}{3}\delta^{a_1a_3}\delta^{a_2a_4} - d^{ma_1a_3}d^{ma_2a_4}\right) \nonumber \\
	+&\metrica{1}{4}\metrica{2}{3}\left(\frac{1}{3}\delta^{a_1a_4}\delta^{a_2a_3} - d^{ma_1a_4}d^{ma_2a_3}\right).
\end{align}

With this tensor basis, we construct the general structure with three symmetric form factors as
\begin{equation}
	\Gamma^{a_1a_2a_3a_4}_{\nu_1\nu_2\nu_3\nu_4} = V'_{\Gamma^{(0)}}(p^2){\Gamma^{(0)}}_{\mu_1\mu_2\mu_3\mu4}^{a_1a_2a_3a_4} + V'_G(p^2) G^{a_1a_2a_3a_4}_{\mu_1\mu_2\mu_3\mu_4} + V'_X(p^2) X^{a_1a_2a_3a_4}_{\mu_1\mu_2\mu_3\mu_4}.
\end{equation}
with scalar form factors $V'_i$ depending on the single momentum scale $p$.
This in turn is related to the complete correlation function by the contraction with four external propagators. 
To extract each form factor from the lattice we again apply the trace operation in the colour space.
This operation involves the structures in \cref{eq:4gluon colour possible tensors} which make for more intricate operations than the one found for the three gluon vertex. For these the group identities in \cref{apend:liegroups} were used.
Using the notation
\begin{equation}
	\Tr\left[G_{\mu_1\mu_2\mu_3\mu_4}\right] = D_{\mu_1\nu_1}(p_1)D_{\mu_2\nu_2}(p_2)D_{\mu_3\nu_3}(p_3)D_{\mu_4\nu_4}(p_4)\sum_{\substack{a_i\\
			i \in {1,2,3,4} }}
		\Tr\left(t^{a_1}t^{a_2}t^{a_3}t^{a_4}\right)\Gamma^{a_1a_2a_3a_4}_{\nu_1\nu_2\nu_3\nu_4}
\end{equation}
with the arguments of $G_{\mu_1\mu_2\mu_3\mu_4}(p_1,p_2,p_3,p_4)$ and $\Gamma_{\mu_1\mu_2\mu_3\mu_4}(p_1,p_2,p_3,p_4)$ omitted, and after performing the three non-vanishing Lorentz contractions we obtain
\begin{align}
	&\metrica{1}{2}\metrica{3}{4}\Tr\left[G_{\mu_1\mu_2\mu_3\mu_4}\right] = 6 A_n V_{\Gamma^{(0)}} + 15G_n V_G + 3(4X_n+X'_n)V_X \label{eq:binosi contract1} \\
	&\metrica{1}{3}\metrica{2}{4}\Tr\left[G_{\mu_1\mu_2\mu_3\mu_4}\right] = -12 A_n V_{\Gamma^{(0)}} + 15G_n V_G + 3(2X_n+3X'_n)V_X \label{eq:binosi contract2} \\
	&\metrica{1}{4}\metrica{2}{3}\Tr\left[G_{\mu_1\mu_2\mu_3\mu_4}\right] =  6 A_n V_{\Gamma^{(0)}} + 15G_n V_G + 3(4X_n+X'_n)V_X \label{eq:binosi contract3}
\end{align}
where the $V_i$ are related to the pure vertex form factors by
\begin{equation}
	V_i(p^2) = V'_i(p^2)D(p^2)^3 D(9p^2),
\end{equation}
and the following colour coefficients resulting from the trace and sum operation are
\begin{align}
	&A_n = \frac{N_c^2(N_c^2-1)}{8}, \\
	&G_n = \frac{N_c^2-1}{4N_c^2}(2N_c^2 - 3),\\
	&X_n = \frac{1}{3}\frac{(N_c^2 - 1)^2}{4N_c} - \frac{(N_c^2-1)(N_c^2-4)^2}{8N_c^2}, \\
	&X'_n = -\frac{1}{3}\frac{(N_c^2 - 1)^2}{4N_c} - \frac{(N_c^2-1)(N_c^2-4)}{2N_c^2}.
\end{align}

Our interest is to obtain each form factor $V$ independently, however by looking at \cref{eq:binosi contract1,eq:binosi contract2,eq:binosi contract3} we see that only two contractions are linearly independent and thus only two objects can be extracted. 
Hence, following \cite{Binosi_2014} the $X$ structure will be disregarded. With this further approximation the equations simplify to
\begin{align}
&\metrica{1}{2}\metrica{3}{4}\Tr\left[G_{\mu_1\mu_2\mu_3\mu_4}\right] = 6 A_n V_{\Gamma^{(0)}} + 15G_n V_G \label{eq:binosi contract1 no X} \\
&\metrica{1}{3}\metrica{2}{4}\Tr\left[G_{\mu_1\mu_2\mu_3\mu_4}\right] = -12 A_n V_{\Gamma^{(0)}} + 15G_n V_G \label{eq:binosi contract2 no X}
\end{align}
and each form factor is obtained by
\begin{align}
	&V_{\Gamma^{(0)}} = \frac{1}{18 A_n}\left( \metrica{1}{2}\metrica{3}{4}\Tr\left[G_{\mu_1\mu_2\mu_3\mu_4}\right] - 
	\metrica{1}{3}\metrica{2}{4}\Tr\left[G_{\mu_1\mu_2\mu_3\mu_4}\right] \right), \\
	&V_G = \frac{1}{45 AG_n}\left( 2 \metrica{1}{2}\metrica{3}{4}\Tr\left[G_{\mu_1\mu_2\mu_3\mu_4}\right] + 
	\metrica{1}{3}\metrica{2}{4}\Tr\left[G_{\mu_1\mu_2\mu_3\mu_4}\right] \right).
	\label{eq: form factors 4gluon}
\end{align}
These complete form factors are obtained in lattice Monte-Carlo simulations by computing the corresponding linear combinations of the complete correlation function $G_{\mu_1\mu_2\mu_3\mu_4}$.
In \cref{sec:result_4gluon}, Monte-Carlo results for this kinematic configurations will be presented.

\chapter{Results}\label{chap:results}

In this chapter we investigate lattice tensor representations of the gluon propagator by considering the tensor structures introduced in the previous chapter. 
In addition we study the IR behaviour of the three gluon correlation function and report a first computation of the lattice four gluon correlation function.
All results were obtained in a Landau gauge, 4-dimensional pure $SU(3)$ Yang-Mills theory from the Wilson action, \cref{eq:Wilson action}.

\begin{table}[htb!]
	\centering
	\begin{tabular}{ccccccc}
		\toprule
		$a~(\si{fm})$               & $1/a~(\si{GeV})$                  & $\beta$              & $N$ & $V~(\si{fm^4})$ & $\#$config & $p_{\text{min}}~(\si{GeV})$ \\ \midrule
		\multirow{2}{*}{0.1016(25)} & \multirow{2}{*}{1.943(47)} & \multirow{2}{*}{6.0} & 80  & $(8.128)^4$     & 550    & $0.153$              \\
		&                            &                      & 64  & $(6.502)^4$     & 2000   & $0.191$      \\ \bottomrule       
	\end{tabular}
	\caption{Lattice setup for both ensembles used in the computation of the gluon correlation functions.}
	\label{tab:lattices}
\end{table}

The lattice setup used in this work can be seen in \cref{tab:lattices}.
We used two ensembles with the same lattice spacing but different volumes. The smaller volume lattice also has a larger number of configurations.

The results shown are either dimensionless or expressed in terms of lattice units. However, these are shown as a function of the physical momentum, $p = p_\text{lat}a^{-1}\gev$ with $a^{-1} = 1.943(47)\gev$.
Additionally, all results represent bare quantities, i.e. non-renormalized values. Renormalized values would differ only by an overall constant factor which does not affect the conclusions.

A complete $H(4)$ group averaging is applied for all quantities as defined in \cref{sec:z4 averaging}. An average of the quantity is taken over all group equivalent points for each gauge field configuration. Only then the ensemble average is taken. 
Also, the reader should be aware that scalar functions on the lattice have the four $H(4)$ invariants as arguments although represented herein with $p^2$ only.  The exception is the case of the extrapolated values where the dependence is partially corrected.

The error bars shown correspond to a tenfold bootstrap sampling from the original set of configurations.
For H4 corrected data, error bars result from an initial bootstrap, followed by the linear regression propagation. Regarding the correction methods, we will use the following convention through all results (unless explicitly stated) -- $\invmomp{4}$ extrapolated data is shown always as a function of the usual lattice momentum $p$ while momentum cuts are generally reserved for the improved momentum data $\hat p$.

\section{Gluon propagator -- Tensor description}\label{sec:results propagator}

In this section we consider the lattice description of the gluon propagator, compared with the usual continuum tensor structure.
For most of this section we analyse the $80^4$ lattice exclusively. The $64^4$ lattice will be considered in the end in order to search for possible finite volume effects on the results.

\subsection{Discretization correction methods}

We begin by illustrating the correction methods defined in the previous chapter to illustrate its advantages and setbacks. We use the gluon propagator as a test, but the conclusions should be applicable to other correlation functions as well as other tensor structures.

All results shown in this analysis are for the continuum tensor \cref{eq:full_continuum_propagator} with form factor $D(p^2)$ and dimensionless dressing function $d(p^2) = p^2D(p^2)$. 
\begin{equation}
D(p^2) = \frac{1}{(N_c^2 - 1)(N_d-1)}\sum_{\mu}D_{\mu\mu}(p).
\end{equation}
Notice that the extraction of $D(p^2)$ is independent of the use of the normal or improved momentum for the basis.

\begin{figure}[htb!]
	\centering
	\input{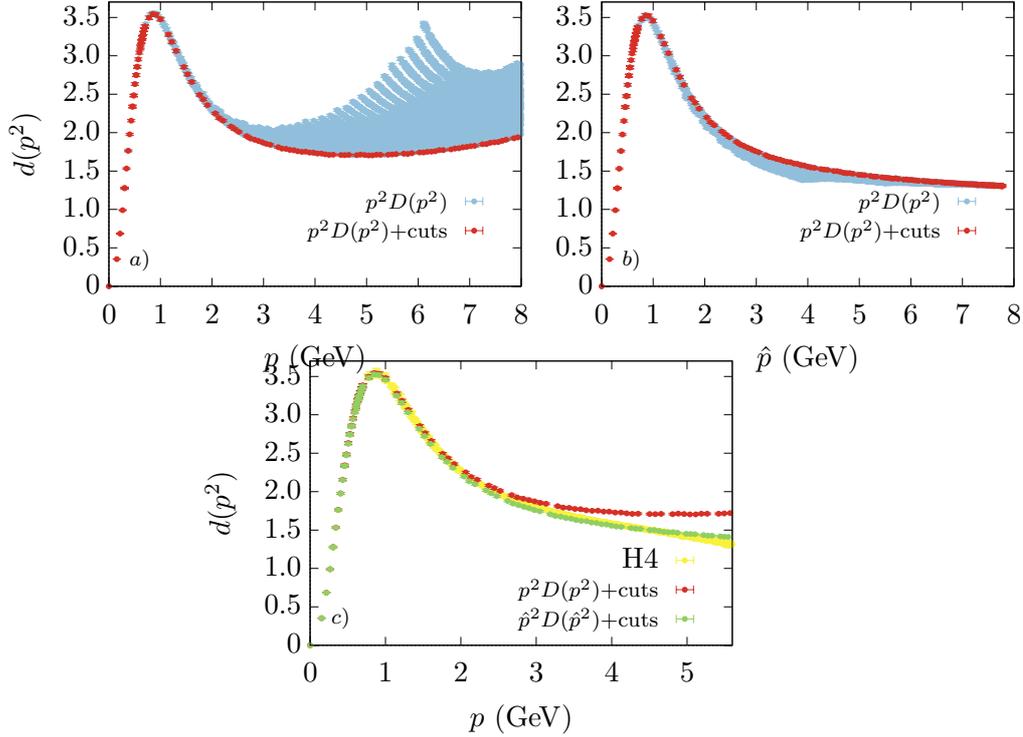}
	\caption{Gluon dressing function $d(p^2)$ from the continuum basis as a function of lattice momentum (top left), and as a function of the improved momentum (top right). The momenta surviving cylindrical and conical cuts are shown for the each plot.
	The comparison between the data in terms of the improved and lattice momenta after complete momentum cuts against the H4 corrected data with lattice momentum is shown in the bottom plot. Results from the $\beta=6.0,~80^4$ lattice.}
	\label{fig:anisotropy_corrections}
\end{figure}

In \cref{fig:anisotropy_corrections} results for the correction methods are shown -- use of the improved momentum; momentum cuts; and the H4 extrapolation. 
In $a)$ and $b)$ the complete data and after momentum cuts is shown in terms of lattice and improved momentum, respectively.
The complete set of data shows structures created by the hypercubic artifacts which are much more pronounced when using lattice momentum.
This is expected since, as introduced in the \cref{sec:latticeartifacts}, $\hat p$ partially accounts for hypercubic errors up to $\order{a^2}$.
The use of the complete momentum cuts (cylindrical and conical) are also shown, and create a much smoother curve. 

The curves in terms of lattice and improved momentum after cuts do not agree for momenta above $\sim 2.5\gev$, this is visible in \cref{fig:anisotropy_corrections} $c)$.
In this plot, the $\invmomp{4}$ extrapolated data is also shown, and we see that it matches the data with cuts as a function of improved momentum for a large range.
An advantage from the extrapolation method is that it offers a higher density of points for a large range when compared with the curve surviving the cuts. 
However, other than the loss of information for lower momentum, the high momentum region is also problematic due to the lack of different $H(4)$ orbits, hence the extrapolation is not reliable. This becomes noticeable for $p\sim 5\gev$ where the discrepancy can be related to the decline in quality of the extrapolation.

\subsection{Lattice basis -- General kinematics}\label{sec:Lattice Basis - general kinematics}

In this section we compare the behaviour of the usual continuum tensor, \cref{eq:full_continuum_propagator}, with two lattice descriptions given in \cref{eq:full_lattice_basis} and \cref{eq:partial_lattice_basis}. 
The most general continuum basis, \cref{eq:general basis continuum}, will also be considered.
We disregard, for now, the generalized diagonal configurations and other kinematics for which the extraction of all form factors is not possible (details in \cref{append_sec:projectors_latticebasis}).

The dimensionless form factors $p^2\Gamma(p^2)$ will be considered due to their appearance in the continuum relations, defined below. These are $p^2E(p^2)$, $p^4F(p^2)$, $p^4H(p^2)$ for the larger basis. The only exception is for the terms $p^4G(p^2)$, and $p^4I(p^2)$ which are expressed in lattice units.

\subsubsection*{Continuum relations}

To probe the accuracy of our results we consider a benchmark result. We use the data published in \cite{Dudal_2018} from a precise continuum basis computation of the propagator using improved momentum and additional cuts. This result comes from a partial Z4 averaging procedure, i.e. only using momentum permutations. This data will always be referred as $D(\hat p^2)$ or $d(\hat p^2) = \hat p^2D(\hat p^2)$ and shown as a function of improved momentum only.

In addition to this benchmark, we consider continuum relations that relate form factors among themselves and also with the continuum tensor basis result, $D(p^2)$. These relations are expected to be properly satisfied for the infrared region where hypercubic effects are smaller\footnote{Note that this does not guarantee that we are extracting proper continuum physics for the IR region. There are still finite volume and finite spacing effects -- see \cite{Oliveira_2012}.}.
The reproduction of the continuum basis, \cref{eq:full_continuum_propagator}, by the extended basis, \cref{eq:full_lattice_basis}, for low momentum implies
\begin{align}
&E(p^2) \rightarrow D(p^2) \\
&-p^2F(p^2), ~ -p^2H(p^2) \rightarrow D(p^2)\\
&G(p^2), ~ I(p^2) \rightarrow 0.
\label{eq:continuum relations extended}
\end{align}
while for the reduced lattice basis, \cref{eq:partial_lattice_basis}, the continuum relations are
\begin{align}
&J(p^2) \rightarrow D(p^2) \\
&-p^2K(p^2), ~ -p^2L(p^2) \rightarrow D(p^2)
\label{eq:continuum relations reduced}
\end{align}
In addition, for the most general continuum second order tensor, \cref{eq:general basis continuum}, we obtain
\begin{align}
&A(p^2), ~ -p^2B(p^2) \rightarrow D(p^2).
\label{eq:continuum relations AB}
\end{align}

\begin{figure}[htb!]
	\centering
	\input{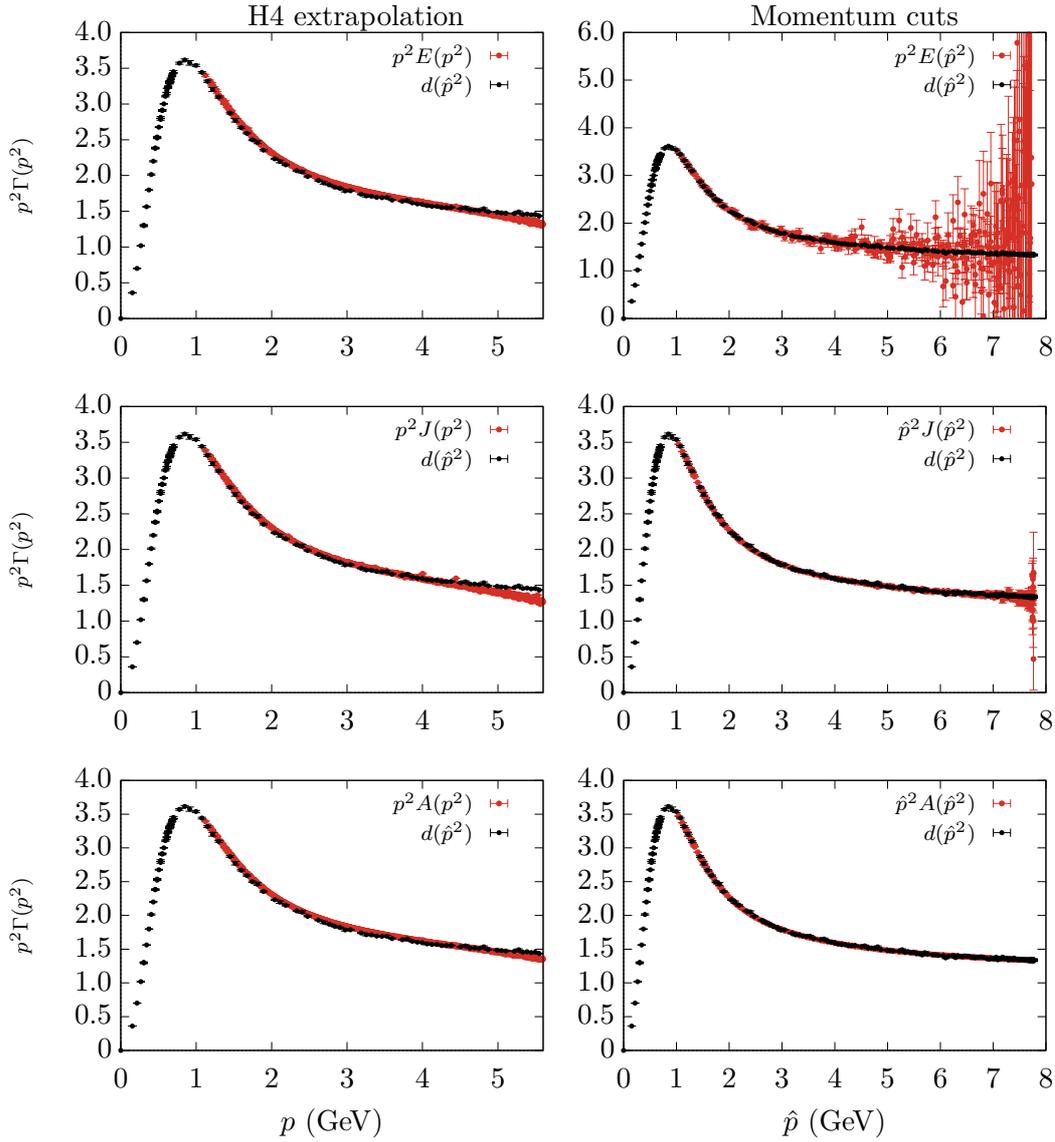}
	\caption{$p^2E(p^2)$, $p^2J(p^2)$, and $p^2A(p^2)$ dressing functions as a function of the lattice momentum after a $\invmomp{4}$ extrapolation (left) and as a function of the improved momentum $\hat p$ after momentum cuts. The results come from the $\beta=6.0,~80^4$ lattice and the benchmark continuum dressing function $\hat p^2D(\hat p^2)$ is plotted as a function of the improved momentum.}
	\label{fig:DEFH_factors}
\end{figure}

\begin{figure}[htb!]
	\centering
	\input{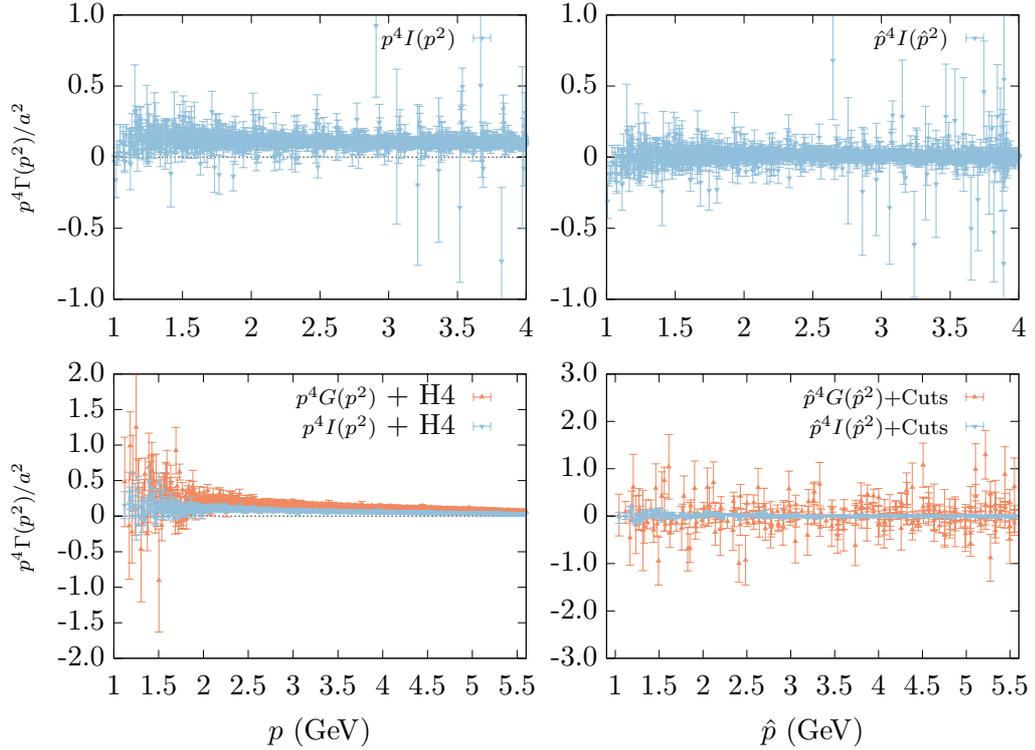}
	\caption{Dimensionless form factors $p^4G(p^2)$ and $p^4I(p^2)$. $G$ is shown only after the correction methods. The original data is shown in the top row for the lattice momentum $p$ (left) and improved momentum $\hat p$ (right) for a restricted range of momenta. Below, $p^4G(p^2)$ and $p^4I(p^2)$ after the corrections are applied are presented, namely the H4 extrapolated results and momentum cuts. All data from the $\beta=6.0,~80^4$ lattice.}
	\label{fig:GI_factors}
\end{figure}

\begin{figure}[htb!]
	\centering
	\input{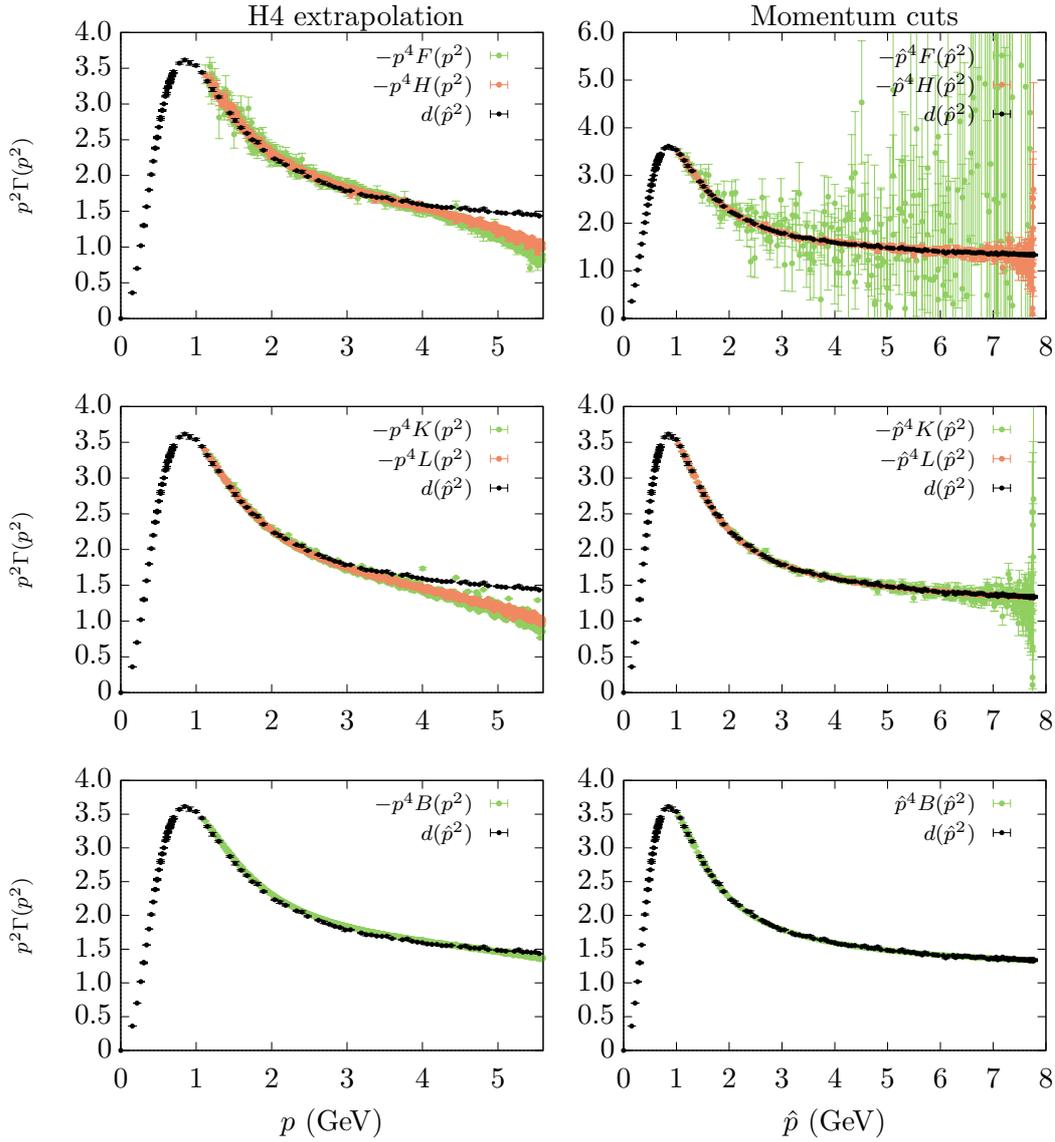}
	\caption{Dressing functions for the different tensor bases as a function of the lattice momentum after a $\invmomp{4}$ extrapolation (left) and as a function of the improved momentum $\hat p$ after momentum cuts. These come from the $\beta=6.0,~80^4$ lattice. The improved continuum tensor form factor $D(\hat p^2)$ is also shown. }
	\label{fig:DEFH_factors_noE}
\end{figure}

The reproduction of these relations can be verified in \cref{fig:DEFH_factors,fig:GI_factors,fig:DEFH_factors_noE} where the form factors are reported as a function of lattice momentum $p$ after a $\invmomp{4}$ extrapolation (left column), and as a function of improved momentum with momentum cuts (right). 
In \cref{fig:DEFH_factors}, we compare only the form factors associated with the metric tensor $E(p^2)$, $J(p^2)$, and $A(p^2)$.

The functions represented in \cref{fig:DEFH_factors,fig:DEFH_factors_noE} are such that in the continuum limit they all should become equal, thus satisfying \cref{eq:continuum relations extended,eq:continuum relations reduced,eq:continuum relations AB}. 
It can be seen for \cref{fig:DEFH_factors,fig:DEFH_factors_noE} that within one standard deviation, continuum relations are satisfied for improved momentum with additional cuts, although with increased fluctuations when compared with the H4 corrected data on the left. 
The latter, however, have a restricted range of compatibility with the benchmark result.
In addition, for \cref{fig:DEFH_factors_noE} the two H4 form factors $F$ and $H$ for the extended basis seem to deviate from the expected behaviour.
The same happens for the smaller lattice basis, and this should be related to the limitations of the extrapolation for low and high momentum.
Despite the fluctuations, the fact that the continuum relations are satisfied for a large range of momentum indicates that the lattice is fine and large enough to obtain results close to continuum.

In \cref{fig:GI_factors}, the form factors $p^4G(p^2)/a^2$ and $p^4I(p^2)/a^2$ are reported. In the bottom row, results are shown after the correction methods are applied for both form factors. 
The appearance of the larger fluctuations for $G$ and $I$ are expected due to its values being closer to zero and the increased mixing among a larger number of form factors when extracting each function.
This is also why $I(p^2)$, which only mixes with $H(p^2)$, shows less fluctuations when compared with $G(p^2)$.

For low momentum, both correction methods and functions satisfy the continuum relations within statistical fluctuations in \cref{fig:GI_factors}. 
However, for momenta above $\sim2\gev$ the H4 extrapolation results deviate from zero. 
This is already visible before the extrapolation is applied. To see this, in the top row $p^4I(p^2)$ is shown for all available configurations without corrections, but for a restricted range of momentum ($p^4G(p^2)$ was disregarded due to having large fluctuations). 
$p^4I(p^2)$ is much closer to zero for the improved momentum basis than for lattice momentum before any correction is applied. This result can be viewed as a another improvement in the tensor description after the change of variables to the momentum $\hat p$ when building the tensor basis. 

\begin{figure}[htb!]
	\centering
	\input{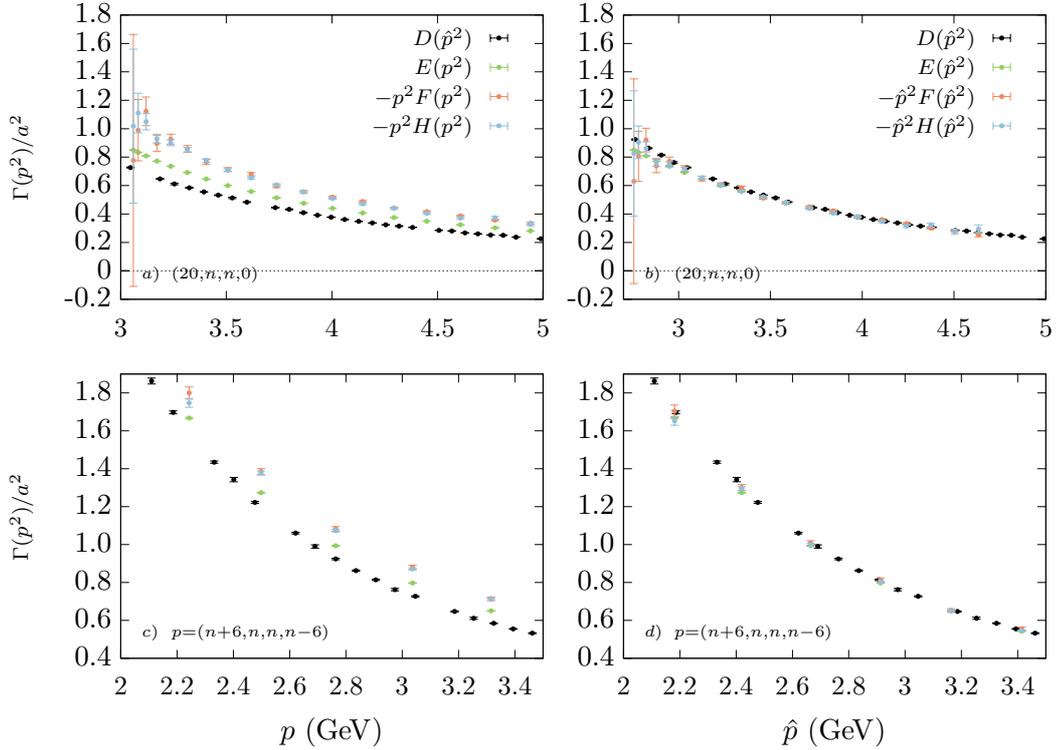}
	\caption{$E(p^2)$, $-p^2F(p^2)$, and $-p^2H(p^2)$ from the improved momentum lattice basis (right) and from the normal momentum lattice basis (left). Data from the $\beta=6.0,~80^4$ lattice. The standard result for $D(\hat p^2)$ is also shown as a function of the improved momentum.}
	\label{fig:DEFH_factors_config}
\end{figure}

In fact, the change of variables from $p$ to $\hat p$ also provides an improvement for the remaining form factors $E(p^2)$, $-p^2F(p^2)$, and $-p^2H(p^2)$. However, this is concealed by the complete set of data, thus specific momentum configurations are helpful in exposing this effect.
In \cref{fig:DEFH_factors_config} these three form factors are shown for two different kinematics for both the normal and improved momentum bases in the left and right columns, respectively.
The continuum relations are much better satisfied for the improved momentum case.
In regards to reproducing the expected result, $D(\hat p^2)$, the form factor $E(p^2)$ shows the best results for lattice momentum. 

The combination of the results from \cref{fig:DEFH_factors,fig:DEFH_factors_noE,fig:GI_factors} means that the continuum relations are properly reproduced for a large range of momenta. This can be interpreted as the survival (at least to some extent) of the Slavnov-Taylor identity and Landau gauge condition on the lattice that fix the form of the gluon propagator to be orthogonal.
This also confirms the improvement obtained from the change of variables $p\rightarrow \hat p$ with respect to the description of lattice correlation functions.

Other than allowing to check the continuum relations, \cref{fig:DEFH_factors,fig:DEFH_factors_noE,fig:GI_factors} allow to compare the three extended tensor bases from the point of view of the general description of the gluon propagator. 
With this analysis we inspect the difference between the reduced and extended lattice bases in regards to reproducing the gluon propagator -- this will be complemented by the reconstruction analysis below.
Turning again to \cref{fig:DEFH_factors,fig:DEFH_factors_noE}, all results portray $\hat p^2D(\hat p^2)$ within one standard deviation, although with increased fluctuations as one increases the basis elements (bottom to top in the right columns). Nonetheless, all three sets of functions seem define a single curve compatible with the benchmark result when represented in terms of the improved momentum $\hat p$. 
However, even with the momentum cuts large fluctuations appear for the larger tensor basis, due to the mixing of different elements in the projection of form factors.
In fact, for $p^4F(p^2)$ in terms of improved momentum in \cref{fig:DEFH_factors_noE} the fluctuations are present through a larger range, starting around $1.5\gev$.

The same form factors, but in terms of the normal momentum bases (left column of both figures) and after the $\invmomp{4}$ extrapolation also reproduce the benchmark result $d(\hat p^2)$ although in a limited range. The H4 extrapolation seems to remove most of the statistical fluctuations when compared to the data in the right column.
For this method there is a clear distinction between the metric, $p^2E(p^2)$, $p^2J(p^2)$, and $p^2A(p^2)$ in \cref{fig:DEFH_factors} and the remaining non-vanishing form factors in \cref{fig:DEFH_factors_noE}. The range of agreement with the benchmark result is larger for the metric form factors with the deviation appearing for $p\sim5\gev$. 
On the other hand, the curves in the left column of \cref{fig:DEFH_factors_noE} have a smaller range of agreement (except for the basis $\{A,B\}$) with deviations starting for lower momenta.
 
 
Regarding the fluctuations appearing for larger tensor bases, this problem can be overcome by using a binning procedure, where points inside each momentum bin are averaged using a weighted average. Although with this we are summing non equivalent points with respect to the group symmetry, this procedure is allowed by noting that the uncertainty in the scale setting (choice of $a$) is around $2.5\%$. This uncertainty allows us to define the bins in which the average is performed. 
For data in terms of lattice momentum, the averaging is taken only for the H4 corrected values.
 
\begin{figure}[htb!]
	\centering
	\input{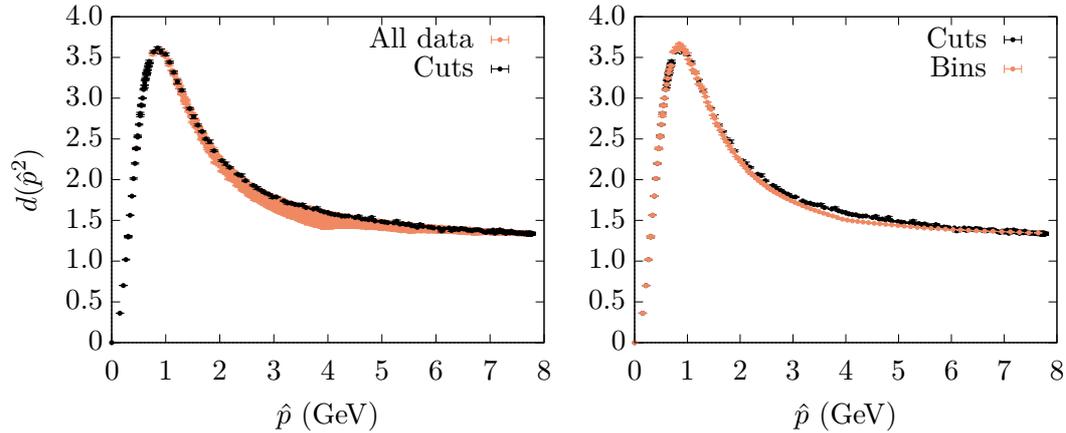}
	\caption{Gluon dressing function $d(\hat p^2)$ as a function of the improved momentum for the continuum basis published in \cite{Dudal_2018}. The left plot shows the complete set of data and the curve surviving momentum cuts. Additionally, the right plot shows the averaged data in each bin -- description in the text.}
	\label{fig:prop_binning_benchmark}
\end{figure}

To understand the reliability of this procedure we start by considering the effect of binning the data for the benchmark result. In \cref{fig:prop_binning_benchmark} the data published in \cite{Dudal_2018} is shown with the usual momentum cuts, as well as the binned results (right plot). 
The binning seems to introduce deviations from the results after cuts for a range between $\hat p \sim 2-5\gev$. This deviation can be accounted for in the following figures since it should be related to the use of the complete set of data in terms of improved momentum which still carries some hypercubic artifacts.

\begin{figure}[htb!]
	\centering
	\input{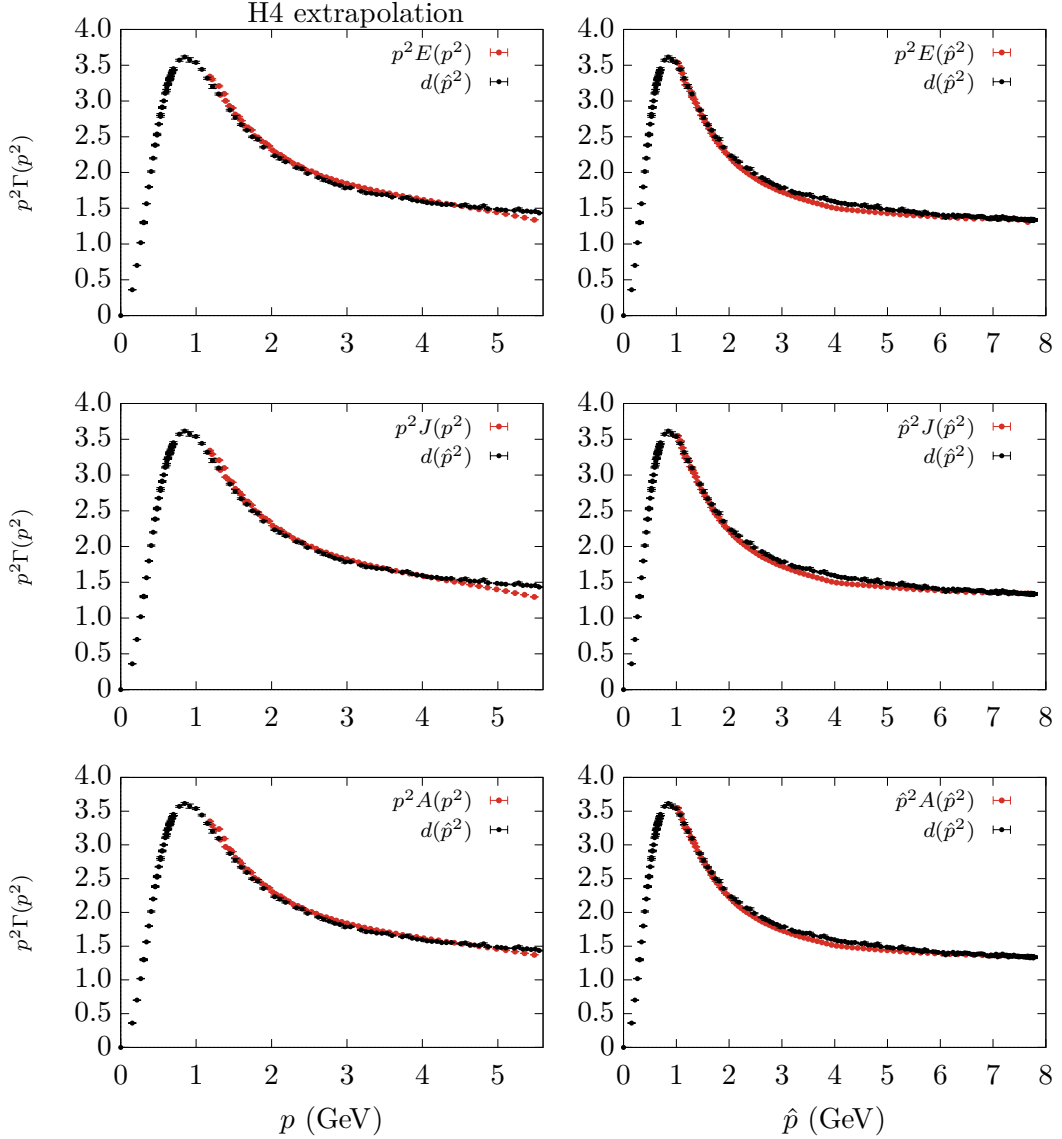}
	\caption{Dressing functions $p^2E(p^2)$, $p^2J(p^2)$, and $p^2A(p^2)$ from the $\beta=6.0,~80^4$ lattice as a function of the lattice momentum after a $\invmomp{4}$ extrapolation (left) and as a function of the improved momentum $\hat p$. The data is shown after a binning of $2.5\%$ in momentum was performed. The continuum dressing function $\hat p^2D(\hat p^2)$ is shown with momentum cuts.}
	\label{fig:DEFH_factors_bin}
\end{figure}

\begin{figure}[htb!]
	\centering
	\input{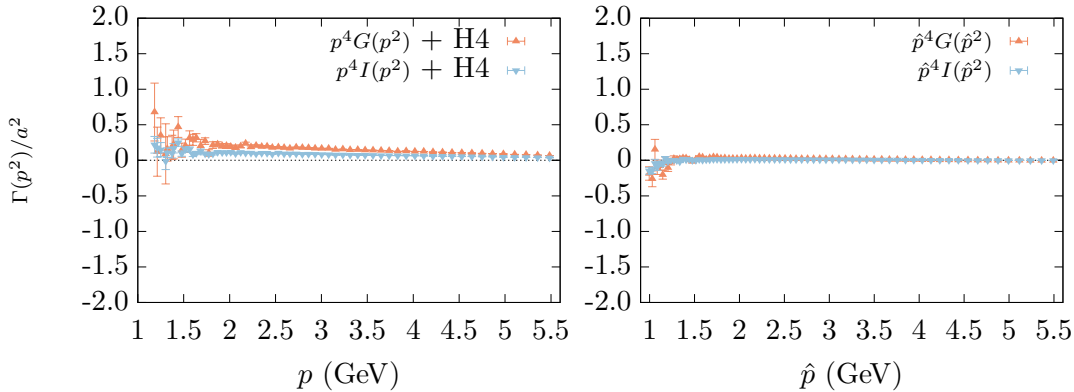}
	\caption{Form factors for the higher order terms of the extended basis $p^4G(p^2)$ and $p^4I(p^2)$ in terms of the usual momentum after the $\invmomp{4}$ extrapolation (left) and as a function of the improved momentum (right) without any correction applied. Both cases are shown after a $2.5\%$ binning is applied in the momentum axis. Data from the $\beta=6.0,~80^4$ lattice.}
	\label{fig:GI_factors_bin}
\end{figure}

\begin{figure}[th!]
	\centering
	\input{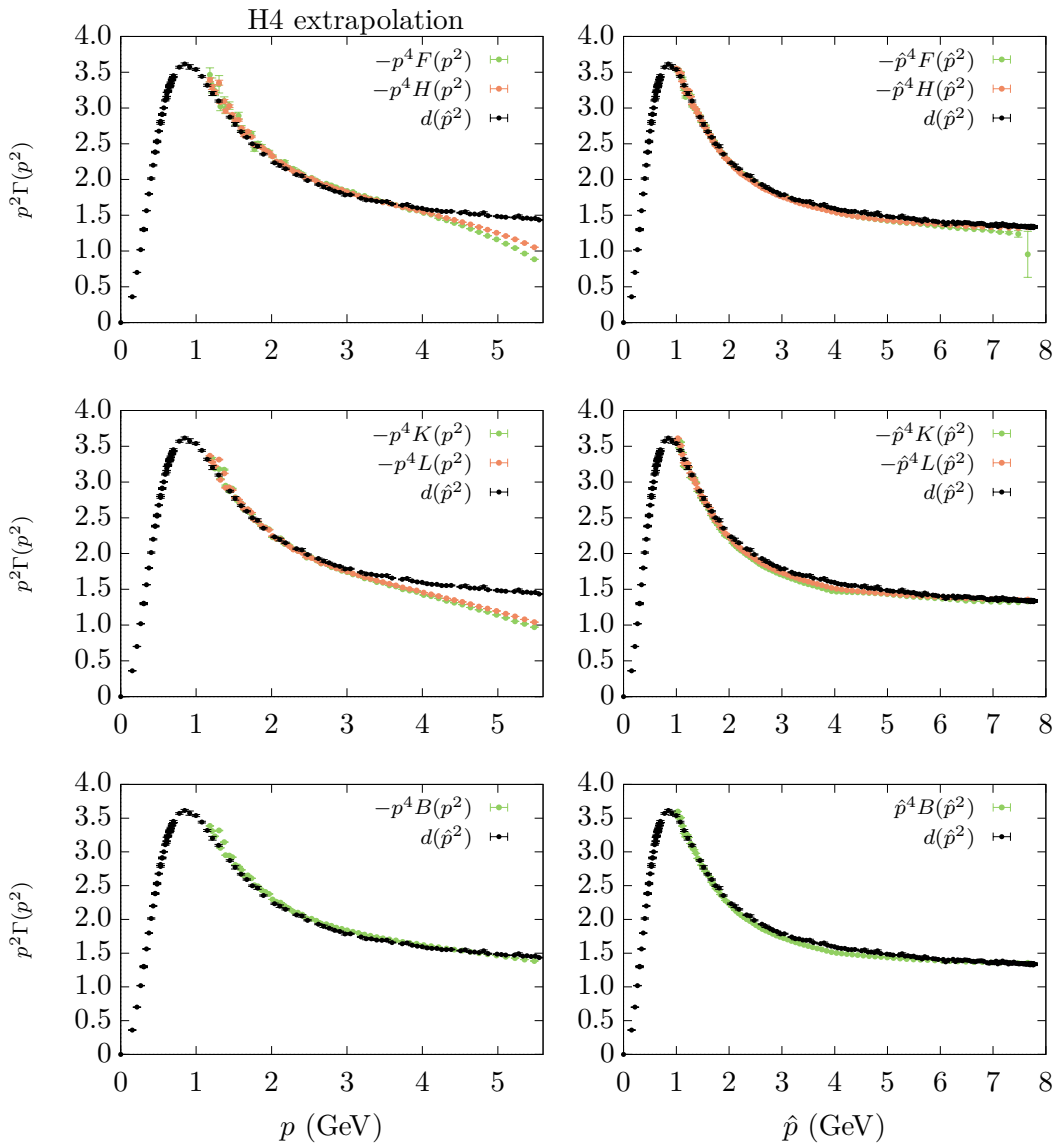}
	\caption{$\beta=6.0,~80^4$ lattice non-metric dressing functions for three tensor bases as a function of the lattice momentum after a $\invmomp{4}$ extrapolation (left) and as a function of the improved momentum $\hat p$, both after a $2.5\%$ binning procedure applied to the momentum. The continuum dressing function $\hat p^2D(\hat p^2)$ is shown with momentum cuts.}
	\label{fig:DEFH_factors_noE_bin}
\end{figure}

The binned versions of \cref{fig:DEFH_factors,fig:DEFH_factors_noE,fig:GI_factors} are shown in \cref{fig:DEFH_factors_bin,fig:DEFH_factors_noE_bin,fig:GI_factors_bin}. The binning of the data defines smoother curves with smaller statistical errors which allow for better analysis of the deviations from the benchmark result. 
For \cref{fig:DEFH_factors_noE_bin} some small fluctuations are noticed for $p\sim1.2\gev$ for the extrapolated data.
This should be related to the fluctuations noticeable in the non-binned counterpart, \cref{fig:DEFH_factors_noE}.

The data as a function of the improved momentum in the right columns of \cref{fig:DEFH_factors_bin,fig:DEFH_factors_noE_bin} shows a good agreement with $d(\hat p^2)$ while the large statistical fluctuations have been absorbed by the averaging procedure. The visible deviation for the mid range of momentum do not appear in non-binned results and should be associated with the binning procedure. 
For the H4 corrected data, the binning procedure results in a reduction of fluctuations and allows to better recognize the deviations from the benchmark result.
In general, for the extrapolated data, the best agreement with the expected result seems to be obtained by the smaller tensor basis $\{A,B\}$. 
For the improved momentum bases the situation is not so clear, the best match with $d(\hat p^2)$ seems to be obtained for $p^2H(p^2)$.

For the form factors $G(p^2)$ and $I(p^2)$, also shown after a binning procedure in \cref{fig:GI_factors_bin}, the interpretation given for \cref{fig:GI_factors} is now much clearer. Large fluctuations for low momentum are expected due to the smallness of $\Delta_1,~\Delta_2$ in the extraction of both terms -- \cref{append_sec:projectors_latticebasis}. 
The improved momentum basis shows a better agreement with the continuum relations, while the normal momentum after the extrapolation shows deviations for higher momenta.

From the above analysis we would conclude that the use of larger bases does not improve the description of the gluon propagator. In fact, the use of larger bases introduces fluctuations in the computations.
This, together with the fact that the continuum relations are obtained through the complete  range of momentum restrains us from considering further additions to the lattice basis. The use of a more complete tensor basis would require an increase in the statistics to counteract the fluctuations coming from the mixing with a larger number of terms.

Regarding the results obtained in \cite{vujinovic2019tensor} using a similar approach, the continuum relations are only satisfied for low momentum (or close to diagonal configurations) while in our case the relations are satisfied through all range of momentum, namely when using $\hat p$.
Note, however that the referred work uses only 2 and 3-dimensional $SU(2)$ lattices with a larger lattice spacing, and thus the comparison is to be taken with care.

\subsubsection{Completeness of the tensor bases}

The analysis of the form factors alone does not offer the full picture for how the lattice bases affect the description of the tensor\footnote{It is important to distinguish the description of the gluon propagator $D(p^2)$, from the description of the original lattice tensor $D_{\mu\nu}(p)$ which is the focus when exploring the completeness of a basis.}. 
Indeed, form factors alone do not allow to perceive how faithful the tensor description with a given basis is. 
The most evident case is for the continuum description which returns the exact same form factor using normal or improved momentum while the latter reproduces the original tensor with greater accuracy. This will be analysed below.

We consider the reconstruction introduced in \cref{sec:reconstruction of tensors} applied to the tensor bases that have been studied, namely the extended and reduced lattice bases, \cref{eq:full_lattice_basis,eq:partial_lattice_basis}, and also the continuum basis with a single form factor $D(p^2)$. 
The reconstruction ratio
\begin{equation}
\mathcal{R} = \frac{\sum_{\mu\nu}|\Gamma^\text{\tiny orig}_{\mu\nu}|}{\sum_{\mu\nu}|\Gamma^\text{\tiny rec}_{\mu\nu}|}
\label{eq:ratio_1}
\end{equation}
is computed using the previously shown form factors.

\begin{figure}[htb!]
	\centering
	\input{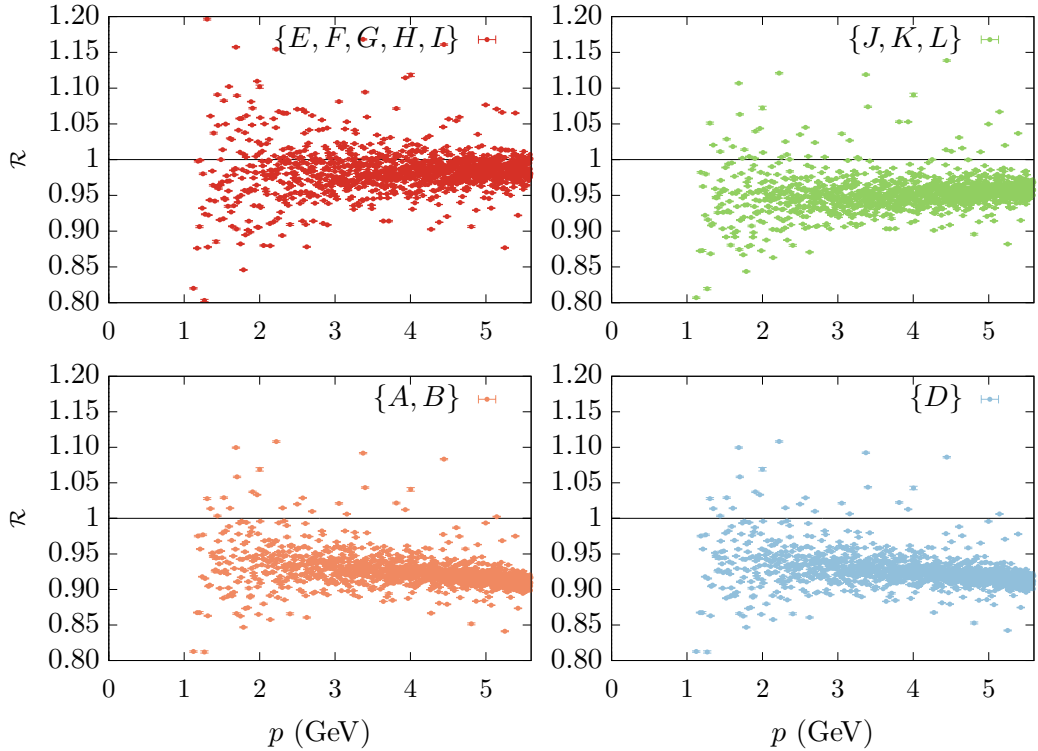}
	\caption{Reconstruction ratio for the normal momentum bases after the H4 extrapolation. Each plot is labelled by the corresponding form factors for each basis. Data from the $\beta=6.0,~80^4$ lattice.}
	\label{fig:recons_H4}
\end{figure}

We begin by consider H4 corrected data shown in \cref{fig:recons_H4}. 
From its analysis we notice an improvement in the reconstruction when adding tensor elements. In fact, the larger basis has the best result when compared to the other three structures, with the results being in general closer to one. 
The comparison between the two continuum tensors is not very informative since the differences appear to be negligible.

To understand the differences in tensor descriptions from the lattice bases we consider specific momentum configurations to evaluate the reconstruction ratio in \cref{eq:ratio_1}. 
The use of specific momentum configurations also helps to reinforce the existence of special kinematics for which the continuum description is approached. 

\begin{figure}[htb!]
	\centering
	\input{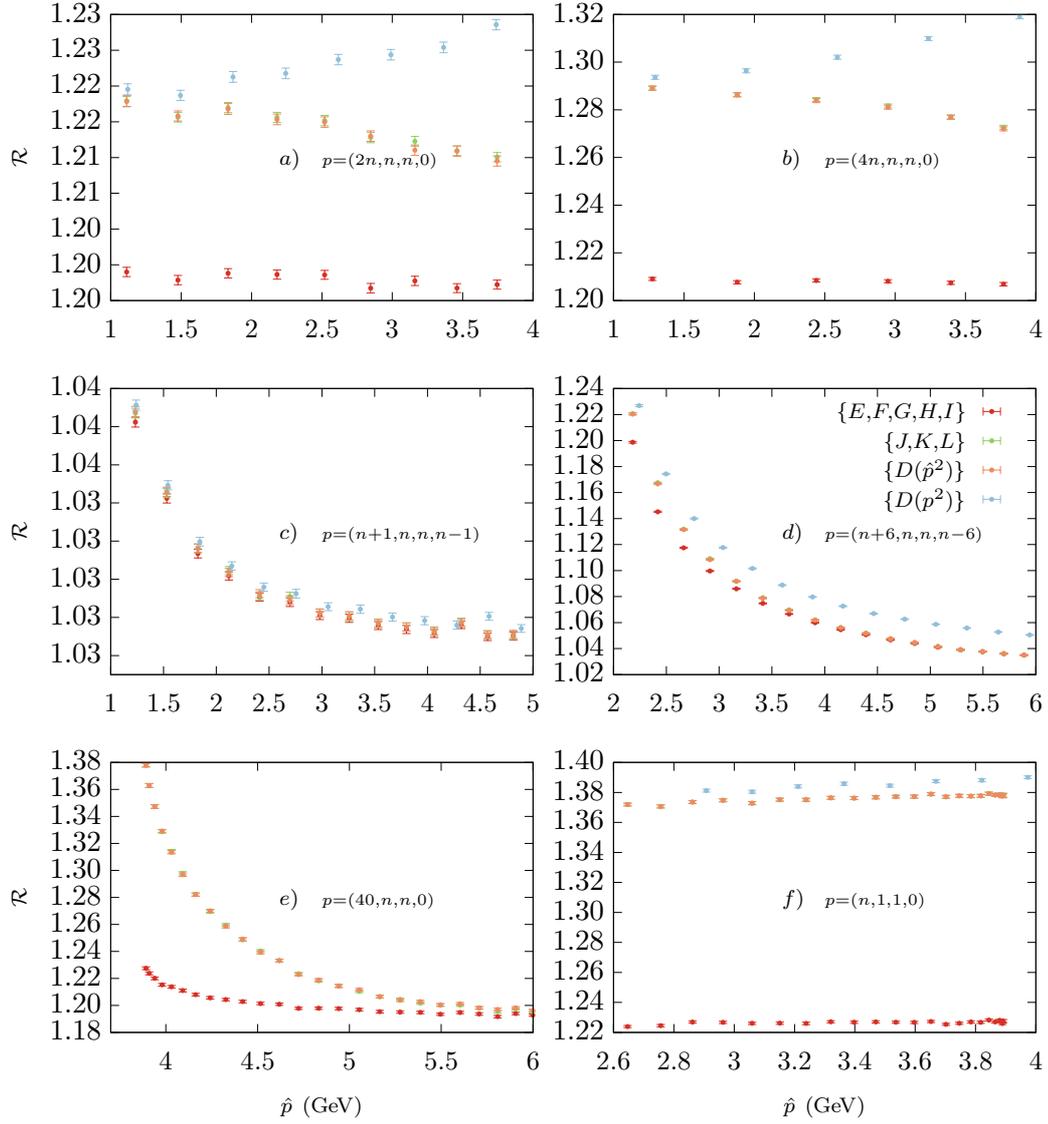}
	\caption{Reconstruction ratio $\mathcal{R}$ for various single scale momentum configurations using two lattice bases, \cref{eq:full_lattice_basis,eq:partial_lattice_basis}, and the continuum tensor \eqref{eq:full_continuum_propagator} using the improved momentum and lattice momentum. Results from the $\beta=6.0,~80^4$ ensemble.}
	\label{fig:recons_general_config}
\end{figure}

In fig. \ref{fig:recons_general_config} the ratio for six different momentum configurations is shown. The range of momentum was chosen for each plot in order to evidence the essential behaviour for each kinematics. 
The continuum basis $\{A,B\}$ is not shown since the results exactly match the ones from the single form factor basis. This could be explained by the orthogonality of the propagator on the lattice, that further restricts the $\{A,B\}$ basis, ending with a single effective form factor.
In addition, to study the differences in using improved or lattice momentum we consider the usual continuum basis in terms of both momenta.
Conversely, both lattice tensors are shown as a function of $\hat p$ only.

The general behaviour in \cref{fig:recons_general_config} shows that the most complete lattice basis is  better at portraying the original tensor, having lower ratios across most of the configurations and for a large range of momentum. There are, however, special kinematic points for which the remaining tensor bases  match the result from this basis. 

Another striking feature comes from the comparison between the two continuum bases using normal and improved momentum. The latter shows better ratios and thus a better description of original tensor\footnote{Notice that although the extraction of $D(p^2)$ is independent of the use of $p$ ou $\hat p$, the use of both momenta changes the description of the full tensor.}.

The first row in \Cref{fig:recons_general_config} displays two similar kinematics, only distinguished by its distance from the diagonal, with $(4n,n,n,0)$ being farther from it. The same general behaviour is obtained for both kinematics, although with a significant improvement for the left case whose $\mathcal{R}$ values are closer to 1 for the whole range of momenta. 

The second row in fig. \ref{fig:recons_general_config} also represents two similar configurations, again with the one on the left being closer to the diagonal, thus having an overall better ratio among all bases. 
Additionally, there is an effect common to both, namely the angle from the diagonal is not constant through all momenta. Instead, it depends on $n$ like $\theta = \arccos\sqrt{1/(1+1/(2n^2))}$. This dependence dictates the behaviour of the ratio, decreasing for increasing $n$.

The bottom row shows two distinct configurations. The case $(40,n,n,0)$ has an expected minimum for large $n$, when approaching the configuration $(40,40,40,0)$ from the left. 
The one on the right has a constant ratio, but very different descriptions among the basis with the extended lattice basis having a much lower ratio. 

In general, we conclude that with respect to the description of the gluon propagator tensor, $D_{\mu\nu}(p)$ the use of more complete bases provides a better result. In addition, the improved momentum is again reinforced as the better momentum vector to use. Note that the purpose of considering larger bases is not only to obtain a better description of the scalar functions characterizing the propagator, but also to properly understand its lattice tensor structure, and how it deviates from the continuum form (these deviations should be more evident for coarser lattices, with a larger lattice spacing). 

In addition, our analysis provides results differing from those in \cite{vujinovic2019tensor}.
Namely, in this work the reconstruction from the three form factor lattice basis\footnote{The extended tensor basis with five form factors was not considered in this previous work.} shows better reconstruction results than in our case. This, however is related to the use of a lower dimensional lattice for which the tensor is fully described by less form factors\footnote{The gluon propagator is described in general by $N_d(N_d + 1)/2$ independent tensor structures, depending on the dimension of the lattice $N_d$.}. This results in the structure $\{J,K,L\}$ being a more complete basis for $N_d<4$ than for our 4-dimensional case.
Again, comparisons with these results should be considered with care.

\subsubsection*{Orthogonality of the tensor basis}

The Landau gauge condition is expressed by the orthogonality of the gluon field, $p_\mu A_\mu(p)= 0$. 
This condition, together with the Slavnov-Taylor condition, constrains the tensor form of the gluon propagator in the continuum. It is important to study how this condition affects the form of the two gluon correlation function on the lattice. 

It is also relevant to notice that the gauge fixing on the lattice cannot be implemented with infinite precision. In our simulations the condition satisfies $|\partial A| \lesssim 10^{-7}$. It is also worth referring that we have explicitly tested orthogonality of the gluon fields by computing the correlation functions after applying the projection operator
\begin{equation}
A_\mu^\text{ort} = \left(\delta_{\mu\nu} - \frac{p_\mu p_\nu}{p^2}\right)A_\nu(p)
\end{equation}
where $A_\mu(p)$ are the original gauge fields. Yet, the analysis after this demand does not change neither the form factors nor the ratios $\mathcal{R}$. This serves as a good test of the orthogonality on the lattice.

Also, in lattice simulations for general kinematics the Landau gauge condition is much better realized for the improved momentum rather than normal momentum, $\hat p_\mu A_\mu(p) \ll p_\mu A_\mu(p)$, with the results differing by several orders of magnitude. The exception occurs for kinematics having a single momentum scale for which we can establish $\hat p_\mu A_\mu(p) \propto p_\mu A_\mu(p)$, with the proportionality constant given by $\sin(n)/n$.

In the continuum, the orthogonality of the propagator is ensured by its tensor structure by the transverse form $(\delta_{\mu\nu} - p_\mu p_\nu/p^2)$. However, for the extended bases this is not the case, and the orthogonality should manifest in relations among the form factors.
For the extended lattice basis, the following relation is expected
\begin{align}
&\sum_{\mu}p_\mu D_{\mu\nu}(p) = 0 \nonumber \\
&= E(p^2) + p_\nu^2F(p^2) + p_\nu^4G(p^2) + (p^2-p_\nu^2)H(p^2) + \left(p^{[4]} + p^2p_\nu^2 - 2p_\nu^4\right)I(p^2)
\label{eq:orthogonality_general}
\end{align}
for momentum $p_\nu \neq 0$.

\begin{figure}[ht!]
	\centering
	\input{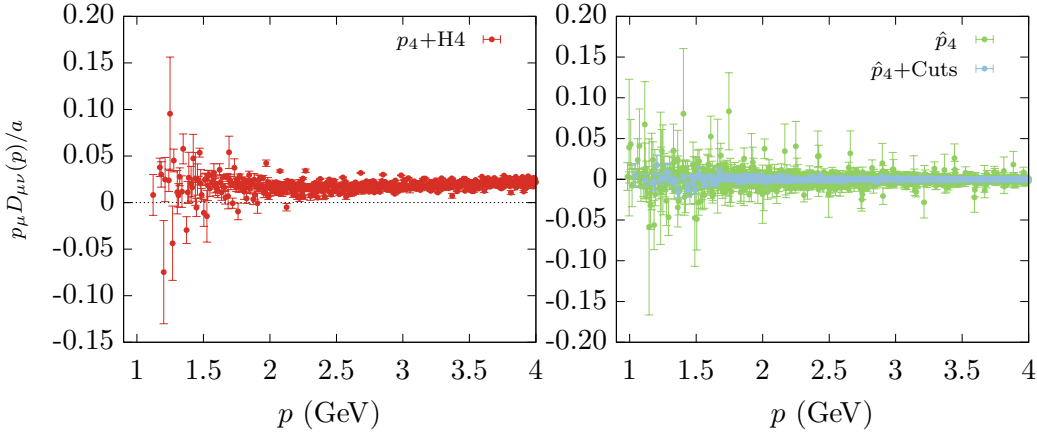}
	\caption{Orthogonality condition, \cref{eq:orthogonality_general} shown for the normal momentum basis after H4 extrapolation from the $\beta=6.0,~80^4$ lattice. Right plot shows the result using the improved basis result without corrections and also with momentum cuts in terms of the improved momentum. For all data the $p_4$ component was considered.}
	\label{fig:orthog_h4_vs_imp}
\end{figure}

We look for deviations from this relation which, following the previous discussion, are expected to be more perceptible for the lattice momentum $p$. In \cref{fig:orthog_h4_vs_imp} the orthogonality condition is shown for the fourth component of momentum, $p_4$ (the conclusions from the remaining components are the same).
The orthogonality relation, \cref{eq:orthogonality_general}, is shown for the H4 extrapolated data (left) where we see that the condition is satisfied only for lower momenta although with increased fluctuations. 
Contrarily, the improved basis (right) shows a much better realization of the orthogonality for the full momentum range. 
The low momentum region involves higher statistical fluctuations that can be partially eliminated by cutting momenta farther from the diagonal. 

Note that this analysis of the orthogonality serves also as a complementary verification of the continuum relations and the completeness of the basis. Indeed, imposing $G,~I\rightarrow0$ and $-p^2F,-p^2H \rightarrow E$ the relation \eqref{eq:orthogonality_general} is immediately satisfied. 

\subsection{Lattice basis -- Generalized diagonal configurations}

Throughout the previous analysis we excluded the generalized diagonal kinematics for which the complete set of lattice form factors is not possible to obtain. 
However, it was hinted that these are special regarding the description by the continuum tensor and for the orthogonality condition.
In this section these configurations are studied, and some quantitative arguments are laid to support previous claims. The generalized diagonal configurations were introduced in \cref{sec:tensor representations of the lattice}. These are defined by a single scale, thus include on-axis momenta with a single non-vanishing component, full diagonal momenta $(n,n,n,n)$, and mixed configurations $(n,n,0,0)$ and $(n,n,n,0)$.

\begin{figure}[htb!]
	\centering
	\input{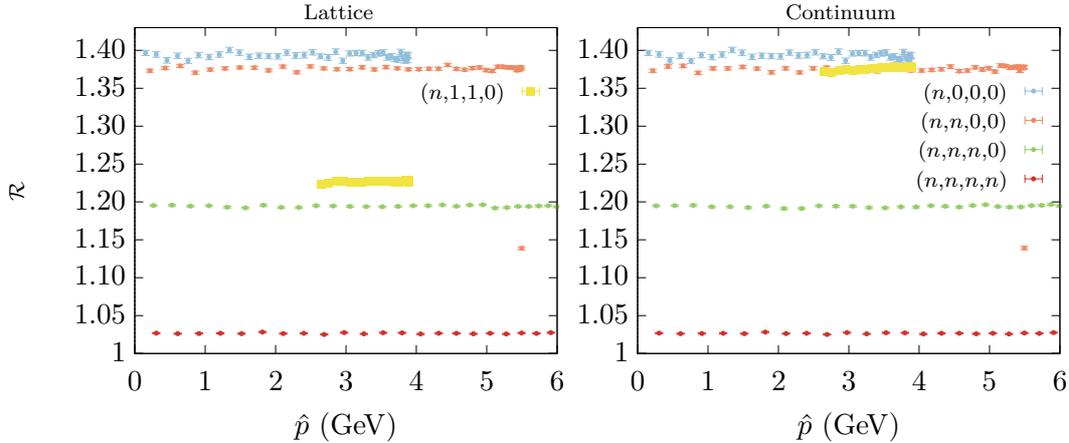}
	\caption{Reconstruction ratio for all four generalized diagonal configurations from the $\beta=6.0,~80^4$ lattice considering the most complete lattice basis (left) and the usual continuum tensor basis (right). Also shown is the reconstruction for the kinematics $(n,1,1,0)$ using the same two bases.}
	\label{fig:recons_diag}
\end{figure}

We start by analysing the reconstruction results for the four generalized diagonal configurations in \cref{fig:recons_diag}.
Firstly, there is a clear hierarchy in the faithfulness in the description among the four configurations. The closer to the diagonal, the better description. This should be related to softer discretization artifacts along the diagonal, as opposite to the ones farther from it. The ratio deviates considerably from unity, reaching differences of about $40\%$ for on-axis momenta.

The other striking feature is the correspondence between both bases. Although neither basis is complete, it would be expected that having more independent terms would result in a better description. This apparent conflict can be explained by the special properties of these kinematics. Although we are using five form factors, the degeneracy of the tensor allows only to extract a reduced number (two or three depending on the configuration -- see \cref{apend_chap:lattice_tensors}) hence reducing the freedom in the tensor description. In addition, the combination of the gauge condition and Slavnov-Taylor identity on the lattice further restricts the tensor by establishing relations among the form factors. Therefore, for these kinematics, both bases provide the same effective degrees of freedom.

In fig. \ref{fig:recons_diag} a momentum configuration close to on-axis momentum is also shown. It represents the same configuration as in fig. \ref{fig:recons_general_config} $f)$. It should be noticed that for this kinematic configuration, the complete extraction of 5 form factors is possible. 
The ratio for $(n,1,1,0)$ is  much smaller when using the lattice basis than for the continuum structure which is closer to the result from $(n,0,0,0)$ and again shows that the lattice basis is better at describing the original tensor for a general configuration.

\subsubsection*{Continuum relations}

In the above analysis we referred that the diagonal kinematics are special regarding its reproduction of the continuum relations. 
To sustain these claims, we verify that these are exactly satisfied for these kinematics.
We consider the full diagonal momenta $p = (n,n,n,n)$, for which only two objects may be extracted,
\begin{align}
&E(p^2) + n^2F(p^2) + n^4G(p^2) = \frac{1}{N_d}\sum_{\mu}D_{\mu\mu}(p) \label{eq:formfactor1_extract_fulldiagonal} \\
&n^2H(p^2) + 2n^4I(p^2) = \frac{1}{N_d(N_d - 1)}\sum_{\mu\neq\nu}D_{\mu\nu}(p).
\label{eq:formfactor2_extract_fulldiagonal}
\end{align}

Since we want to establish relations among the continuum and lattice parametrizations, we consider the right side of \cref{eq:formfactor1_extract_fulldiagonal,eq:formfactor2_extract_fulldiagonal} expressed by the continuum tensor $D^c_{\mu\nu} = D(p^2)(\delta_{\mu\nu} - p_\mu p_\nu/p^2)$.
By carrying out this replacement, the expressions reduce to,
\begin{align}
4 E(p^2) + p^2F(p^2) + p^4G(p^2) &= 3 D(p^2)  \\
-p^2H(p^2) - \frac{1}{2}p^4I(p^2) &= D(p^2)
\end{align}
which by considering $G,~I \rightarrow 0$ precisely reduce to the continuum relations
\begin{align}
E(p^2), -p^2F(p^2), -p^2H(p^2)   &=  D(p^2).
\end{align}
In fact, this last step was unnecessary since due to the form of the basis, $p^2F(p^2) + p^4G(p^2)$ could just be replaced by a new form factor $p^2F'(p^2)$. 
In this case it is irrelevant how the form factor is defined since only the combination of the two can be extracted.
An analogous argument can be made for the off-diagonal terms.
Thus, for diagonal momenta, the extended lattice basis exactly reduce to the continuum description.
In fact, this is the rationale for the argument given above on the decrease in independent form factors in the case of diagonal kinematics.

For on-axis momenta only diagonal terms can be attained
\begin{equation}
D_{\mu\mu}(p) = E(p^2) + p_\mu^2F'(p^2)
\end{equation}
where we used the simpler notation, $F'(p^2) = F(p^2) + n^2G(p^2)$. For this configuration the continuum parametrization has the following form
\begin{equation*}
D^c_{\mu\mu} = 
\begin{cases}
D(p^2) &  \mu = 2,3,4 \\
0      &  \mu = 1.
\end{cases}
\end{equation*}
Extracting each lattice form factor with \cref{appen_eq:on axis E,appen_eq:on axis F} and replacing the tensor elements by the continuum parametrization gives
\begin{align*}
&E(p^2) = \frac{1}{3}\sum_{\mu}D^c_{\mu\mu}(p) =  D(p^2) \\
&p^2F(p^2) = D^c_{11}(p) - E(p^2) = -D(p^2),
\end{align*}
thus confirming the continuum relations for this configuration. The treatment for the mixed configurations $(n,n,0,0)$ and $(n,n,n,0)$ is analogous and does not alter the conclusions -- it can be seen in \cref{appen_sec:mixed_continuum_relations}.

We confirm that the continuum relations are satisfied for single scale configurations and thus the description with the lattice or continuum tensor is equivalent.
Hence, we see that if we want to have a proper description of lattice objects the continuum tensor basis provides a good result if one focus on the diagonal kinematics. This serves also to again validate the conventional approach to the computation of the propagator using momentum cuts.

\begin{figure}[htb!]
	\centering
	\input{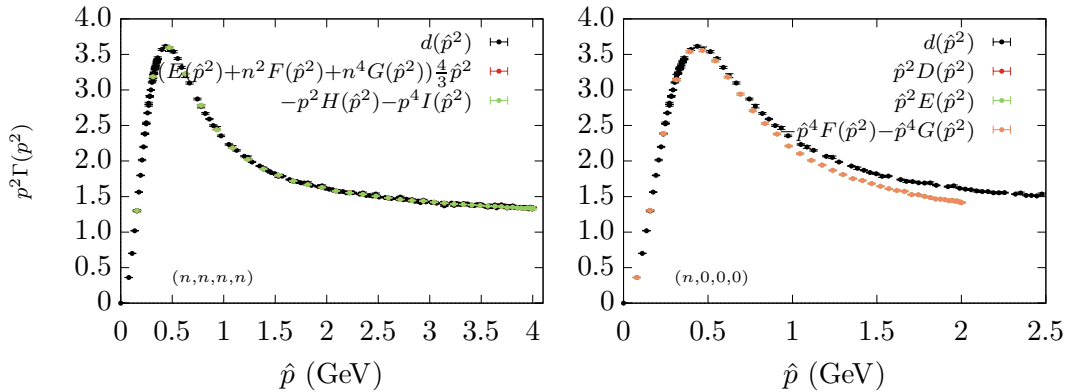}
	\caption{Form factors from the lattice basis for the diagonal configuration $p=(n,n,n,n)$ (left) and for the on-axis momentum $p=(n,0,0,0)$ (right) both as a function of improved momentum. Results from the $\beta=6.0,~80^4$ lattice. Shown for comparison is the benchmark result $d(\hat p^2)$.}
	\label{fig:continuum_relations_diag}
\end{figure}

We confirm this numerically in \cref{fig:continuum_relations_diag} which shows the previous continuum relations. 
The three expressions show a very good agreement.
The left plot shows the two possible form factors for $(n,n,n,n)$ which other than satisfying the continuum relations among them also have a very good agreement with the benchmark result $d(\hat p^2)$. 
For on-axis momentum the continuum relations are also confirmed among the two lattice form factors and the continuum scalar $D(p^2)$. However, hypercubic artifacts render this configuration problematic from the perspective of the reproducing the expected result\footnote{Note that the benchmark result consists of data surviving momentum cuts, and on-axis momenta do not survive the cuts. This is the reason the result deviates quite considerably for momentum above $\sim 0.5\gev$.}. 

Regarding the orthogonality for generalized diagonal configurations, these are the same as continuum relations. In fact, for the case $(n,n,n,n)$ the orthogonality condition is
\begin{equation*}
p_\mu D_{\mu\nu}(p) = n(E(p^2) + n^2F(p^2) + n^4G(p^2)) + 3n^3(H(p^2) + 2n^2I(p^2)) = 0
\end{equation*}
which is the same as obtained above for the continuum relations. Thus, following the previous conclusions, both orthogonality and continuum relations are guaranteed when studying the generalized diagonal kinematics.

\subsection{Finite volume effects}

We explore possible finite volume effects by analysing results from a $64^4$ lattice with the same inverse coupling, $\beta=6.0$. Having a larger ensemble (2000 configurations) results in lessened statistical fluctuations. On the other hand, a smaller volume restricts the access to low momenta.

Due to the momentum restriction on the extraction of the five form factors for a general kinematics, we cannot reach the lowest momentum points where the finite volume effects should be noticeable. For the rest of momentum range the continuum relations for the form factors show the same general behaviour as the $80^4$ lattice, \cref{fig:DEFH_factors,fig:DEFH_factors_noE,fig:GI_factors}, as thus we do not consider its analysis.

\begin{figure}[htb!]
	\centering
	\input{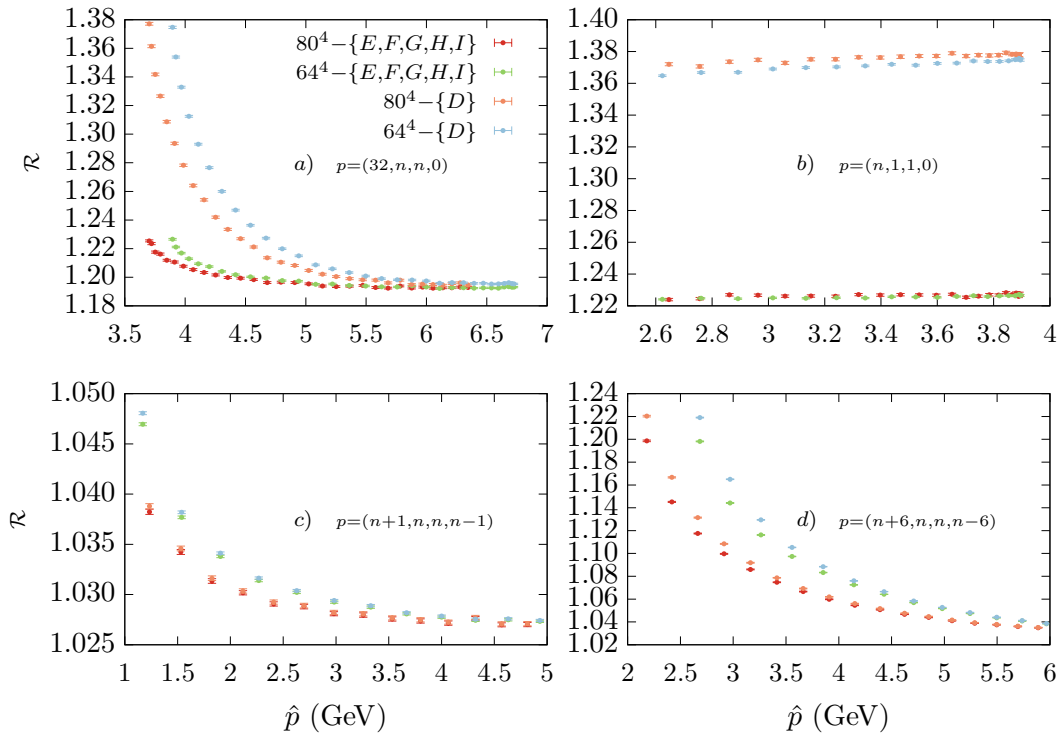}
	\caption{Reconstruction ratio for the extended lattice basis and the usual continuum description both in terms of the improved momentum. These are shown for the two different lattices with $80^4$ and $64^4$ sites, and same spacing $1/a = 1.943(47)~\gev^{-1}$. Four distinct momentum configurations are shown.}
	\label{fig:64_recons_config}
\end{figure}

We turn our attention to the reconstruction -- the finite volume of the lattice is not taken into account in the basis construction and thus it could affect the reconstruction of the original tensor.
The comparison among the two lattices is shown in \cref{fig:64_recons_config} with the extended and continuum basis shown in terms of the improved momentum. 
The first thing to notice is that the reconstruction is better for the $80^4$ lattice, showing a smaller ratio, except for special points such as diagonal kinematics. 
This is perceptible for the high momentum region of $a)$, $c)$, and $d)$.
In $b)$, both lattices show the same ratio for the extended basis while the continuum basis shows a slight difference with the $80^4$ ensemble having a higher ratio.

Despite both lattices provide similar results for special kinematic points, the remaining configurations differ, and the completeness of the bases seems to be reduced for the smaller volume lattice.
In fact, in \cref{fig:64_recons_config} $c)$ and $d)$ even the $80^4$ continuum tensor provides a better reconstruction than the $64^4$ extended lattice basis.

\begin{figure}[htb!]
	\centering
	\input{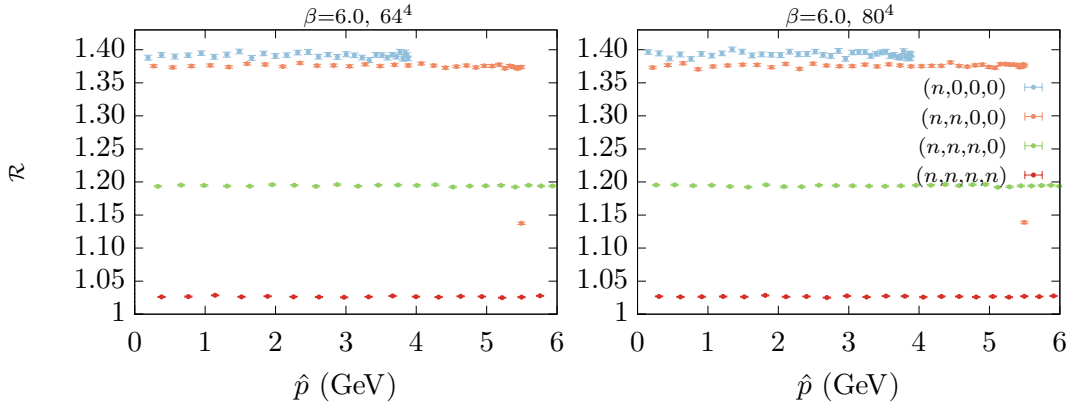}
	\caption{Reconstruction ratio for all four generalized diagonal configurations considering the most complete lattice basis for the $(6.502~\si{fm})^4$ lattice (left) and the $(8.128~\si{fm})^4$ lattice (right). Both lattices having the same lattice spacing $1/a = 1.943(47)~\gev^{-1}$.}
	\label{fig:recons_diag_volume}
\end{figure}

To complete the reconstruction analysis it is worth to reproduce \cref{fig:recons_diag} for the two different lattices, see \cref{fig:recons_diag_volume}. We consider only the largest basis and confirm that the reconstruction for diagonal kinematics is independent of the lattice volume. 
Therefore, other than having a better description by the continuum form, these kinematics seem also to be insensitive to the volume of the lattice regarding its tensor description.
With this analysis we confirm that the momentum cuts, namely choosing the diagonal momenta seems to be an appropriate methodology for lattice computation of correlation functions.

\section{Three gluon vertex}\label{sec:results 3_gluon}	

The focus of this section is the analysis of the three gluon correlation function. 
In particular, we look for a possible sign change and subsequent logarithmic divergence which are expected to occur in the infrared region for some specific kinematic limits and for some form factors of the three gluon correlation function. The zero-crossing and IR divergence are related to the concept of dynamical mass generation \cite{Aguilar_2014,Binosi_2012_mass,cornwall_mass} whereby the gluon acquires an effective momentum dependent mass $m(p^2)$, while the ghost seems to be transparent to this process thus remaining effectively massless. 
This property should also affect different gluon correlation functions, particularly the IR form of the gluon propagator \cite{Fischer_2006,Duarte_2016}.

This behaviour has been predicted by various DSE analysis employing different truncation schemes and approximations for the three gluon vertex \cite{pelaez_3gluon,Blum_2015,Eichmann_3gluon,Cyrol_threegluon}. The basic mechanism for the appearance of the zero-crossing and subsequent logarithmic divergence in the three gluon vertex is reviewed in \cite{Aguilar_2014}. 
It boils down to the appearance of a diverging ghost loop in the Dyson-Schwinger equation for the propagators which in turn affects the three gluon vertex -- see \cite{Eichmann_3gluon} for a thorough analysis. 
From a qualitative point of view we can justify the divergence due to the supposedly ghost masslessness and its loop contributing with a term of the form $\sim\ln(q^2)$, which diverges for $p^2\rightarrow0$. 
On the other hand, the gluon loop is associated with a term $\sim \ln(q^2+m^2)$, remaining IR finite due to the momentum dependent effective gluon mass\footnote{Note that in these schemes the divergence occurs in a theory with a finite gluon propagator $D(0)\geq 0$ and finite ghost propagator (as is the case of lattice results). Therefore, the origin of the divergences is not related to the inherently divergent `scaling' solutions appearing in the DSE formalism. These solutions and its properties are discussed in \cite{Fischer_2006}.}, $m(0) > 0$.

Since the DSE formalism requires approximations for the propagators/vertices entering the truncated equations, its results require validation, usually coming from lattice simulations. However, the study of the IR region is constrained by the finite volume of the lattice and also by large statistical fluctuations associated with the vertices. 
Although the zero-crossing and the three gluon vertex divergence have been observed for 3-dimensional $SU(2)$ theory, its degree of divergence seems to be lower than the one expected from the DSE framework \cite{maas2020threegluon}. Other lattice investigations in both $SU(2)$ and $SU(3)$ and in three and four dimensions \cite{anthony_2016,ATHENODOROU2016444,bocaud_refining_zerocrossing,Cucchieri_2008} suggest the presence of the zero-crossing albeit failing to observe the divergence. 
Contrarily, a recent analytical study of the gluon and ghost propagators using lattice data suggest the presence of a mass regularizing the ghost propagator in the deep IR \cite{alex2020analytic}. This could in turn remove the infrared divergence for the three gluon vertex.

The zero-crossing provides a non-trivial constraint on the behaviour of gluon vertices which due to its logarithm divergence makes the effect difficult to observe\footnote{In three dimensions the corresponding effect is a $\sim 1/p$ divergence favouring its detection \cite{cucchieri_3d_2006,cucchieri_3d_2008}} in small volume lattices. This effect also strongly depends on the kinematic configuration. In this work we focus on the `asymmetric' configuration with a vanishing momentum $(p_1,p_2,p_3)=(p,0,-p)$ for which we extract a single form factor $\Gamma(p^2)$ that is expected to display the sign change in the IR region. This kinematic was considered in other lattice studies \cite{ATHENODOROU2016444,bocaud_refining_zerocrossing,parrinelo_3gluon} as well as continuum approaches \cite{DSE_3gluon_binosi, Eichmann_3gluon}. In \cite{Aguilar_2014} the ratio
\begin{equation}
	R(p^2) = \frac{{\Gamma^{(0)}}^{a_1a_2a_3}_{\mu_1\mu_2\mu_3}(p,0,-p) G^{a_1a_2a_3}_{\mu_1\mu_2\mu_3}(p,0,-p)}{{\Gamma^{(0)}}^{a_1a_2a_3}_{\mu_1\mu_2\mu_3}(p,0,-p)D^{a_1b_1}_{\mu_1\nu_1}(p)D^{a_2b_2}_{\mu_2\nu_2}(0)D^{a_3b_3}_{\mu_3\nu_3}(p){\Gamma^{(0)}}^{b_1b_2b_3}_{\nu_1\nu_2\nu_3}(p,0,-p)} = \frac{\Gamma(p^2)}{2}
\end{equation}
was  related to the diverging ghost loop appearing in the DSE for the gluon propagator (under the chosen truncation scheme).

Other than $(p,0,-p)$, other kinematics are generally considered in the literature, namely the `symmetric' configuration ($p_i^2 = p^2,~ p_i\cdot p_j = -p^2/2,~i\neq j$) \cite{ATHENODOROU2016444,bocaud_refining_zerocrossing} for which the zero-crossing is easier to observe due to smaller fluctuations, thus having a more defined range for the sign change. The asymmetric configuration, on the other hand, is associated with increased statistical fluctuations due to the vanishing momentum component $p_2=0$ \cite{bocaud_refining_zerocrossing}.

Therefore we aim at investigating the possible occurrence of the zero-crossing and narrowing the range of momentum where it is expected to occur under both possible hypothesis for the ghost behaviour, namely the existence or absence of a dynamical ghost mass that regularizes the vertex. 
In addition we look for possible signs of the divergence for vanishing momentum.
This work follows the investigation from \cite{anthony_2016} albeit with increased statistics due to the use of a larger configuration ensemble and also due to the use of the full group symmetry -- complete Z4 averaging. 


For the three gluon vertex we restrict the analysis to the larger lattice, with 550 configurations, see \cref{tab:lattices}. The reason is the need of deep IR momentum points to study the structures introduced before. 
The larger ensemble has a smaller volume and thus its smallest momentum is higher than the corresponding for the $80^4$ lattice.
This ensemble will be considered as comparison for the general behaviour of the data in the IR.
The reader should also be aware that all quantities shown below are not renormalized, which again amounts to a constant factor.

\subsection{Three gluon correlation function}

We start by analysing the complete correlation function, i.e. the vertex with external propagators, extracted with the following contraction
\begin{align}
G(p^2) &\equiv \delta_{\mu_1\mu_3}p_{\mu_2}\expval{\Tr\left[A_{\mu_1}(p)A_{\mu_2}(0)A_{\mu_3}(-p)\right]} \nonumber \\ 
&= V \frac{N_c(N_c^2-1)}{4}D(p^2)D(0)D(p^2)\Gamma(p^2)p^2.
\label{eq:complete 3gluon vertex}
\end{align}

\begin{figure}[htb!]
	\centering
	\input{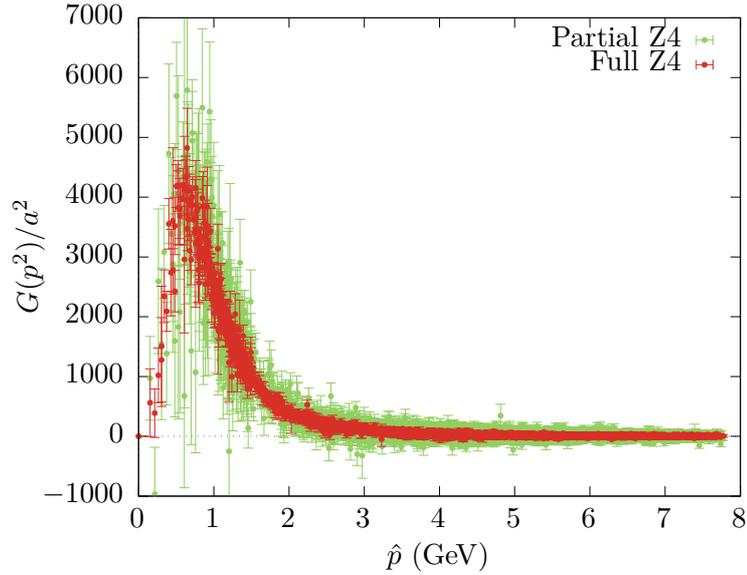}
	\caption{Three gluon correlation function from the $\beta=6.0,~80^4$ ensemble contracted with, and as a function of the improved momentum. All data is shown without correction methods using a partial Z4 averaging with permutations only, and also for the complete Z4 averaging.}
	\label{fig:3gluon_average_partialZ4}
\end{figure}

It is important to notice the difference in the statistical accuracy obtained by considering the complete Z4 averaging as opposed to the partial (permutation only) case. 
A look at \cref{fig:3gluon_average_partialZ4} allows to perceive the change induced by the use of all $H(4)$ equivalent points for the averaging, which enhances the signal to noise ratio. Statistical fluctuations are lessened through all range of momentum for the complete Z4 case and the data defines a smoother curve, with decreased error bars.
Given the lessened statistical precision found in lattice computation of vertices when comparing with the results for the gluon propagator in the last section, it is crucial to consider possible ways of increasing the statistics.
For this reason, the rest of this section considers the complete Z4 averaged data.

Regarding the $\invmomp{4}$ extrapolation, we notice that this procedure can be extended to a higher momentum than the one used for the gluon propagator without loss of integrity of the method. 
The H4 method uses the $H(4)$ orbits to `reconstruct' the continuum object -- extrapolating data to $\invmomp{4}\rightarrow0$. 
While for the gluon propagator the structures formed by the orbit points are well defined and with small uncertainty associated, the three gluon orbit structures are concealed by large fluctuations. Hence, the extrapolated function for the three gluon maintains a momentum dependence close to the original data but with increased precision.
Notice, however that this is not an advantage of the method for the three gluon vertex, but a consequence of the reduced precision associated with this vertex which allows us to extend the range, within the original uncertainty.

\begin{figure}[htb!]
	\centering
	\input{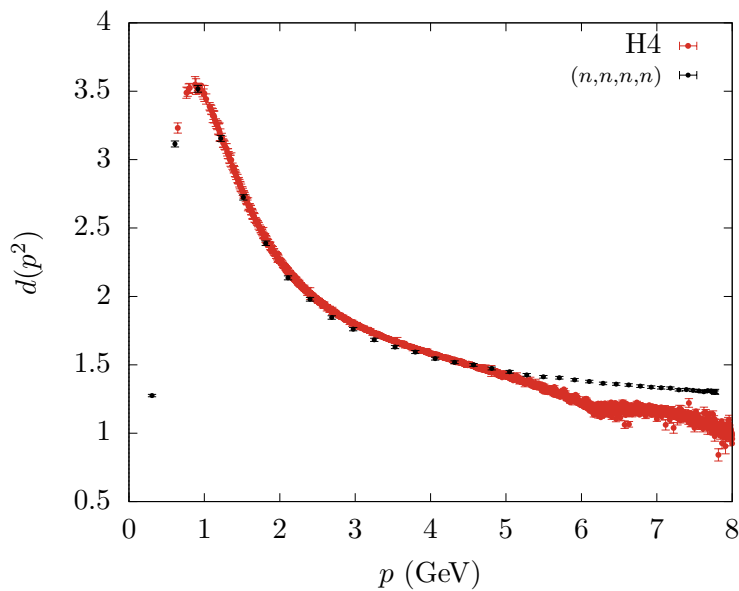}
	\caption{H4 extrapolated data for the gluon propagator dressing function $d(p^2)$ compared with full diagonal momenta $(n,n,n,n)$ as a function of improved momentum. Data from the $\beta=6.0,~80^4$ ensemble.}
	\label{fig:prop_h4vsdiagonal}
\end{figure}

To support these claims on the extension of the method we compare the effect of extending the extrapolation for both the gluon propagator and the three gluon vertex. 
In \cref{fig:prop_h4vsdiagonal} the H4 extrapolation for the propagator was extended to all momentum and compared with diagonal configurations due to its lessened hypercubic artifacts. The dressing function for $(n,n,n,n)$ momentum is shown as a function of improved momentum as it was observed in the previous section to produce a better match with the expected behaviour.
We see that for momenta above $p\sim5\gev$ the difference between both results is large, evidencing the inaccuracy of the extrapolation for this momentum scale. In fact, the extrapolation for momenta above $p\sim6\gev$ becomes unstable, producing a less smooth curve.

\begin{figure}[htb!]
	\centering
	\input{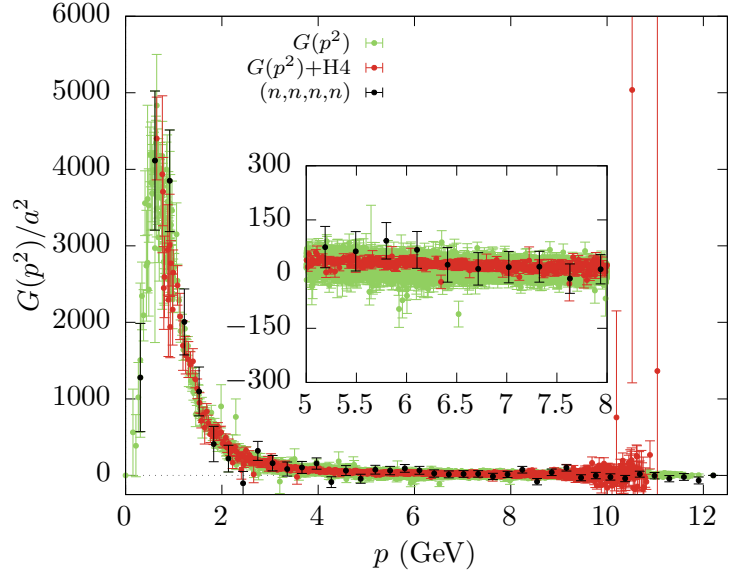}
	\caption{Original and $\invmomp{4}$ extrapolated data for the three gluon correlation function from the $\beta=6.0,~80^4$ ensemble as a function of the lattice momentum $p$. The H4 correction was applied for the full momentum range. The configuration $(n,n,n,n)$ is shown for comparison.}
	\label{fig:3gluon_h4_detail}
\end{figure}

Contrarily to this case, if we extend the $\invmomp{4}$ extrapolation for the three gluon vertex, the disagreement is only obtained for larger momenta. In \cref{fig:3gluon_h4_detail} the H4 corrected vertex is again plotted against the diagonal kinematics. We see that the general behaviour of the curve is maintained after the correction (with additional precision), and that it follows the diagonal curve. 
Therefore, for the three gluon vertex an extension of the extrapolation is possible within the statistical accuracy. Notice however that the extension is not complete since for momenta above $p\sim 8\gev$ large fluctuations arise and the extrapolation is not reliable. In fact, for the highest momenta, the extrapolation is not possible due to the lack of $H(4)$ orbit elements, analogously to the IR region.

\subsection*{Perturbative UV prediction}

Although we are interested in the infrared behaviour of the correlation function, we begin by probing how the continuum perturbative predictions match lattice results for high momenta. To perform this comparison we apply the H4 extrapolation as well as conical cuts with improved momentum.
Following \cite{anthony_2016}, to study the ultraviolet region of our results we use the one-loop renormalization group improved result for the propagator  
\begin{equation}
D(p^2) = \frac{Z}{p^2}\left[ \ln\left(\frac{p^2}{\mu^2}\right) \right]^{-\gamma}
\label{eq:prop_perturbative}
\end{equation}
with $Z$ a global constant, $\mu = 0.22\gev$ and $\gamma = 13/22$ the gluon anomalous dimension. For the three gluon vertex a similar expression is obtained,
\begin{equation}
\Gamma(p^2) = Z'\left[ \ln\left(\frac{p^2}{\mu^2}\right) \right]^{\gamma_{3g}}
\label{eq:3gluon perturb gamma}
\end{equation}
with the anomalous dimension $\gamma_{3g} = 17/44$. These two expressions can be combined to construct the corresponding three gluon correlation function computed above, \cref{eq:complete 3gluon vertex}
\begin{equation}
G_\text{UV}(p^2) = \frac{Z''}{p^2}\left[ \ln\left(\frac{p^2}{\mu^2}\right) \right]^{\gamma'}
\label{eq:3gluon_perturb}
\end{equation}
with $\gamma' = \gamma_{3g} - 2\gamma = -35/44$ the overall anomalous dimension. This result is expected to be valid for high momentum.

\begin{figure}[htb!]
	\centering
	\input{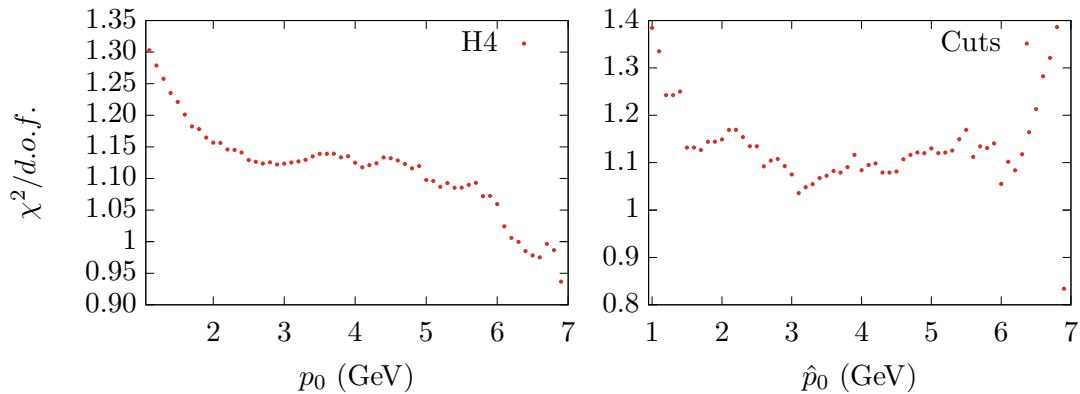}
	\caption{$\chi^2/d.o.f.$ obtained from the fit of the functional form \eqref{eq:3gluon_perturb} to the $\beta=6.0,~80^4$ lattice data as a function of the momentum range cut off, $p>p_0\gev$. Left plot shows the result of the fit for the H4 corrected data while the right plot with diagonal momenta as a function of the improved momentum.}
	\label{fig:3gluon_fits}
\end{figure}

\begin{figure}[htb!]
	\centering
	\input{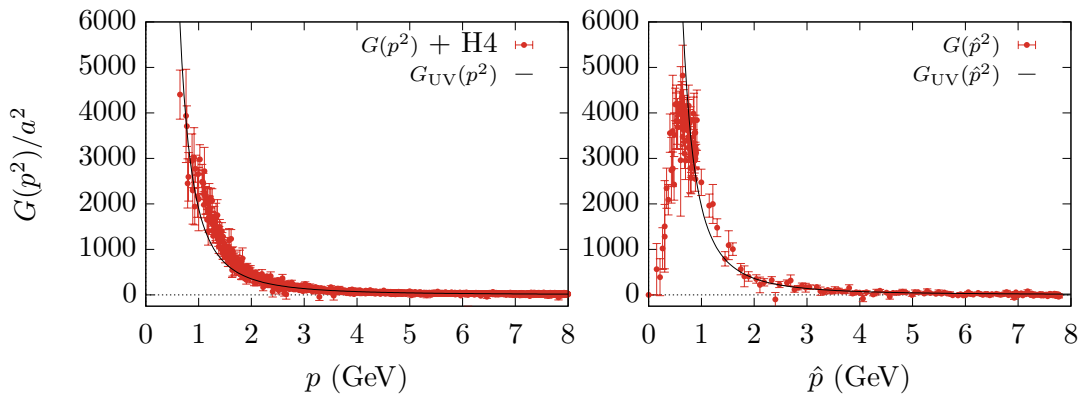}
	\caption{Three gluon correlation function $G(p^2)$ after the H4 extrapolation as a function of the lattice momentum (left) and as a function of the improved momentum after cuts for $\hat p > 1\gev$. The perturbative prediction, \cref{eq:3gluon_perturb} is also represented after a fit to the extrapolated and diagonal configurations, respectively. All results shown are from the $\beta=6.0,~80^4$ ensemble.}
	\label{fig:3gluon_perturb}
\end{figure}

To better understand the validity of the perturbative prediction, the fits were performed with \textit{Gnuplot} \cite{Gnuplot_52} for various momentum ranges $[p_0,8]\gev$ with varying $p_0$. The upper bound at $8\gev$ is considered also for H4 corrected data due to large errors in the lattice data.
The fit was applied to H4 corrected data as a function of lattice momenta, and also for the data as a function of improved momentum.
To evaluate its quality we compute the $\chi^2/d.o.f.$\footnote{This function measures the deviation of the approximated curve obtained by the fit to the data points. It is defined as,
	$$\chi^2 = \sum_{i}\left( \frac{G_i - f(p_i)}{\delta G_i} \right)$$
	where $G_i$ and $\delta G_i$ are the data points and corresponding error, while $f(p_i)$ is the fitted curve evaluated at the momentum of $G_i$. The degrees of freedom ($d.o.f.$) are the number of data points to be adjusted deducted by the number of adjustable parameters. A good fit to the data is obtained by a reduced $\chi^2$ close to unit, i.e. $\chi^2/d.o.f.\sim1$.} 
taking into account the uncertainty in the data, and which ought to be minimized for various values $p_0$, this is shown in \cref{fig:3gluon_fits}.

For H4 corrected data, the best fit is obtained for momentum $p\sim6.5\gev$. However, for momenta above $p\sim 2.5\gev$ the fit already shows a stable match with the lattice data. Above this scale the fit maintains a $\chi^2/d.o.f.$ below $\sim1.15$. 
The fit for $p_0=2.5\gev$ is shown in the left plot of \cref{fig:3gluon_perturb} for which $\chi^2/d.o.f. = 1.14$. The data seems to follow the perturbation theory prediction for $p$ above $\sim2.5\gev$.

The fits for the data as a function of improved momentum surviving the cuts show similar $\chi^2/d.o.f.$ values for most fitting ranges. However the values seem to oscillate less smoothly, and in fact become high for $p_0$ above $6\gev$.
In the right plot of \cref{fig:3gluon_perturb} the fit for $p_0>3\gev$ is shown, having $\chi^2/d.o.f. = 1.09$. 
This curve also shows a good agreement with the lattice data thus validating the perturbative prediction for high momenta.

\begin{figure}[htb!]
	\centering
	\input{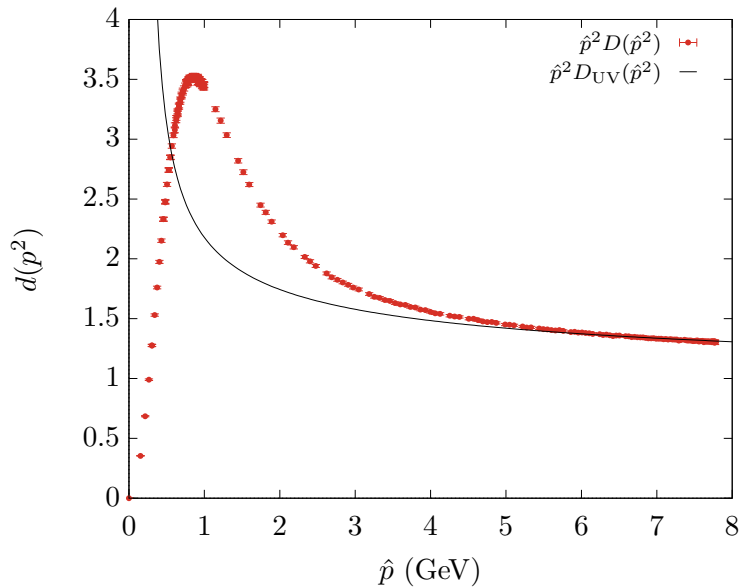}
	\caption{Gluon propagator $D(p^2)$ from the $\beta=6.0,~80^4$ lattice as a function of the improved momentum after cuts abover $1\gev$. The renormalization group improved perturbative result, \cref{eq:3gluon_perturb} was fitted to the data for $p\in [5,8]\gev$, resulting in a fit with $\chi^2/d.o.f. = 1.10$.}
	\label{fig:prop_perturb_improved}
\end{figure}

To compute the pure three gluon vertex we need to explicitly remove the contribution of the external propagators by dividing by its form factor $D(p^2)$, \cref{eq:complete 3gluon vertex}.
Hence, we also compare the lattice computation of $D(p^2)$ with the perturbative result, \cref{eq:prop_perturbative}. 
The increase in accuracy for this object allows only a fit to higher momenta and in addition, we do not consider the extrapolated data due to its restrictions to high momentum for the propagator.
This is shown in \cref{fig:prop_perturb_improved} as a function of the improved momentum. A good match with the lattice data is obtained, with $\chi^2/d.o.f. = 1.10$ for the range $p>5\gev$. Again, the perturbative result is confirmed for sufficiently high momentum.

\subsection{Three gluon one particle irreducible function}

Although the possible sign change associated with the three gluon vertex should be noticeable for the complete correlation function shown before, this carries high statistical fluctuations for momenta below $p\sim1\gev$, hindering the IR analysis of the curve. 
In addition, since continuum investigations work with the 1PI function we need to remove the propagators if we want to properly compare lattice and continuum results.
In this way we isolate the pure one particle irreducible function, which for the $(p,0,-p)$ kinematics and the tensor basis considered is described by $\Gamma(p^2)$, \cref{eq:3gluon_gamma}.

Firstly, we notice that the comparison with the UV perturbative prediction from \cref{eq:3gluon perturb gamma} is not possible for $\Gamma(p^2)$ due to large statistical fluctuations dominating the high momentum region. 
These arise due to the high momentum form of the gluon propagators, where for a general kinematic configuration they behave as $D(p^2) \sim 1/p^2$. 
This induces a $p^6$ factor in $\Gamma(p^2)$ when dividing by $D(p^2)$\footnote{The poor signal to noise ratio for $\Gamma(p^2)$ for high momentum is a common complication for general lattice computed 1PI functions with more than two external legs. This problem is not completely solved by the increase in the number of configurations since it is inherently associated with the high momentum behaviour of the propagators.}. 
In turn, this factor enlarges the uncertainty associated with $\Gamma(p^2)$ --  this can be noticed by a simple Gaussian error propagation, see \cite{anthony_2016}.
For the kinematics in consideration the factor is softened to $p^4$ due to the vanishing momentum $p_2= 0$, $D(0) > 0$. However, the $p^4$ factor combined with large fluctuations in $D(0)$ create strong fluctuations in the ratio $p_\mu G_{\nu\mu\nu}(p,0,-p)/D(p^2)^2D(0)$ for high momenta.

Regarding the detection of the zero-crossing this is not a problem since $D(p^2)$ is essentially constant for the deep IR region and thus the signal has a more stable behaviour and higher precision. Additionally, the H4 extrapolation is not useful for it disregards points in this region.

\begin{figure}[htb!]
	\centering
	\input{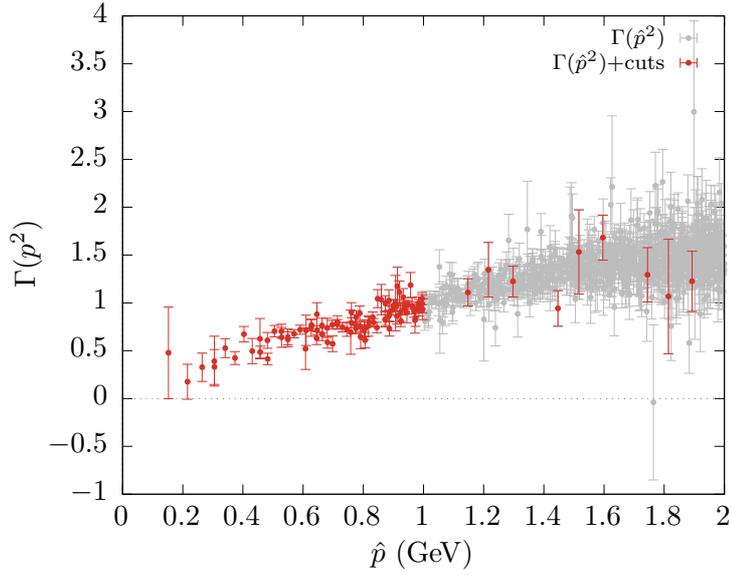}
	\caption{Complete set of data from the $\beta=6.0,~80^4$ lattice for the three-gluon 1PI, $\Gamma(p^2)$ as a function of the improved momentum. The data surviving momentum cuts above $1\gev$ is also shown.}
	\label{fig:3gluon_gamma}
\end{figure}

In \cref{fig:3gluon_gamma} both the complete set of data for  $\Gamma(p^2)$, and the points surviving momentum cuts after $1\gev$ are shown as a function of improved momentum.
This result matches the momentum dependence obtained in other lattice studies, namely it follows the results from \cite{anthony_2016} although with an improved signal to noise ratio.
As expected, large statistical fluctuations arise for momenta above $\sim 1.5\gev$. 
The two lowest momentum points are both compatible with zero within one standard deviation. The lowest non on-axis momentum is compatible with zero within the uncertainty, $\Gamma(p = 0.216\gev) = 0.176(182)$, while the lowest on-axis momentum is also compatible with zero although having a larger error associated $\Gamma(p=0.152\gev)=0.477(479)$.
However, these two points do not provide a statistically relevant signal of the possible zero-crossing.

In order to improve the analysis of the infrared behaviour of the 1PI function we consider three different functional forms to fit the data in \cref{fig:3gluon_gamma},
\begin{align}
	&\Gamma_1(p^2) = a_1 + z_1 \ln(\frac{p^2}{\mu^2}), ~(a_1,z_1) \label{eq:toymodel}\\
	&\Gamma_2(p^2) = a_2 + z_2 \ln(\frac{p^2 + m^2}{\mu^2}), ~(a_2,z_2,m) \label{eq:log with mass} \\
	&\Gamma_3(p^2) = 1 + cp^{-d}, (c,d); \label{eq:power law ansatz}
\end{align}
the adjustable parameters appear in parenthesis.
The first functional form, \cref{eq:toymodel}, comes from a simple Landau gauge, four-dimensional QCD toy model for asymptotically low momentum \cite{Aguilar_2014,bocaud_refining_zerocrossing}.
The second logarithm, \cref{eq:log with mass} has an additional constant $m^2$ to account for the possible dynamical ghost mass predicted in \cite{alex2020analytic}. This mass could in principle remove the three gluon divergence by regularizing the ghost loop, nonetheless a sign change is possible depending on the value of the parameters.
Both constants $a_1,a_2$ serve to partially take into account the non-leading terms which become relevant for higher momenta.

The third form for $\Gamma(p^2)$, \cref{eq:power law ansatz}, is a power law ansatz \cite{maas2020threegluon} which allows to study the degree of the possible divergence in the IR and also estimate the position of the zero-crossing.
In \cite{ATHENODOROU2016444,Aguilar_2014,bocaud_refining_zerocrossing} more appropriate curves, obtained by solving the DSEs for this momentum configuration are considered and fitted to lattice data.

To better understand the validity of the functional forms, the range of the fit was tested for the limits $[p_i,p_f]$ with variable $p_f$ while $p_i$ is the lowest, non-zero momentum value. The value of $p_f$ was restricted to $2\gev$, above which $\Gamma(p^2)$ is involved in large fluctuations, in fact these are noticeable already in the upper momenta of \cref{fig:3gluon_gamma}. As a lower bound, we consider $p_f$ above $ 0.5\gev$ since not enough data exists below this threshold.

\begin{figure}[htb!]
	\centering
	\input{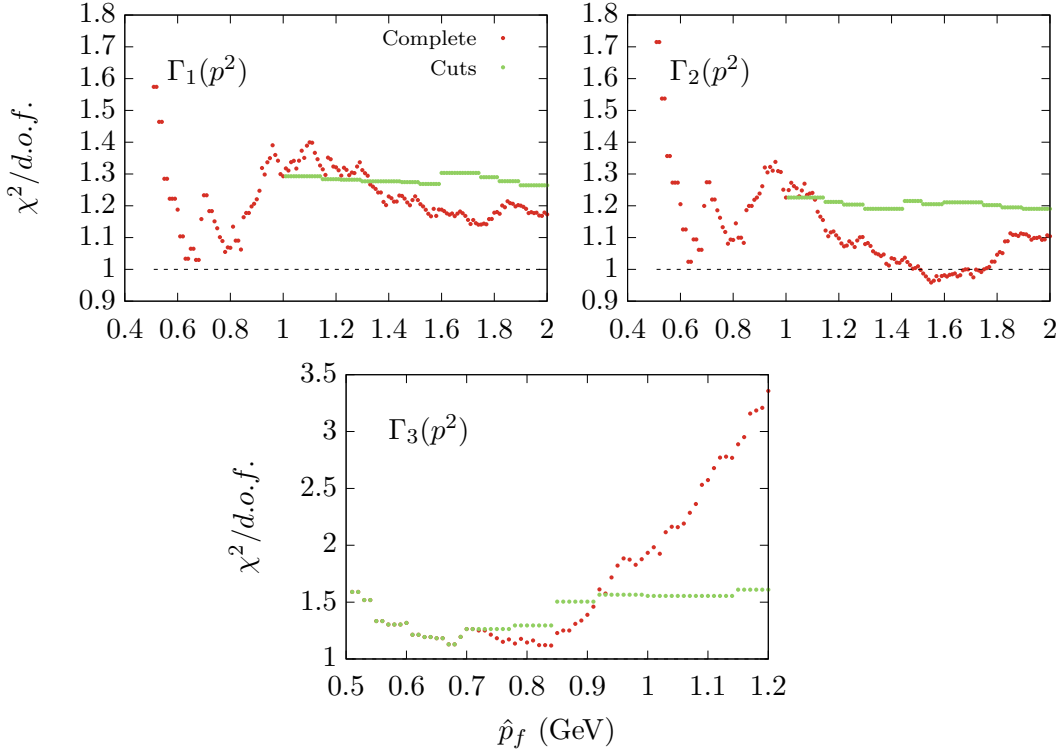}
	\caption{$\chi^2/d.o.f.$ of the three fits from \cref{eq:toymodel,eq:log with mass,eq:power law ansatz} (top left, top right and bottom, respectively) for the varying momentum range $p\in[p_i,p_f]$. Both fits with and without momentum cuts were considered.}
	\label{fig:3gluon_chi_3fits}
\end{figure}

Since we want to explore the quality of the fit with varying range $p_f$ we consider the analysis for the complete set of data in \cref{fig:3gluon_gamma}.
In addition, we compare the result of the fits with the data surviving momentum cuts above $1\gev$ to try to overcome the problem of large fluctuations for higher momenta.
The quality of the fit was controlled with the $\chi^2/d.o.f.$ shown for all functional forms and both sets of data in \cref{fig:3gluon_chi_3fits}.

The results for the $\chi^2/d.o.f$ as a function of the fitting range, in \cref{fig:3gluon_chi_3fits} are similar for both logarithms, $\Gamma_1$ and $\Gamma_2$. 
The quality of the fit seems to be highly dependent on the range for $p_f$ below $0.8\gev$, with $\chi^2$ rapidly oscillating.
Above $1\gev$ the momentum cuts are applied and thus the results for both sets of data become different.
The reduced $\chi^2$ oscillates around $\chi^2/d.o.f=1.3$ for the complete data in the range $p_f\sim1-1.4\gev$. For larger momentum ranges, $p_f > 1.4\gev$, the fit with the complete data provides reduced $\chi^2$ values closer to one, indicating a better match to the data.

Although the quality of the fit has a similar behaviour for both logarithms, the one with an additional mass shows $\chi^2/d.o.f$ values closer to unity. This value remains between $0.9-1.1$ for $p_f>1.1\gev$ for the complete data using $\Gamma_2$ while for the form $\Gamma_1$ the reduced $\chi^2$ stabilizes around $1.2$ for this range.

For the data surviving momentum cuts the behaviour is simpler. Both functional forms provide a stable $\chi^2$ around $\chi^2/d.o.f= 1.3$ for $p_f > 1\gev$. 
This should be a good indication of the smoothness of the data created by the cuts, and also that the curves match the results, within the uncertainty.
It is important also to notice that although in general the complete data provides a fit with better quality, the data after momentum cuts is associated with lessened lattice artifacts and thus this prediction should also be considered.

The behaviour of the fit for the third functional form $\Gamma_3$ is different than the one described above.
From the bottom panel in \cref{fig:3gluon_chi_3fits} we see that the best fit is obtained for $p_f$ in the range $0.6-0.8\gev$ and that the reduced $\chi^2$ grows rapidly for momenta above this region.
Since the quality of the fit becomes worse above $0.9\gev$ the momentum cuts were applied for $p>0.7\gev$ instead.
Notice that in addition, the fit was restricted to $p_f=1.2\gev$, above which the fits become worse.
In fact, since this functional form is considered to probe the degree of the possible divergence in $\Gamma(p^2)$ it should be valid for lower momentum\footnote{This was thoroughly explored in \cite{maas2020threegluon} for both 3 and 4-dimensional cases and found that the power law is compatible with the data for momenta below $\sim1\gev$ only.} when compared with the first two models. This is why the quality of the fit rapidly decreases when reaching $p_f\sim 1\gev$.
The quality from the data with cuts remains practically constant above $0.9\gev$ with a value around $\chi^2/d.o.f=1.5$.

\begin{figure}
	\centering
	\input{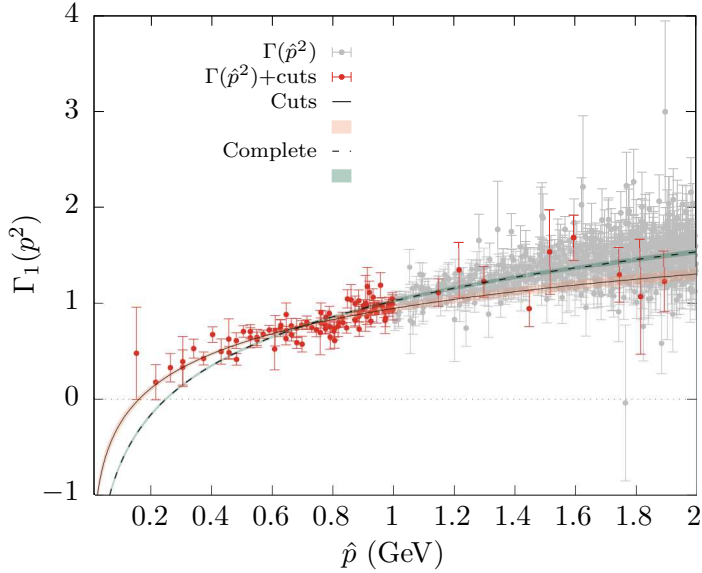}
	\caption{$\Gamma(p^2)$ from the $\beta=6.0,~80^4$ ensemble as a function of improved momentum. The data after momentum cuts is also shown. Two fits using \cref{eq:toymodel} and $p_f = 1.7\gev$ were adjusted considering the complete data, and the set after momentum cuts.}
	\label{fig:3gluon_gamma_diag_diverg1_cairo}
\end{figure}

To better understand how each form matches the lattice data we analyse each model independently and show the result of the fits for a specific value of $p_f$.
We choose $p_f$ above $1\gev$ in order to distinguish between the complete data and the one surviving momentum cuts.
For the $\Gamma_1$ logarithm the choice $p_f = 1.7\gev$ provides fits with $\chi^2/d.o.f. = 1.14$ and $\chi^2/d.o.f. = 1.28$ for the complete and the data after cuts, respectively.
It is important to refer that the parameters of this curve and the corresponding uncertainty do not vary significantly for the range $1.3<p_f<2\gev$ which further supports the quality of the fit -- more on this below.
The resulting curves and corresponding uncertainty (computed assuming Gaussian propagation of the error) are shown in \cref{fig:3gluon_gamma_diag_diverg1_cairo}. 
The fit for the data surviving momentum cuts seems to provide a better match with the three gluon vertex $\Gamma(p^2)$  for the lowest momentum range, namely for $p\sim0.2-0.8\gev$. 
However, the uncertainty in the curve parameters is slightly higher.
The use of the complete lattice data seems to shift the position of the possible sign change for higher momenta, with  $p_0 = 0.249(3)\gev$ and $p_0 = 0.160(12)\gev$ for the complete data and for the data after momentum cuts, respectively. 

\begin{figure}
	\centering
	\input{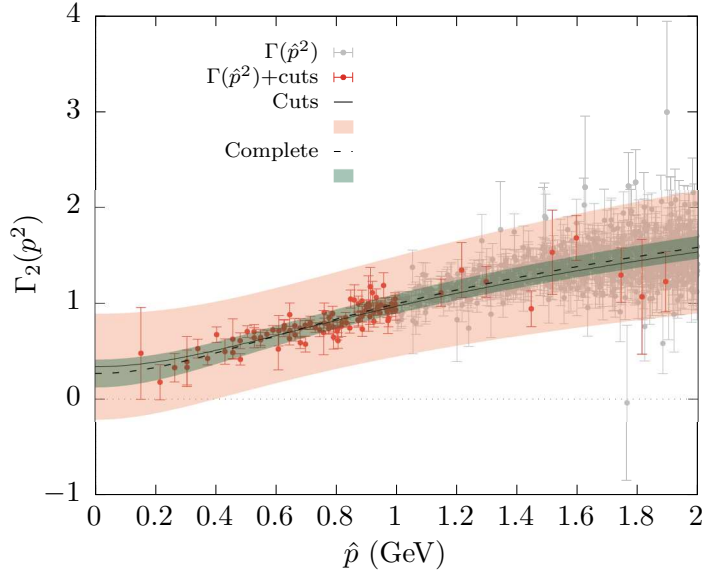}
	\caption{$\Gamma(p^2)$ from the complete set as a function of improved momentum from the $\beta=6.0,~80^4$ ensemble. The data after momentum cuts are applied is also shown. The functional form in \cref{eq:log with mass} with range $p_f = 1.7\gev$ was adjusted to the complete and partial data.}
	\label{fig:3gluon_gamma_diag_mass_cairo}
\end{figure}

For the second logarithmic form, \cref{eq:log with mass}, a similar reasoning is considered for the choice of $p_f$. 
The range $p_f = 1.7\gev$ provides a good fit to the data with $\chi^2/d.o.f. = 0.984$ and $\chi^2/d.o.f. = 1.21$ for the complete set and the data after cuts, respectively. 
The corresponding curves are shown in \cref{fig:3gluon_gamma_diag_mass_cairo}.
Although the quality of the fit indicated by the $\chi^2$ seems to be better for the logarithm with additional mass, the uncertainty in the parameters is larger.
Nonetheless, both curves in \cref{fig:3gluon_gamma_diag_mass_cairo} have a similar form and suggest a good match with the data for the full range of momenta.

Regarding the possible sign change, the fit with the complete data suggests a positive IR value for $\Gamma(0)$ and an absent sign change, within the uncertainty of the curve.
On the other hand the curve using momentum cuts allows for a possible sign change. 
However, although we predict that within this model $p_0$ should occur below $0.35\gev$, the existence of a sign change is not guaranteed by the predictions made from the curve and the substantial uncertainty carried by the resulting curve does not allow further conclusions.

\begin{figure}
	\centering
	\input{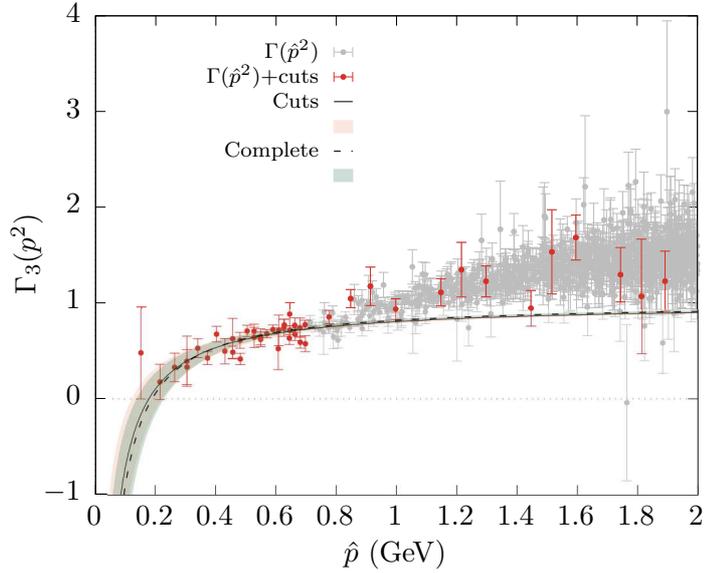}
	\caption{$\Gamma(p^2)$ for the complete kinematics as a function of improved momentum from the $\beta=6.0,~80^4$ ensemble. The set of points surviving momentum cuts is also shown. The functional form in \cref{eq:power law ansatz} with $p_f = 0.85\gev$ was adjusted to the complete and partial data.}
	\label{fig:3gluon_gamma_diag_diverg3_cairo}
\end{figure} 

For the power law form, \cref{eq:power law ansatz}, a good balance in the quality of the fit and a reasonable uncertainty is obtained for $p_f = 0.85\gev$ for which the complete data provides a better fit with $\chi^2/d.o.f. = 1.12$ as opposed to $\chi^2/d.o.f. = 1.29$ for the data surviving momentum cuts.
The analysis of the corresponding curves in \cref{fig:3gluon_gamma_diag_diverg3_cairo} shows that both fits have a comparable form, barely changed by the change in the set of data (this is expected due to the small range considered above $0.7\gev$, above which cuts were applied).
Both results are compatible with a sign change, with $p_0=0.189(31)\gev$ for the curve using the complete data and $p_0=0.179(48)\gev$ for the other set.

Since this last functional form is expected to match the data for low momentum only, where the divergence is supposed to occur, the curve fails to match lattice data for momenta above $\sim1\gev$.
For lower momenta the curve seems to provide a good match with the data, although with decreased precision when compared with the results from \cref{fig:3gluon_gamma_diag_diverg1_cairo}.
The exponents $d$ from the fits are $d = 0.940(135)$ and $d = 1.01(10)$ for the complete and partial sets, respectively. These seem to be compatible with previous findings for both $SU(2)$ and $SU(3)$ lattice investigations \cite{maas2020threegluon,Sternbeck_3gluon}.
However, since we do not find a clear numerical evidence for the divergence due to the lack of points in the deep IR region, this result is not reliable and should be taken with care.


\begin{figure}[htb!]
	\centering
	\input{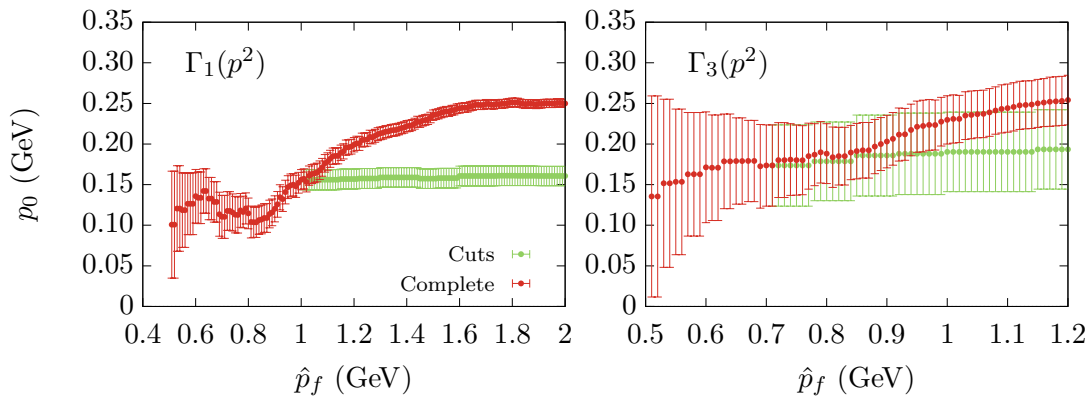}
	\caption{Prediction for the sign change $p_0$ from the fits using \cref{eq:toymodel} (left) and \cref{eq:power law ansatz} (right) for varying fitting ranges $[0,p_f]$.}
	\label{fig:3gluon_zerocrossing}
\end{figure}

Both the first and last functional forms, \cref{eq:toymodel,eq:power law ansatz}, are considered in order to study the possible zero-crossing with subsequent divergence. Despite not having a clear signal on the divergence, we can study how the estimated position and uncertainty for $p_0$ varies with different fitting ranges\footnote{Although a sign change can also be observed for the form \eqref{eq:log with mass}, as seen in \cref{fig:3gluon_gamma_diag_mass_cairo}, its existence strongly depends on the momentum range of the fit. Besides, the uncertainty associated is much larger and therefore its explicit computation as a function of $p_f$ is not shown.}.
The $p_0$ values for $\Gamma_1$ and $\Gamma_3$ are shown in \cref{fig:3gluon_zerocrossing} as a function of $p_f$ for both the complete and partial sets of data.
From the analysis of this figure we notice that $p_0$ is associated with smaller uncertainty when computed with the first form, \cref{eq:toymodel} and using the complete set of data.
In addition, the complete data seems to shift the position of the zero-crossing for higher momentum when compared to the partial data. 

For the logarithmic case, $p_0$ varies very little for the range $p_f<1\gev$, showing values around $0.1-0.15\gev$. 
Above $1\gev$ the data surviving momentum cuts maintains a constant value around $p_0=0.15\gev$.
This prediction lies in a region where in fact the lattice results are compatible with zero within the uncertainty.
On the other hand the prediction from the complete data grows for $p_f>1\gev$ reaching a seemingly constant value of $p_0=0.25\gev$ above $p_f=1.6\gev$.
We see that both sets of data seem to approach a constant value for large fitting ranges, however the values are not compatible within one standard deviation.

For the power law, right plot in \cref{fig:3gluon_zerocrossing}, although the same tendency as for $\Gamma_1$ is observed for $p_0$, the uncertainty in this model is much larger. 
The result from the data surviving cuts seems to remain constant for the whole range of momenta, while the complete result increases for larger $p_f$.
However, in this case the intervals predicted by the two sets are compatible within the uncertainty.
The combination of these results indicates a possible value for the zero-crossing position at an interval $0.1-0.25\gev$.

Although a similar analysis for the form $\Gamma_2$ is not possible, it is important to refer that the fit with \cref{eq:log with mass} maintains a stable behaviour, similar to the one found in \cref{fig:3gluon_gamma_diag_mass_cairo} for a large range of $p_f$.
This is a good indication of the model describing the data. However, an increase in the precision of the results is needed to better understand the possibility of the sign change and IR finiteness of the three gluon vertex.


\subsection*{Finite volume effects}

\begin{figure}[htb!]
	\centering
	\input{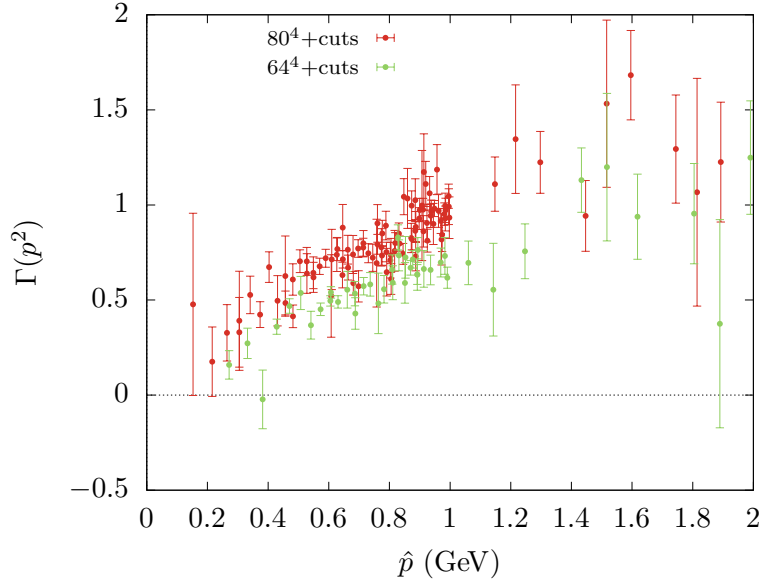}
	\caption{$\Gamma(p^2)$ from the $\beta = 6.0, 80^4$ ensemble compared with the results from \cite{anthony_2016} using the $\beta = 6.0, 64^4$ lattice with 2000 configurations. Above $1\gev$ only data surviving momentum cuts is shown.}
	\label{fig:3gluon_volume}
\end{figure}

To complete the analysis of the three gluon vertex we compare the results obtained from the $80^4$ lattice using 550 configurations and those from the $64^4$ lattice with 2000 configurations and partial Z4 averaging\footnote{The data from the $64^4$ was previously computed in \cite{anthony_2016} using momentum cuts above $1\gev$.}. 
Since both lattices have the same spacing, this comparison allows to search for possible finite volume effects for the three gluon vertex.

The dimensionless form factor $\Gamma(p^2)$ is shown for both lattices in \cref{fig:3gluon_volume} where momentum cuts were applied above $1\gev$.
Although the $80^4$ lattice data is noisier and shows larger error bars, as a result of the difference in the size of the ensembles, both sets of data seem to have the same general behaviour approaching the infrared region.
However, the current data suggests a possible shift enhancing the $\Gamma(p^2)$ for the $80^4$ lattice in comparison with the $64^4$ results.
The curve produced by the $80^4$ lattice data seems to be above the $64^4$ results for momenta below $1.5\gev$, above which the fluctuations become larger and the results become compatible within the uncertainty. 
This enhancement could result from the difference in lattice sizes and suggests a finite volume effect for low momentum. 

Finite volume effects for the gluon propagator were studied in \cite{Oliveira_2012}, which was found to have an IR decrease with the increase of lattice size at a fixed spacing $a$.
However, the relevant momentum scales for this effect seem to be different for the three gluon vertex, with the enhancement extending to higher momenta than for the propagator.
If we consider this effect for the propagator, and disregard a possible, independent finite volume effect on the complete three gluon correlation function $G(p^2)$, the pure vertex $\Gamma(p^2)$ is enhanced for low momentum when dividing by the product $D(p^2)^2D(0)$.
Indeed, the lattice data seems to be compatible with an increase for low momentum, however this is a rather rough estimate of the effect and we should have in mind that the finite volume can also directly affect the complete correlation function.

\begin{table}[htb!]
	\centering
	\begin{tabular}{ccccc}
		\toprule
		\multirow{2}{*}{} & \multicolumn{2}{c}{$64^4$}  & \multicolumn{2}{c}{$80^4$}  \\
		\cmidrule{2-5}
		& $\chi^2/d.o.f.$ & $p_0~(\si{GeV})$     & $\chi^2/d.o.f.$ & $p_0~(\si{GeV})$     \\
		\cmidrule{2-5}
		$\Gamma_1$        & 1.09            & 0.180(14) & 1.28            & 0.156(18) \\
		$\Gamma_2$        & 1.06            &           & 1.19            &           \\
		$\Gamma_3$        & 1.12            & 0.209(43) & 1.18            & 0.180(43) \\
		\bottomrule
	\end{tabular}
	\caption{Fit parameters for the $64^4$ and $80^4$ lattice using the three models in \cref{eq:toymodel,eq:log with mass,eq:power law ansatz}.}
	\label{tab:fit_volume}
\end{table}

\begin{figure}[htb!]
	\centering
	\input{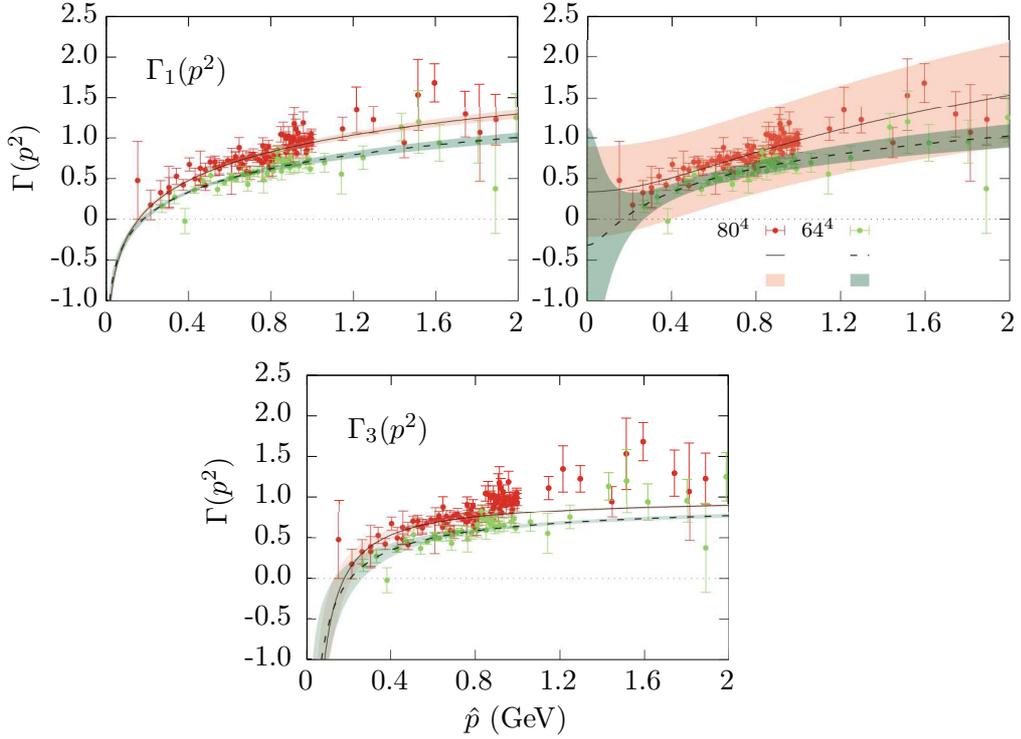}
	\caption{$\Gamma(p^2)$ with momentum cuts above $1\gev$ for the $80^4$ and $64^4$ lattice. The curves result from the fits with \cref{eq:toymodel} (top left), \cref{eq:log with mass} (top right), and \cref{eq:power law ansatz} (bottom plot) with fitting ranges $p_f = 1.7\gev$ for the first two, and $p_f=0.85\gev$ for the latter.}
	\label{fig:3gluon_3fits_volume}
\end{figure}

Regarding the position of a possible sign change, assuming the previous hypothesis for the finite volume effect, the change in the propagator amounts to an overall multiplicative factor and thus the position of the zero-crossing is untouched.
However, again we notice that the complete effect on the three gluon correlation function may induce further changes and can in fact change this value.
Besides, since no statistically relevant signal of the zero-crossing is found for neither of the ensembles, we cannot probe how the volume affects this property.

To better understand the possible finite volume effect we reproduce the fits with the three models, \cref{eq:toymodel,eq:log with mass,eq:power law ansatz} for the same momentum ranges as in the previous analysis for each corresponding model. The results are shown in \cref{fig:3gluon_3fits_volume} for the three models and the fit parameters are summarized in \cref{tab:fit_volume}. We see that in general the $\chi^2$ is lower for the $64^4$ due to the smoothness of the data computed from a larger ensemble.
Moreover, the position of the possible zero-crossing for both $\Gamma_1$ and $\Gamma_3$ seem to be shifted for slightly higher momenta in the $64^4$ lattice, however both estimates for the sign change are compatible within the uncertainty.
The form $\Gamma_2$ seems to have lower $p_0$ for the $64^4$ lattice, however a large uncertainty is associated with the results for momenta below $\sim0.3\gev$ which hinders the analysis of a possible sign change.

\section{Four gluon vertex}\label{sec:result_4gluon}

In this section we report on the four gluon correlation function computed from the two ensembles in \cref{tab:lattices}. 
As referred in \cref{sec:intro_fourgluon_bases}, on a lattice simulation we have access to the full Green's functions only. However, the four point correlation function involves, besides the pure four gluon 1PI function, also the disconnected terms contributions and those associated with the three gluon irreducible diagrams. All these contributions can be removed by a proper choice of the kinematics. 

Even after discarding these contributions, a lattice simulation returns the four gluon Green function that combines the corresponding irreducible diagram with external gluon propagators, \cref{eq:pure four gluon vertex}.
Then, to measure the four point 1PI function the full Green's function requires the removal of the gluon propagators. However, this operation enhances the fluctuations, specially at large momenta, where the propagator becomes small, and adds a further difficulty to the measurement that we aim to perform.
Due to increased fluctuations for the pure vertex we only show the complete correlation function.

Regarding previous investigations on the IR properties of the four gluon vertex only continuum studies have been conducted \cite{Huber_2015,Binosi_2014}, also establishing a possible zero-crossing for some form factors. Some qualitative relations may be established between lattice and continuum results. However, these comparisons should be considered with care due to a weak signal conveyed by the lattice four gluon correlation function.

In general, the fluctuations of higher order functions in a Monte-Carlo simulation are larger and the computation necessarily calls for the use of large ensembles of configurations. 
To try to overcome the problem of statistical fluctuations, in all cases we perform a Z4 average, as done in the previous sections.
Unfortunately, although increasing the quality of the Monte-Carlo signal, the Z4 averaging is not sufficient to produce results with small or relatively small statistical errors for the statistics that we are using. Certainly, an increase in the number of gauge configurations will allow to overcome, at least partially, the problem of the statistical fluctuations. 

Additionally, only a restricted class of momentum points will be shown, namely the generalized diagonal kinematics. These allow to reach lower momentum values and carry lessened hypercubic artifacts.
However, of the four types of diagonal momenta only the mixed cases will be shown. The reason is again related with the effort to increase the signal to noise ratio. On-axis momenta are disregarded for involving higher hypercubic artifacts, and generally larger error bars due to smaller statistics. On the other hand, fully diagonal kinematics of the form $(n,n,n,n)$ are disregarded due to having a smaller set of possible distinct $H(4)$ averaging points. 
Both $(n,n,n,0)$ and $(n,n,0,0)$ retain a good balance in `non--equivalent' Z4 averaging points while not being strongly affected by $H(4)$ artifacts when compared with on-axis momenta.

As a starting point we are interested only in obtaining a proper signal of the four gluon correlation function. A detailed analysis of the infrared behaviour of the functions is difficult due to the uncertainty associated with the data. 
The $64^4$ lattice with 2000 configurations provides a much better result and will be analysed. The $80^4$ lattice with 550 configurations allows access to lower momenta, however substantial fluctuations in the data inhibit its analysis. For the latter, only points above a given momentum will be shown and compared with the results from the larger ensemble.

\subsection{Four gluon correlation function}


\begin{figure}[htb!]
	\centering
	\input{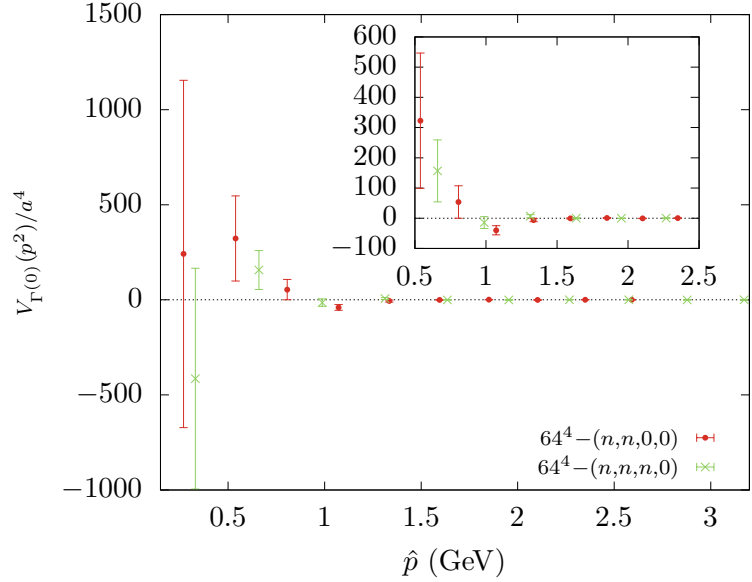}
	\caption{Four gluon vertex form factor $V_{\Gamma^{(0)}}(p^2)$ with external propagators from the $\beta=6.0,~64^4$ lattice. Only mixed diagonal configurations are considered. The smaller plot shows a restricted range of momentum to better visualize the mid momentum region. All data was rescaled by a factor of 1000.}
	\label{fig:4gluon_Gamma0}
\end{figure}

\begin{figure}[htb!]
	\centering
	\input{analysis_results/four_gluon/plots/4gluon_G.tex}
	\caption{Four gluon vertex form factor $V_G(p^2)$ with external propagators from the $\beta=6.0,~64^4$ lattice. Only mixed diagonal configurations are considered. The smaller plot shows a restricted range of momentum to better visualize the mid momentum region. All data was rescaled by a factor of 1000.}
	\label{fig:4gluon_G}
\end{figure}

We now show the results for the four gluon correlation function from the $\beta=6.0,~64^4$ and $80^4$ ensembles. As introduced in \cref{sec:intro_fourgluon_bases}, for the configuration $(p,p,p,-3p)$ only two form factors are possible to extract, $V_{\Gamma^{(0)}}(p^2)$ and $V_G(p^2)$ associated with the tree-level and the $G$ tensor, respectively. 

For this  particular kinematics the results for the $64^4$ lattice are shown in \cref{fig:4gluon_G,fig:4gluon_Gamma0}. Only the two mixed diagonal configurations are shown with $V_G(p^2)$ and $V_{\Gamma^{(0)}}(p^2)$ on the first and second figure, respectively. Notice these are not the pure, dimensionless form factors due to the presence of the external propagators, i.e. we are using 
\begin{equation}
V_i(p^2) = V'_i(p^2)(D(p^2))^3 D(9p^2),
\end{equation}
where $V'_i(p^2)$ corresponds to the pure vertex form factor, as defined in \cref{sec:intro_fourgluon_bases}.
 
A smaller plot is shown in each figure with a narrower range to facilitate the analysis of the behaviour of the function for the mid-momentum range.
Both sets of data $(n,n,0,0)$ and $(n,n,n,0)$ seem to follow a similar curve although with enlarged statistical fluctuations in the IR region.
The fact that two sets of non-equivalent kinematics produce similar curves should be an evidence of this result being a proper signal of the four gluon correlation function. 

$V_{\Gamma^{(0)}}(p^2)$ shown in \cref{fig:4gluon_Gamma0} seems to oscillate quite smoothly near $1.1\gev$ where it reaches a minimum. It subsequently grows for low momentum and seems to approach a finite value near the origin. However, the considerable amount of uncertainty associated with the first two points  hinders the interpretation of the IR behaviour.

The values for $V_G(p^2)$ in \cref{fig:4gluon_G} have larger uncertainty compared to $V_{\Gamma^{(0)}}(p^2)$. Nonetheless, both kinematics seem to follow the same behaviour, suggesting a local maximum for $p\sim 1\gev$ (see the small plot), followed by a minimum around $p = 0.6\gev$ with a possible growth for low momentum.
Notice that the uncertainty involved does not allow to properly confirm this. 

\begin{figure}[htb!]
	\centering
	\input{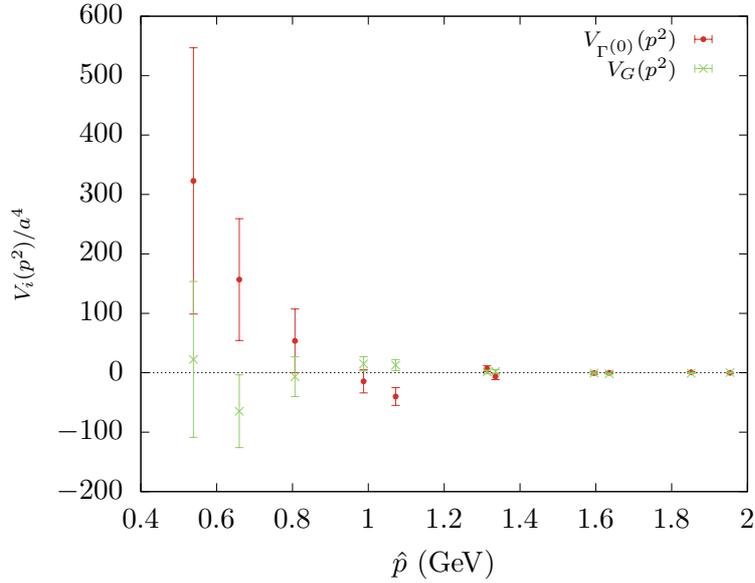}
	\caption{Four gluon vertex form factors $V_{\Gamma^{(0)}}(p^2)$ and $V_G(p^2)$ with external propagators from the $\beta=6.0,~64^4$ lattice. Only mixed diagonal configurations are shown and the lowest momentum points disregarded due to large fluctuations.}
	\label{fig:4gluon_Gamma0_G}
\end{figure}

From the comparison of both form factors in \cref{fig:4gluon_Gamma0_G} for the same momentum configurations we  notice that the contribution from $V_{\Gamma^{(0)}}(p^2)$ is slightly larger than the contribution from $V_{G}(p^2)$ for the range $0.5-1.5\gev$. This possible difference in the weight of the contribution from each structure was also explored in \cite{Huber_2015} with the results following the same pattern. 
Again, the large uncertainty affecting lattice results allows only for a qualitative and limited comparison.

\begin{figure}[htb!]
	\centering
	\input{analysis_results/four_gluon/plots/4gluon_Gamma0_80-4.tex}
	\caption{Four gluon vertex form factor $V_{\Gamma^{(0)}}(p^2)$ with external propagators from the $\beta=6.0,~80^4$ (red) and $64^4$ (green) ensembles. Only mixed diagonal configurations are considered and the lowest momentum points were disregarded. All data was rescaled by a factor of 1000.}
	\label{fig:4gluon_Gamma0_80.4}
\end{figure}

\begin{figure}[htb!]
	\centering
	\input{analysis_results/four_gluon/plots/4gluon_G_80-4.tex}
	\caption{Four gluon vertex form factor $V_G(p^2)$ with external propagators from the $\beta=6.0,~80^4$ (red) and $64^4$ (green) ensembles. Only mixed diagonal configurations are considered and the lowest momentum points were disregarded. All data was rescaled by a factor of 1000.}
	\label{fig:4gluon_G_80.4}
\end{figure}

A further evidence for this result being a proper signal of the four gluon correlation function is found from the comparison with the $80^4$ lattice.
In \cref{fig:4gluon_Gamma0_80.4,fig:4gluon_G_80.4} both $V_{\Gamma^{(0)}}(p^2)$ and  $V_G(p^2)$ are shown for mixed diagonal configurations $(n,n,0,0)$ and $(n,n,n,0)$ and for both lattices. A smaller range of momentum was considered discarding the two lowest momenta (these show large fluctuations, mainly for the larger lattice).

The form factor $V_{\Gamma^{(0)}}(p^2)$ is compared for both lattices in \cref{fig:4gluon_Gamma0_80.4}.
Looking only at the $80^4$ data we notice a possible similar structure to that found in \cref{fig:4gluon_Gamma0} (see the small plot). The $80^4$ results suggest a decrease for negative values and a subsequent growth for lower momentum. However, a discrepant point appears around $p=0.8\gev$ and the errors associated with the data are much larger than those from the $64^4$ lattice.
In addition, if we compare both sets of data in \cref{fig:4gluon_Gamma0_G} from both lattices we notice a  shift in the momentum scales where these structures are found. The possible minimum occurs for higher momentum in the $64^4$ lattice.
Although the general structure of the curve seems to provide the same oscillation, the shift in the data and the large uncertainty in the $80^4$ results could be a sign of inconsistent data and restrains us from making further claims.

The data for $V_G(p^2)$ in \cref{fig:4gluon_G_80.4} also suggests an agreement between the results from both lattices. However, albeit the curves created by both sets of data are compatible and have the same general structure within the uncertainty, the error bars associated with the $80^4$ lattice are large and thus this comparison is unreliable.
In this case, we do not observe a shift in the structure of the curve\footnote{Notice that the momentum points do not perfectly match due to the different lattice size, $N$. The definition of lattice momentum is $ap = 2\pi n/N$.}. 
Both the local highest point, around $p=0.9\gev$ and the minimum near $p=0.6\gev$ seem to occur at the same scales in both ensembles. However, while the minimum for the $64^4$ lattice seems to have a negative value, the same cannot be claimed for the larger lattice due to the large error bars.

Despite the large uncertainty, it is remarkable that two distinct lattices seem to provide the same general behaviour for the form factors with similar structures for the curves. This should be an evidence that we are indeed computing a valid (albeit weak) signal of the four gluon correlation function. 
Nonetheless, a significant increase in the precision of the signal is required to establish reliable conclusions.

\subsubsection*{Comparison with continuum results}

\begin{figure}[htb!]
	\centering
	\includegraphics[scale=0.8]{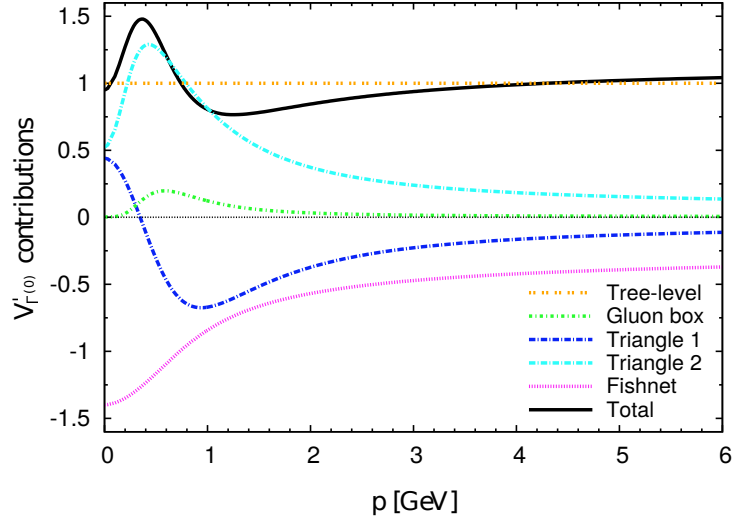}
	\caption{Original data from \cite{Binosi_2014} for the DSE computation of the pure four gluon vertex associated with the tree-level tensor $V'_{\Gamma^{(0)}}(p^2)$. The `total' result in black is the relevant structure for comparison.}
	\label{fig:Gamma0}
\end{figure}

\begin{figure}[htb!]
	\centering
	\includegraphics[scale=0.8]{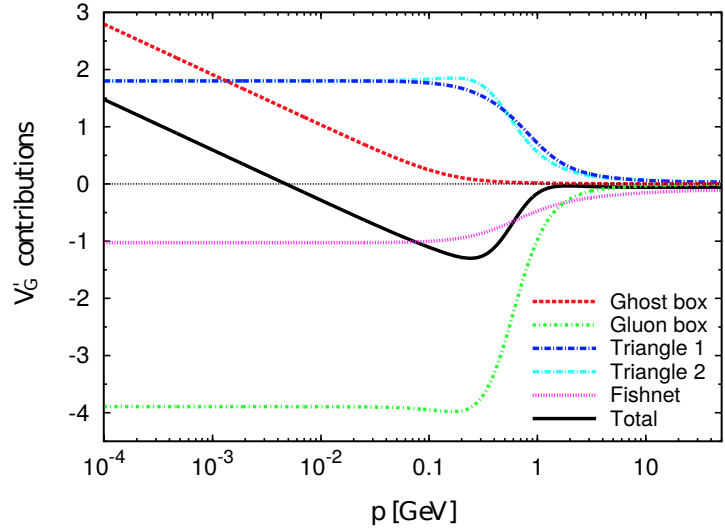}
	\caption{Original data from \cite{Binosi_2014} for the DSE computation of the pure four gluon vertex associated with the tree-level tensor $V'_{G}(p^2)$. The `total' result in black is the relevant structure for comparison.}
	\label{fig:G}
\end{figure}

Despite the large statistical fluctuations, we try to compare our results with previous continuum predictions -- these are currently the only source of possible comparison. For this we compare only the smaller, $64^4$ lattice having a higher precision.

The four gluon vertex was studied in a DSE analysis employing the same tensor basis and kinematic configuration, \cite{Binosi_2014} where it was argued that only the form factor $V_G(p^2)$ shows a possible divergent behaviour in the IR, while $V_{\Gamma^{(0)}}(p^2)$ remains finite.
The original data for the pure vertex form factors $V'_i(p^2)$ from this investigation is shown in \cref{fig:Gamma0,fig:G}. We are interested in comparing our results with the black curves, representing the complete contribution (within the truncation scheme)\footnote{The remaining curves are the individual contributions from one-loop diagrams in the DSE formalism.}.

Although on the lattice we can only access the complete vertex with some reasonable statistical accuracy, we can establish some general comparisons with the continuum results by considering the smooth, and practically constant behaviour of the gluon propagators in the IR.
In addition to this approximation, both the large uncertainty associated with lattice results and the approximations involved in the DSE approach call for careful conclusions from the following comparisons.

Comparing the results for the tree-level form factor in \cref{fig:4gluon_Gamma0,fig:Gamma0} we notice a discrepant shift in the overall functions, namely the DSE curve sets in at unit values for large momenta, while the lattice data seems to approach zero. Notice, however that this could be an effect of the external propagators. 
Nonetheless, the general structure of the lattice data seems to follow the behaviour of the continuum prediction within the large uncertainty. Namely, the pattern of oscillations is similar, showing what seems like a local minimum for $p\sim1\gev$ followed by a sign change for positive values below $1\gev$. 
For smaller momentum the data is less reliable due to larger uncertainty, however it seems to approach a finite IR value, which again follows the continuum prediction.

Another DSE study, using the tree-level tensor only, \cite{Huber_2015} obtained a similar result to that in \cref{fig:Gamma0}. Due to the orthogonality between both tensors $\Gamma^{(0)}$ and $G$, \cref{eq:orthogonality 4gluon tensors}, the results assuming only the tree-level tensor for the basis should have the same general behaviour as the one found in \cref{fig:4gluon_Gamma0}. 
Therefore, this serves as a further connection between lattice and continuum results due to the same qualitative structure in $V_{\Gamma^{(0)}}(p^2)$.

The results for $V_G(p^2)$ in \cref{fig:4gluon_G,fig:G} are also compatible within the large uncertainty of the lattice results. In this case no shift is observed between continuum and lattice data.
The form factor computed from the lattice shows a decrease to negative values for $p\sim 0.6\gev$ in the $64^4$ ensemble, which is also noticeable in the DSE result around the same momentum scales.
For lower momentum the data suggests a possible sign change and subsequent IR growth, again compatible with previous continuum results.
Notice, however that the error bars for low momenta provide limited confidence in these observations.
Also, the finite volume of the lattice does not allow to reach sufficiently low momenta to better evaluate the IR behaviour.

\chapter*{Conclusion}\label{chap:conclusion}
\addcontentsline{toc}{chapter}{\nameref{chap:conclusion}}

In this thesis we computed and analysed three different gluon correlation functions of the pure Yang-Mills theory in Landau gauge using the lattice formalism of QCD.
Two lattices were considered with the same lattice spacing and different physical volumes -- see \cref{tab:lattices}.


In the first part of the work we investigated the gluon propagator to understand how the use of continuum tensor bases affects the knowledge of lattice computed tensors in a 4-dimensional theory. To date, only 2 and 3-dimensional studies have been conducted on this topic \cite{vujinovic2019tensor,Vujinovi_2019}.
To this end we constructed suitable lattice tensor bases respecting the corresponding lattice symmetries.

Continuum relations among lattice and continuum form factors were identified and evaluated for every tensor structure. We found that, within the uncertainty, continuum relations are satisfied for a large range of momentum which seems to indicate that the lattice data is compatible with the Slavnov-Taylor identity.
Furthermore, to probe the quality of our results we used the data from a precise lattice computation \cite{Dudal_2018} as a comparison.
The results obtained with various bases match this benchmark result although with increased fluctuations for larger bases.

The completeness of each tensor basis in describing the lattice tensor $D_{\mu\nu}(p)$ was studied.
Specific kinematics were considered independently for a detailed analysis and
we found that, in general, the most complete bases (larger number of form factors) provide a better reproduction of the original lattice tensor and the use of a continuum tensor basis for the propagator leads to non-negligible loss of information of the lattice correlation function.
The orthogonality of the propagator using lattice tensors was also studied and it serves as a complementary analysis of the completeness for each basis. 

The analysis of the reconstruction for specific kinematics hinted about the existence of special points for which the continuum basis matches the description from lattice bases.
These are single scale momenta which were then investigated exclusively.
Although for these points the continuum and lattice tensors provide the same quality in the description of the tensor, the results are substantially better for configurations closer to the diagonal of the lattice.
Moreover, continuum relations are exactly satisfied by these kinematics and constrain the number of independent form factors describing the tensor. This is in turn related with the similar completeness from lattice and continuum bases.

With this work we provide additional validation for the traditional method to compute vertex functions using points near the diagonal of the lattice.
We conclude that diagonal data not only reduce hypercubic artifacts in the form factors (lattice scalars) but also in the tensor structures that form the basis. This is noticeable in the good reconstruction results obtained for diagonal configurations.
We also confirm that, in general, the use of improved momentum provides a better description of lattice objects than the naively discretized lattice momentum. 
In fact, this change of variables improves also the fulfilment of both continuum and orthogonality conditions, as well as the match with the benchmark result.

Although we did not consider a fully complete tensor to describe the gluon propagator, we found that an increase in the degrees of freedom is accompanied by a considerable rise in statistical fluctuations in the form factors.
This restricts the number of independent tensor structures used due to limited statistics.

The effect of a finite volume lattice was also explored. We found that the generalized diagonal configurations seem to be insensible to the finite volume regarding its reconstruction. 
For the remaining configurations we observed that, in general, the larger lattice provides lower ratios for the reconstruction for both continuum and lattice bases. 

The finiteness of the space was not taken into account in the construction of lattice tensors, and the search for proper bases with respect to the symmetries as well as the size of the lattice should improve the description of the propagator.
Moreover, mixed terms involving both improved and lattice momentum could be considered as well as continuum vanishing terms, depending explicitly on the lattice spacing.
Identically, the behaviour of different tensor bases with varying spacings could be explored.
Finally, proper tensor structures respecting lattice symmetries for higher order correlation functions are yet to be constructed, and would allow to probe how the use of continuum bases affects its description.


In the second part we analysed the three gluon correlation function from the $80^4$ lattice.
We began by showing that the use of the complete set of group transformations (Z4 average) provides an improved signal to noise ratio. This is crucial for the computation of higher order functions.
A comparison with the perturbative prediction for high momenta was performed for both two and three gluon correlation functions, and was confirmed by fitting both curves for sufficiently high momentum.


We analysed the IR behaviour for the three gluon 1PI function.
Two different hypothesis were considered, namely a possible zero-crossing occurring for low momenta with a subsequent IR divergence. 
The effect is interpreted using the concept of dynamical mass generation for the gluon which acquires a momentum dependent mass, whereas the ghost is supposed to remain massless thus inducing a possible divergence.
This hypothesis is advocated by various continuum studies, however it is highly dependent on the approximations employed.
Conversely, an analytic investigation of the gluon and ghost two point functions suggest a possible dynamical ghost mass which should regularize the vertex and thus remove the IR divergence \cite{alex2020analytic}.

Since the IR data provides no clear evidence of the sign change, let alone the possible divergence for lower momenta, we analysed the behaviour of the data by considering three different functional models.
The first form contains an IR unprotected logarithm, \cref{eq:toymodel}, which other than the zero-crossing also allows an subsequent divergence.
Both the complete set of data, and the points surviving momentum cuts above $1\gev$ provide good quality fits.
The results of the fits for various ranges indicate a zero-crossing around $0.15-0.25\gev$ from this functional form.
Notice, however, that while we try to model the zero-crossing, the divergence is not sustained by lattice data, hence predictions for this property are less reliable.

The second functional form, \cref{eq:log with mass}, represents the case of a non-vanishing dynamical ghost mass which is included in the logarithm and removes the IR divergence while still allowing for a sign change. 
In this case the complete data provides a good fit with the curve for the range of momenta considered. It is consistent with a positive IR value for the vertex and an absent sign change.
On the other hand, although the data after momentum cuts also matches the data, this curve is associated with a larger uncertainty.
In this case a sign change is possible below $0.4\gev$ but it is not guaranteed by the curve.

The last model, \cref{eq:power law ansatz} is a power law ansatz whose purpose is to probe the degree of the possible divergence for low momentum, and thus the functional form is restricted to lower momentum. 
This can be noticed by the decline in the quality of fit for momenta above $1\gev$ and by the poor match between the curve and lattice data for this region. On the other hand, for momenta below $\sim1\gev$ the curve shows a good agreement with lattice data.
However, since the possible divergence lacks confirmation from lattice data, no reliable conclusions can be established.


Although we do not yet have precise IR data to validate the zero-crossing, we tried to establish a momentum range for the sign change using the analysis of the three models.
However, the possible divergence is currently out of our grasp due to the lack of data in the very deep IR. 
The search for this property calls for a larger lattice, however to obtain a sufficient amount of statistics with a large lattice requires a substantial increase in the computational resources.
This difficulty could be overcome by using large ensembles of high volume but coarser lattices.
This should be possible due to the seemingly negligible effect of the discretization in the infrared region for this vertex found in \cite{maas2020threegluon}. 

For a possible finite IR value for the vertex, the data seems to be compatible with the model despite the large uncertainty hindering a more detailed analysis.
A better description of the deep infrared region is necessary to make a more accurate study.

To conclude the study of the three gluon vertex, a comparison between the $80^4$ and $64^4$ lattice data was conducted to search for possible finite volume effects.
The results from the $80^4$ lattice seem to be enhanced relatively to those from the $64^4$ lattice, creating a shift for momenta below $\sim1.4\gev$. 
This can be partially explained by previous investigations of the gluon propagator, which was found to decrease in the IR with increasing volume, and thus inducing an enhancement in the three gluon vertex when divided by the propagators.
We also compared the predictions from the three models in \cref{eq:toymodel,eq:log with mass,eq:power law ansatz} with the results from the $64^4$ lattice.
While the curves are modified due to the shift in the data, remarkably the prediction for the zero-crossing seems to remain unchanged within the uncertainty.
This is compatible with the finite volume effect amounting to a multiplicative factor such as the one induced by the division of the external propagators.
However, in order to properly understand the effect, a detailed analysis of both the complete and pure three gluon functions is necessary for different lattice volumes.
For the second model, \cref{eq:log with mass}, the fit with the $64^4$ data follows the same behaviour but with increased precision.
The sign change seems to be predicted for lower momenta, however this is not unambiguously confirmed within the error bars.

While we explored a single kinematic configuration, additional configurations could be considered to analyse its IR behaviour.
The use of different volume lattices for other kinematics would also allow to improve the knowledge on the possible finite volume effect.
Another extension of this work could be related to the large statistical fluctuations affecting the high momentum region of the three gluon 1PI function. However, as discussed in \cref{sec:results 3_gluon} this is not achievable by an increase in the number of gauge-field configurations and thus other alternatives should be envisioned.




For the final topic we computed the four gluon correlation function.
As a higher order function, it is associated with larger statistical fluctuations which hinder the attainment of a discernible signal.
In fact, current precision allows only to study the complete correlation function, while the 1PI function carries large fluctuations.
Using a suitable kinematic configuration we isolated the contribution of the pure four gluon 1PI function with external propagators. In addition to the choice of kinematics, an approximation of the Lorentz tensor basis reduced the number of possible structures to three.
However, for the kinematics $(p,p,p,-3p)$ and the approximation employed, only two form factors are possible to extract.

To improve the signal quality, we analysed the correlation function only for configurations $(n,n,n,0)$ and $(n,n,0,0)$.
The points from both kinematics seem to define a single and smooth curve, except for low momentum due to fluctuations and large error bars.
Additionally, the results from both ensembles have seemingly matching curves within the uncertainty, with exception of some momentum points in the $V_{\Gamma^{(0)}}(p^2)$ form factor, which show some discrepancies for the $80^4$ data.
Notice, however, that the $80^4$ lattice provides reduced statistics and the comparison is to be taken with care.

To complete the analysis we compared lattice results against the pure four gluon vertex from previous continuum investigations \cite{Binosi_2014,Huber_2015}. 
This is a very delicate comparison due to the impossibility of the computation of the lattice four gluon 1PI function.
Hence, only a very qualitative connection between the continuum and lattice curves was established.
Nonetheless, this should be a good indication of the signal obtained.

Although the results are an evidence that we are indeed peeking at the four gluon correlation function, the statistical relevancy of the signal is still very small and the signal should be improved in order to properly analyse the vertex.
From the previous analysis, the main structures observed in the form factors should be noticeable for a reasonable range of momentum achievable by our current lattices.  
Thus, an increase in statistics for the current ensembles should help providing a clearer curve. 
Besides, the pure 1PI form factors may only be computed accurately with increased precision.

\printbibliography

\appendix
\begin{appendices}
	
\chapter{$SU(N)$ generators and identities}\label{apend:liegroups}

$SU(N)$ is the special unitary group of degree $N$ whose elements $U$ are $N\times N$ unitary matrices, $U^\dagger U = \mathds{1}$, satisfying $\det(U) = 1$. It is a Lie group, with its elements being continuously generated by real parameters $\theta^a \in \mathds{R}$. Each element can be written as
\begin{equation}
U = e^{i\theta^at^a}
\end{equation}
where $t^a$ are the $N^2-1$ group generators, corresponding to each parameter $\theta^a$. The generators are hermitian and traceless matrices
\begin{align}
	(t^a)^\dagger = t^a, && \tr(t^a) = 0,
\end{align}
that span a vector space underlying the corresponding Lie algebra, $\mathfrak{su}(N)$. The generators obey the commutation relation
\begin{equation}
[t^a,t^b] = if^{abc}t^c
\end{equation}
where $f^{abc}$ are the antisymmetric structure constants, specific for each group and non-zero for a non-abelian group.
A fundamental property of Lie groups is the \textit{Jacobi identity}
\begin{equation}
[t^a,[t^b,t^c]] +  [t^b,[t^c,t^a]] + [t^c,[t^a,t^b]]= 0
\end{equation}
implying 
\begin{equation}
f^{ade}f^{bcd} + f^{bde}f^{cad} + f^{cde}f^{abd} = 0.
\end{equation}

There are two main irreducible representations of the groups $SU(N)$. 
The fundamental representation consists of $N$-dimensional complex vectors, with the group as well as the algebra elements being $N\times N$ matrices. For QCD, $N=3$, this corresponds to the representation of the 3-spinor quark field. 
The usual choice of the normalization of the generators is
\begin{equation}
	f^{acd}f^{bcd} = N\delta^{ab}
\end{equation}
from which we can derive for the fundamental representation,
\begin{equation}
	\Tr\left(t^at^b\right) = \frac{\delta^{ab}}{2}.
\end{equation}
The structure constants may be written as
\begin{equation}
	f^{abc} = -2i\tr([t^a,t^b]t^c)
\end{equation}
and the product of two generators has the general form,
\begin{equation}
	t^at^b = \frac{\delta^{ab}}{2N} + \frac{1}{2}d^{abc}t^c + \frac{1}{2}if^{abc}t^c
\end{equation}
where the totally symmetric object is defined as $d^{abc} = 2\Tr\left(t^a\{t^b,t^c\}\right)$, making use of the anti-commutator defined as
\begin{equation}
	\{t^a,t^b\} = \frac{\delta^{ab}}{N} + d^{abc}t^c.
\end{equation}
Additional identities may be obtained
\begin{align}
	&\Tr\left(t^at^bt^c\right) = \frac{1}{4}(d^{abc} + if^{abc})\\  
	&f^{abc}f^{abc} = N(N^2 - 1) \\
	&f^{abm}f^{cdm} = \frac{2}{N}\left( \delta^{ac}\delta^{bd} - \delta^{ad}\delta^{bc} \right)
					+ d^{acm}d^{dbm} - d^{adm}d^{bcm} \\
	&f^{abm}d^{cdm} + f^{acm}d^{dbm} + f^{adm}d^{bcm} = 0
\end{align}
with a further relation for $N=3$,
\begin{equation}
	\delta^{ab}\delta^{cd} + \delta^{ac}\delta^{bd} + \delta^{ad}\delta^{bc} = 
		3\left( d^{abm}d^{cdm} + d^{acm}d^{dbm} + d^{adm}d^{bcm} \right).
\end{equation}


The other important representation is the adjoint representation to which the generators belong and acts on the vector space spanned by the generators themselves -- it is an $N^2-1$ dimensional representation. 
In QCD, the 8 gluon fields live on the adjoint representation of the group $SU(3)$ and transform accordingly.
The representation matrices of the generators are given by the structure constants
\begin{equation}
	(t^b)_{ac} = if^{abc}.
\end{equation}
A useful relation is the trace of four generators in the adjoint representation
\begin{equation}
	\Tr(t^at^bt^ct^d) = \delta^{ad}\delta^{bc} + \frac{1}{2}\left( \delta^{ab}\delta^{cd} + \delta^{ac}\delta^{bd} \right) +  \frac{N}{4}\left( f^{adm}f^{bcm} + d^{adm}d^{bcm} \right).
\end{equation}
In this representation, the covariant derivative
\begin{equation}
	D_\mu \eta(x) = (\partial_\mu - igA_\mu^at^a)\eta(x)
\end{equation}
takes the component form
\begin{align}
(D_\mu \eta(x))_a &= \partial_\mu \eta_a(x) - igA_\mu^b(t^b)_{ac}\eta_c(x) \\
&= \partial_\mu \eta_a(x) + gf^{abc}A_\mu^b\eta_c(x).
\end{align}

\chapter{Lattice tensors}\label{apend_chap:lattice_tensors}

\section{Construction of the lattice basis} \label{append_sec:construction}

\subsection{Momentum polynomial under a transposition}\label{append_sec:polynomial transform}

We consider a brief proof of the transformation of a polynomial of a vector $p$ under a transposition is given. A transposition is defined by an exchange of two components of a vector, $\sigma \leftrightarrow \rho$, under the operation $T^{\sigma\rho}$. 
A matrix form for this operator is

\begin{equation}
	\begin{cases}
		T^{(\sigma\rho)}_{\mu\nu} = \delta_{\mu\nu}, ~\mu\neq\sigma,\rho \\
		T^{(\sigma\rho)}_{\sigma\nu} = \delta_{\rho\nu} \\
		T^{(\sigma\rho)}_{\rho\nu} = \delta_{\sigma\nu}
		\label{appen_eq:transposition}
	\end{cases}
\end{equation}
which reproduces the correct transformation on the vector $p$:
\begin{equation}
	\begin{cases}
		p'_\nu = p_\nu,~\nu \neq \sigma,\rho  \\
		p'_\sigma = p_\rho,  \\
		p'_\rho = p_\sigma.
	\end{cases}
\end{equation}
Considering the transformation for an arbitrary order of $p$
\begin{equation}
	(p'_\mu)^n = p'_\mu...p'_\mu = T^{(\sigma\rho)}_{\mu\nu_1}p_{\nu_1}...T^{(\sigma\rho)}_{\mu\nu_n}p_{\nu_n}
\end{equation}
and considering the case $\mu\neq\sigma,\rho$ the correct transformation is immediate since all components are left unchanged,
\begin{equation}
	(p'_\mu)^n = p_\mu...p_\mu = (p_\mu)^n = T^{(\sigma\rho)}_{\mu\nu}(p_{\nu}).
\end{equation}
For $\mu = \sigma,\rho$, the transformation is
\begin{align}
	(p'_\sigma)^n &= T^{(\sigma\rho)}_{\sigma\nu_1}p_{\nu_1}...T^{(\sigma\rho)}_{\sigma\nu_n}p_{\nu_n} \nonumber \\
	 &= \delta_{\sigma\nu_1}p_{\nu_1}...\delta_{\sigma\nu_n}p_{\nu_n} \nonumber \\
	 &= T^{(\sigma\rho)}_{\sigma\nu}(p_{\nu})^n = (p_\rho)^n.
\end{align}
This is the same transformation as for the vector $p$, and thus the polynomial transforms accordingly.

\subsection{Second order tensors under $H(4)$ symmetry}\label{appen_sec:Second order tensors under $H(4)$ symmetry}

Here we show that there is no mixing among the diagonal and off-diagonal elements under a general $H(4)$ transformation, using the fact that these transformations can be formed by products of transpositions and inversions.

The transposition operator for the exchange of components $\sigma \leftrightarrow \rho$ was defined in \ref{appen_eq:transposition}. For the inversion of the component $\rho$, we define the operator as
\begin{align}
&P^{\rho}_{\mu\nu} = \delta_{\mu\nu}, ~\mu\neq\rho \\
&P^{\rho}_{\rho\nu} = -\delta_{\rho\nu}.
\label{appen_eq:inversion}
\end{align}
The transformation for a second order tensor under transpositions and inversions is
\begin{align}
	&D'_{\mu\nu} = T^{(\sigma\rho)}_{\mu\tau}T^{(\sigma\rho)}_{\mu\varepsilon}D_{\tau\varepsilon}, \\
	&D'_{\mu\nu} = P^{(\rho)}_{\mu\tau}P^{(\rho)}_{\mu\varepsilon}D_{\tau\varepsilon}.
	\label{appen_eq:inversion_transform_D}
\end{align}

Now we consider the transformation of diagonal elements $\mu=\nu$. For transpositions there are three distinct situations,
\begin{equation}
	\begin{cases}
		D'_{\sigma\sigma} = \delta_{\rho\tau}\delta_{\rho\varepsilon}D_{\tau\varepsilon} = D_{\rho\rho} \\
		D'_{\rho\rho} = D'_{\sigma\sigma} \\
		D'_{\mu\mu} =  D_{\mu\mu},~\mu\neq\rho,\sigma
	\end{cases}
\end{equation}
and we see that no off-diagonal terms appear.

A similar analysis can be considered for the inversions using \ref{appen_eq:inversion_transform_D}
\begin{equation}
	\begin{cases}
		D'_{\rho\rho} = (-\delta_{\rho\tau})(-\delta_{\rho\varepsilon})D_{\tau\varepsilon} = D_{\rho\rho} \\
		D'_{\mu\mu} =  D_{\mu\mu},~\mu\neq\rho
	\end{cases}
\end{equation}
and again for this transformation, no off-diagonal terms appear for the diagonal transformation.

We now consider the off-diagonal transformation, $\mu\neq\nu$. For the transpositions there are again three distinct cases
\begin{equation}
	\begin{cases}
		D'_{\sigma\nu} = \sum_{\tau,\varepsilon}\delta_{\rho\tau}\delta_{\nu\varepsilon} = D_{\rho\nu}  \\
		D'_{\rho\nu} = D'_{\sigma\nu} \\
		D'_{\rho\sigma} =  D_{\sigma\rho}
	\end{cases}
\end{equation}
and no diagonal terms are involved. On the other hand for inversions there are two cases
\begin{equation}
	\begin{cases}
		D'_{\mu\nu} = -D_{\mu\nu},~ &\mu=\rho \wedge \nu=\rho \\
		D'_{\mu\mu} =  D_{\mu\mu},~ &\mu\neq\rho \wedge \nu\neq\rho.
	\end{cases}
\end{equation}
We conclude that a general $H(4)$ transformation does not mix the diagonal and off-diagonal elements for second order tensors.

\section{General construction for projectors}\label{append_sec:projectors_latticebasis}
	
The projectors $\mathcal{P}^k$ are necessary to extract form factors corresponding to each basis element. Here we describe the general form of constructing projectors, for an arbitrary vector space. 

Given a general tensor $\Gamma$, this object will be described by a basis of $N$ tensor elements $\tau^j$,
\begin{equation}
	\Gamma = \sum_{j = 1}^{N}\gamma^j\tau^j
\end{equation}
where $\gamma^j$ are the corresponding dressing functions.
Suppose we want to extract one of the form factors $\gamma^k$ by acting on $\Gamma$ with an operator $\mathcal{P}^k$ (this operation involves the necessary index contractions to build a scalar). The operation is of the form,
\begin{equation}
	\mathcal{P}^k\Gamma = \mathcal{P}^k\left(\sum_{j = 1}^{N}\gamma^j\tau^j\right) = \gamma^k.
\end{equation}
From this we may extract the relation
\begin{equation}
	\mathcal{P}^k\tau^j = \delta^{kj}.
	\label{appen:general_projector}
\end{equation}
using the completeness of the basis, and the linearity of the operator.
Considering the most general form of the projector $\mathcal{P}^k$, constructed from basis elements
\begin{equation}
	\mathcal{P}^k = \sum_{i = 1}^{N}A_{ki}\tau^i
	\label{appen_eq:general_projector_basis}
\end{equation}
and substitute this into eq. \ref{appen:general_projector}, to obtain
\begin{equation}
	\sum_{i = 1}^{N}A_{ki}\tau^i\tau^j = \delta^{kj} \Leftrightarrow A_{ki} = (\tau^k\tau^i)^{-1}.
\end{equation}
This reduces the extraction of the form factors to a matrix inversion problem. We need only to build the matrix with elements $A_{ki}^{-1} = \tau^k\tau^i$, where the contraction of indices referred before is assumed, and obtain its inverse $A$
\begin{equation}
	\mathcal{P}^k = \sum_{i = 0}^{N}(\tau^k\tau^i)^{-1}\tau^i.
\end{equation}

With this mechanism, it is  straightforward to understand why it is impossible to build well defined projectors when there are redundant basis elements that can be written as a linear combination of the remaining elements. 
In this case, not all rows will be linearly independent, and it is know from linear algebra that matrices with this property are singular, i.e. non-invertible, and the projectors cannot be defined.

\subsection{Projectors for the lattice bases}

We use the previous mechanism to build the projectors for the tensor bases considered throughout the work.
We begin with the general form for second order tensors in the continuum
\begin{equation}
	D_{\mu\nu}(p) = A(p)\delta_{\mu\nu} + B(p)p_\mu p_\nu
\end{equation}
with the elements $\tau^1 = \delta_{\mu\nu}$ and $\tau^2 = p_\mu p_\nu$. The matrix $A^{-1}$ for a $N_d$ dimensional space is
\begin{equation}
	A^{-1} = \mqty(N_d^2 & p^2 \\ p^2 & p^4),
\end{equation} 
and its inverse
\begin{equation}
	A = \frac{1}{p^4(N_d - 1)}\mqty(p^4 & -p^2 \\ -p^2 & N_d).
\end{equation}
The projectors are built with \cref{appen_eq:general_projector_basis}
\begin{align}
	&\mathcal{P}^1_{\mu\nu} = \frac{1}{N_d - 1}\left(\delta_{\mu\nu} - \frac{p_\mu p_\nu}{p^2}\right)\\
	&\mathcal{P}^2_{\mu\nu} = \frac{1}{N_d - 1}\left(-\frac{\delta_{\mu\nu}}{p^2} + N_d\frac{p_\mu p_\nu}{p^4}\right),
\end{align}
and the extraction of the respective form factors follows immediately
\begin{align}
	&A(p) = \frac{1}{N_d - 1}\left(\sum_{\mu}D_{\mu\mu}(p) - \frac{1}{p^2}\sum_{\mu\nu}p_\mu p_\nu D_{\mu\nu}(p)\right)\\
	&B(p) = \frac{1}{N_d - 1}\left(-\frac{1}{p^2}\sum_{\mu}D_{\mu\mu}(p) + \frac{N_d}{p^2}\sum_{\mu\nu}p_\mu p_\nu D_{\mu\nu}(p)\right).
\end{align}
This procedure can be simplified when considering the tensor form 
\begin{equation}
	D_{\mu\nu}(p) = D(p^2)\left(\delta_{\mu\nu} - \frac{p_\mu p_\nu}{p^2}\right),
\end{equation}
with the form factor extracted with
\begin{equation}
	D(p^2) = \frac{1}{N_d - 1}\sum_{\mu}D_{\mu\mu}(p).
\end{equation}

We consider now the lattice basis \ref{eq:full_lattice_basis}. 
As referred in the construction of the basis, the diagonal elements do not mix with off-diagonal, which allow us to analyse them independently. The reducibility of the group representation splits the five dimensional matrix into two square matrices of size two and three. 
It is thus important to use two different index contractions, one considering only diagonal terms, $ \sum_{\mu}\tau_{\mu\mu}^i\tau_{\mu\mu}^j$, and the second considering only off-diagonal elements $\sum_{\mu \neq \nu}\tau_{\mu\nu}^i\tau_{\mu\nu}^j$. 
Starting with the diagonal elements $\tau^1 = \delta_{\mu\mu}$, $\tau^2 = p_\mu^2$ and $\tau^3 = p_\mu^4$. The contraction matrix $A^{-1}$ is
\begin{equation}
	A^{-1} = \mqty(N_d & p^2 & p^{[4]} \\ 
				   p^2 & p^{[4]} & p^{[6]}\\
				   p^{[4]} & p^{[6]} & p^{[8]}).
\end{equation}
Hence, the diagonal form factors are 
\begin{align}
	E(p) = \frac{1}{\Delta_1}\bigg[\sum_{\mu}D_{\mu\mu}(p^{[4]}p^{[8]} - (p^{[6]})^2) &+ \sum_{\mu}p_\mu^2D_{\mu\mu}(p^{[4]}p^{[6]} - p^2p^{[8]})  \nonumber \\
	&+ \sum_{\mu}p_\mu^4D_{\mu\mu}(p^{2}p^{[6]} - (p^{[4]})^2)\bigg] 
	\label{appen:factorE} \\
	F(p) = \frac{1}{\Delta_1} \bigg[\sum_{\mu}D_{\mu\mu}(p^{[4]}p^{[6]} - p^2p^{[8]}) &+ \sum_{\mu}p_\mu^2D_{\mu\mu}(N_dp^{[6]} - (p^{[4]})^2)  \nonumber \\
	&+ \sum_{\mu}p_\mu^4D_{\mu\mu}(p^{2}p^{[4]} - N_dp^{[6]})\bigg] 
	\label{appen:factorF}\\ 
	G(p) = \frac{1}{\Delta_1} \bigg[\sum_{\mu}D_{\mu\mu}(p^{2}p^{[6]} - (p^{[8]})^2) &+ \sum_{\mu}p_\mu^2D_{\mu\mu}(p^2p^{[4]} - N_dp^{[6]}) \nonumber \\
	&+ \sum_{\mu}p_\mu^4D_{\mu\mu}(p^{2}p^{[4]} - N_dp^{[6]})\bigg] 
	\label{appen:factorG}
\end{align}
with 
\begin{equation}
	\Delta_1 = N_d\left(p^{[4]}p^{[8]} - (p^{[6]})^2\right) + p^2\left(p^{[4]}p^{[6]} - p^2p^{[8]}\right) + p^{[4]}\left(p^{2}p^{[6]} - (p^{[4]})^2\right).
\end{equation}

Similarly we can repeat the procedure for the two dimensional, off-diagonal case, obtaining both form factors,
\begin{align}
&H(p) = \frac{2}{\Delta_2}\bigg[\sum_{\mu\neq\nu}p_\mu p_\nu D_{\mu\nu}(p^{[4]}p^{[6]} - p^{[10]}) - \sum_{\mu\neq\nu}p_\mu^3 p_\nu^3 D_{\mu\nu}(p^{2}p^{[4]} - p^{[6]})\bigg] 
\label{appen:factorH} \\
&I(p) = \frac{1}{\Delta_1} \bigg[\sum_{\mu\neq\nu}p_\mu p_\nu D_{\mu\nu}(p^{[8]} - (p^{[4]})^2) + \sum_{\mu\neq\nu}p_\mu^3 p_\nu^3 D_{\mu\nu}(p^4 - p^{[4]})\bigg] 
\label{appen:factorI}
\end{align}
with
\begin{equation}
	\Delta_2 = 2\left(p^2p^{[4]} - p^{[6]}\right)\left(p^{[8]} - (p^{[4]})^2\right) + 2\left(p^4 - p^{[4]}\right)\left(p^{[4]}p^{[6]} - p^{[10]}\right).
\end{equation}

Having all projectors for the lattice basis, we need to consider the case of the generalized diagonal kinematics where these projectors are not possible to obtain. 
This analysis is done for each individual configuration.
Starting with the diagonal, $(n,n,n,n)$, the gluon propagator is
\begin{align}
	&D_{\mu\mu}(p) = (E(p) + n^2F(p) + n^4G(p))\delta_{\mu\mu} \nonumber \\
	&D_{\mu\nu}(p) = n^2H(p) + 2n^4I(p),~\mu\neq\nu 
\end{align}
and in this case we can only extract two form factors, for the diagonal and off-diagonal terms. These are extracted with
\begin{align} 
	&E(p) + n^2F(p) + n^4G(p) = \frac{1}{N_d}\sum_{\mu}D_{\mu\mu}(p), \\
	&n^2H(p) + 2n^4I(p) = \frac{1}{N_d(N_d - 1)}\sum_{\mu\neq\nu}D_{\mu\nu}(p).
\end{align}
The mixed configurations, $(n,n,0,0)$ and $(n,n,n,0)$ have non-diagonal terms and the gluon propagator reads
\begin{align}
	&D_{\mu\mu}(p) = E(p)\delta_{\mu\mu} + (F(p) + n^2G(p))p_\mu^2 \nonumber \\
	&D_{\mu\nu}(p) = (H(p) + 2I(p)n^2)p_\mu p_\nu,~\mu\neq\nu. 
\end{align}
For these configurations we consider the parameter $k$ representing the number of non-vanishing components.
The contractions of tensor basis elements are summarized by
\begin{align}
A_{\text{diag}}^{-1} = \mqty(N_d & kn^2 \\ kn^2 & kn^4), && A_{\text{Off-diag}}^{-1} = k(k-1)n^4
\end{align} 
with corresponding inverses
\begin{align}
A_{\text{diag}} = \frac{1}{kn^4(N_d-k)}\mqty(kn^4 & -kn^2 \\ -kn^2 & Nd), && A_{\text{Off-diag}} = \frac{1}{k(k-1)n^4}
\end{align} 
With this, the form factors follow easily
\begin{align}
	&E(p^2) = \frac{1}{kn^4(N_d-k)}\sum_{\mu}D_{\mu\mu}(p)\left(kn^4\delta_{\mu\mu} - kn^2p_\mu^2\right) \\
	&F(p^2)+n^2G(p^2) = \frac{1}{kn^4(N_d-k)}\sum_{\mu}D_{\mu\mu}(p)\left(-kn^2\delta_{\mu\mu} + N_d p_\mu^2\right) \\
	&H(p^2)+2n^2I(p^2) = \frac{1}{k(k-1)n^4}\sum_{\mu \neq \nu}D_{\mu\nu}(p)p_\mu p_\nu.
	\label{appen_eq:mixed_formfactors}
\end{align}
Lastly, for on-axis momenta, $(n,0,0,0)$, only diagonal terms survive
\begin{equation}
	D_{\mu\mu}(p) = E(p) + (F(p) + n^2G(p))p_\mu^2,
\end{equation}
and the form factors are extracted with
\begin{align}
	&E(p) = \frac{1}{3}\sum_{\mu\neq 1}D_{\mu\mu}(p), \label{appen_eq:on axis E} \\
	&n^2F(p) + n^4G(p) = D_{11}(p) - E(p).
	\label{appen_eq:on axis F}
\end{align}

\chapter{Results -- Additional figures}

\section{Gluon propagator}\label{appen:extra gluon propagator}

\subsection{Continuum relations -- mixed diagonal configurations}\label{appen_sec:mixed_continuum_relations}
 
In this section the continuum relations for the momentum configurations $(n,n,n,0)$ and $(n,n,0,0)$ are computed. The procedure follows similarly as the other two diagonal kinematics.
For both cases the lattice gluon propagator reads
\begin{align}
	&D_{\mu\mu}= E(p^2)\delta_{\mu\mu} + (F(p^2) + n^2G(p^2))p_\mu^2 \nonumber \\
	&D_{\mu\nu} = (H(p^2) + 2n^2I(p^2))p_\mu p_\nu, ~ \mu\neq\nu.
\end{align}
Using the extraction for the form factors built in \cref{append_sec:projectors_latticebasis} and also the continuum parametrization
\begin{equation*}
	D_{\mu\nu}^c(p) = D(p^2)\left(\delta_{\mu\nu} - \frac{p_\mu p_\nu}{p^2}\right)
\end{equation*}
the proof of the continuum relations is follows simply, 
\begin{alignat*}{2}
	E(p^2) &= \frac{D(p^2)}{kn^4(N_d-k)}\sum_{\mu}\left(\delta_{\mu\mu} - \frac{p_\mu p_\nu}{p^2}\right)\left(kn^4\delta_{\mu\mu} - kn^2p_\mu^2\right) \\
	&=\frac{D(p^2)}{kn^4(N_d-k)}\sum_{\mu}\left(kn^4 - kn^2p_\mu^2 - \frac{kn^4}{p^2}p_\mu^2 + \frac{kn^2}{p^2}p_\mu^4\right) \\
	&=D(p^2) \\
F(p^2) + n^2G(p^2) &= \frac{D(p^2)}{kn^4(N_d-k)}\sum_{\mu}\left(\delta_{\mu\mu} - \frac{p_\mu p_\nu}{p^2}\right)\left(-kn^2\delta_{\mu\mu} + N_d p_\mu^2\right) \\
&= \frac{D(p^2)}{kn^4(N_d-k)}\sum_{\mu}\left(-kn^2 + N_dp_\mu^2 -\frac{kn^2}{p^2}p_\mu^2 + \frac{N_d}{p^2}p_\mu^2\right) \\
&=-\frac{D(p^2)}{p^2} \\
	H(p^2)+2n^2I(p^2) &=\frac{D(p^2)}{k(k-1)n^4}\sum_{\mu \neq \nu}\frac{-p_\mu p_\nu}{p^2}p_\mu p_\nu \\
	&=-\frac{D(p^2)}{p^2}.
\end{alignat*}
Notice that $p^2 = kn^2$ with the parameter $k$ defined in \cref{append_sec:projectors_latticebasis} and $N_d=4$ the dimensionality of the lattice.
In addition, this result is independent of the use o lattice or improved momentum.

\begin{figure}[htb!]
	\centering
	\input{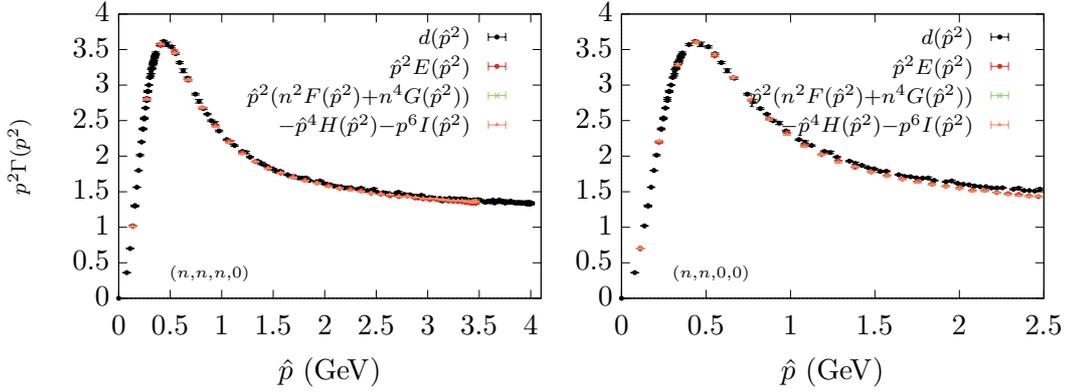}
	\caption{Form factors from the lattice basis for the mixed configurations $p=(n,n,n,0)$ (left) and for $p=(n,n,0,0)$ (right) both as a function of improved momentum. Shown for comparison is the benchmark result $d(\hat p^2)$.}
	\label{fig:continuum_relations_diag_mixed}
\end{figure}

The analysis of the continuum relations for these two configurations is seen in \cref{fig:continuum_relations_diag_mixed}. The continuum relations are exactly satisfied among all three form factors for both configurations. The benchmark result was shown for comparison, and it is noticeable that the further from the diagonal, the worse the correspondence becomes. The configuration $(n,n,0,0)$ deviates from the gluon propagator dressing function for higher momentum, while the result for $(n,n,n,0)$ remains compatible through all range of momenta similarly to the full diagonal momenta.

\end{appendices}

\end{document}